\documentclass[
 longbibliography,
 twocolumn,
 superscriptaddress,
 amsmath,
 amssymb,
 amsthm,
 amsfonts,
 floatfix
]{revtex4-2}

\usepackage[colorlinks=true,urlcolor=blue,citecolor=blue,linkcolor=blue]{hyperref} 
\usepackage{graphicx}
\usepackage{dcolumn}
\usepackage{kotex}
\usepackage{natbib}
\usepackage{color}
\usepackage{amsfonts}
\usepackage{hyperref}
\usepackage{bm}

\usepackage{tabularx}
\usepackage{multirow}

\usepackage{etoolbox}
\makeatletter
\patchcmd{\@makecaption}
  {\small\rmfamily}
  {\footnotesize\rmfamily}
  {}{}
\makeatother

\usepackage{romannum}

\begin{document}

\title{Inferring Stochastic Dynamics Reveals Atypicality and Irreversibility in Social Change}

\author{Youngkyoung Bae}
\email{tardis\_95@snu.ac.kr}
\affiliation{Department of Physics and Astronomy \& Center for Theoretical Physics, Seoul National University, Seoul 08826, Korea}

\author{Hajime Shimao}
\affiliation{Great Valley School of Professional Studies, Penn State, Malvern, PA 19355, USA}

\author{Seungwoong Ha}
\affiliation{University of Virginia, Charlottesville, VA 22903, USA}

\author{Shuo (Luna) Yang}
\affiliation{Department of Business and Economics, Pennsylvania State University Brandywine Campus, Media, PA 19063, USA}

\author{David Wolpert}
\email{david.h.wolpert@gmail.com}
\affiliation{Santa Fe Institute, Santa Fe, NM 87501, USA}
\affiliation{Complexity Science Hub, Vienna}
\affiliation{Arizona State University, Tempe, AZ, USA}
\affiliation{International Center for Theoretical Physics, Trieste, Italy}

\begin{abstract}
Human social systems bear striking similarities, yet similar conditions can give rise to markedly different historical trajectories.
A central challenge for quantitative social science is therefore to use time-series data to identify features common to the evolution of different social systems without obscuring the variation distinguishing individual histories.
Here, we address this challenge by fitting stochastic differential equations (SDEs) to ensembles of observed trajectories through spaces of socially relevant variables. 
Importantly, the SDE allows the average tendency and variability of change to depend on the current state of the system.
This flexibility allows us to use the SDE to analyze myriad aspects of social dynamics.
Specifically, the fitted SDEs identify regions of the state space where stochastic variability is greater, allowing a broader range of plausible subsequent changes.
The fitted SDE also allows us to quantify the dynamical irreversibility of observed trajectories and how that irreversibility depends on the values of the social variables.
In addition, we show how the SDEs can be used to identify statistically atypical transitions in individual trajectories, indicative of exogenous perturbations impacting the associated society. 
We apply these tools to three domains: modern political economy, quantitative history, and cultural analytics. 
This reveals state-dependent change in the organization of modern political-economic systems; a broad but stochastic tendency toward increasing social complexity in historical societies; and time-dependent reorganization of lexical dynamics in English.
Looking forward, we argue that our framework provides a set of powerful quantitative tools that can be used to investigate a broader range of longitudinal social phenomena.
\end{abstract}

\maketitle

Whether human societies exhibit common patterns in how they evolve is a longstanding question across the social sciences~\cite{goldstone1998initial, mcallister2002historical}.
Historical and sociological scholarship has examined how institutions, social relations, and material conditions shape everyday economic activity, political decision-making, and collective protest, and how these actions can, in turn, reproduce or transform those conditions~\cite{north1990institutions, sewell1992theory, ocasio2023institutions}.
The consequences of these actions and of the events to which individuals and groups respond depend on the circumstances in which they occur and on their place within an unfolding historical sequence~\cite{sewell1996three, mahoney2000path}.
Their effects can extend beyond the immediate episode, reinforcing established paths or reshaping the conditions that influence subsequent decisions and outcomes~\cite{david1985clio, pierson2000increasing, capoccia2007study}.
These processes help explain why even societies that appear similar in their institutional or material conditions may follow divergent historical trajectories~\cite{ermakoff2015structure}.
A central task for quantitative analysis is therefore to identify and characterize common dynamical regularities while accounting for the distinct sequences of change realized in individual histories.

The growing availability of digitized historical and cultural records makes this question increasingly amenable to empirical investigation~\cite{ruggles2021revival}.
Longitudinal records of institutions, economies, language, and social organization allow aspects of social change to be represented as measurable trajectories~\cite{michel2011quantitative, turchin2008arise, turchin2015seshat, diebolt2016handbook, jackson2024worldwide}.
Conventional analyses focused on cross-case associations, average effects, or aggregate temporal trends can reveal important empirical patterns~\cite{wooldridge2001econometric, gangl2010causal}.
However, such estimates and summaries do not by themselves capture the sequence of changes within individual histories~\cite{abbott2001time, liao2022sequence}, nor do they specify the state-dependent distribution of possible transitions from which alternative trajectories arise.
Accounting for both common dynamical tendencies and diverse historical trajectories requires characterizing how a system tends to change from a given state and the variability among possible transitions.
Stochastic models provide a natural basis for this description by treating observed trajectories as realizations of a stochastic process whose transition distributions depend on the evolving state of the system~\cite{helbing2010quantitative, wolpert2024past}.

Beyond characterizing common dynamics, understanding social change also requires examining the distinctive features of individual histories.
For example, major disruptions such as financial crises, pandemics, and environmental disasters have been followed by markedly different social responses~\cite{santana2023covid, mcConnell2024rare, lee2025proximity, huang2025built}.
These observations motivate assessing how unusual particular transitions are relative to the range of plausible changes and how such departures appear across successive observations.
A further dimension concerns the temporal asymmetry of historical trajectories, which reveals whether they exhibit a statistical arrow of social change and how this directionality varies over time.
Studies in statistical physics have demonstrated the value of quantifying such asymmetry and related signatures in nonequilibrium dynamics~\cite{skinner2021improved, lynn2021broken, terlizzi2024variance, ishihara2025statespace}.
These aspects of social change call for additional quantitative diagnostics that assess individual histories.

In this paper, we infer stochastic differential equations (SDEs) from ensembles of observed social trajectories to construct a probabilistic description of change within each empirical domain.
The inferred dynamics are characterized by two fields: a drift field describing the conditional mean tendency of change and a diffusion field describing the magnitude and covariance structure of stochastic variability.
We treat each fitted SDE as a \textit{reference process}, a probabilistic baseline that defines distributions over plausible transitions and trajectories.
Relative to this baseline, we assess the atypicality of observed transitions, characterize their immediate patterns across successive observations, and quantify dynamical irreversibility at both transition and trajectory levels.
This approach connects the dynamics of typical change with the statistical features of individual histories within a common probabilistic framework.

We apply our framework across three empirical domains: modern political economy (Democracy--Inequality--Growth, DIG~\cite{bolt2024maddison, coppedge2025vdem, alvaredo2025wid}), quantitative history (Seshat~\cite{turchin2015seshat}), and cultural analytics (Google Books Ngrams~\cite{scheffer2021rise}).
Across these applications, the inferred SDEs reveal state-dependent organization of political-economic tendencies and stochastic variability, a broad tendency toward increasing social complexity despite substantial fluctuations, and a time-dependent reorganization of lexical dynamics around 1980.
At the trajectory level, we find that atypical transitions exhibit different short-term patterns and that dynamical irreversibility varies across individual histories and over time.

\section*{Results}

\subsection*{A Stochastic Framework for Social Dynamics}
\label{sec:framework}

\paragraph*{\textbf{State-space construction.}}

\begin{figure}[!b]
  \centering
  \includegraphics[width=\linewidth]{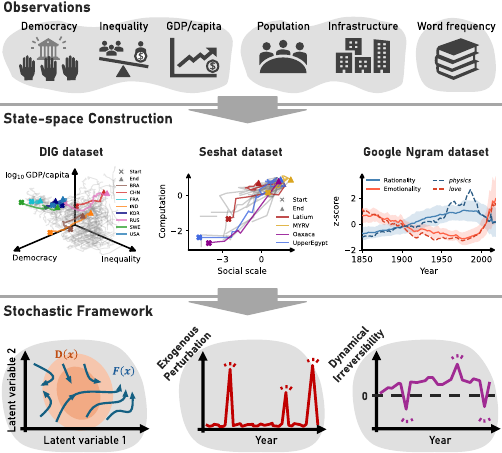}
  \caption{
  Conceptual overview of our framework for analyzing social processes.
  \textbf{top} Observed social indicators, such as democracy, population, and word frequency, are collected across multiple units, including countries, historical regions, and lexical items.
  \textbf{middle} Dataset-specific transformations map these indicators to low-dimensional state trajectories.
  \textbf{bottom} The dynamics of these state variables are modeled by an SDE characterized by a drift field $\bm{F}(\bm{x})$ (blue arrows) and a diffusion field $\mathbf{D}(\bm{x})$ (orange ellipses).
  The inferred SDE serves as a reference process for identifying exogenous perturbations and quantifying the dynamical irreversibility of observed trajectories.
}
\label{fig:fig1}
\end{figure}

We begin by representing each dataset as an ensemble of trajectories in a low-dimensional state space (Fig.~\ref{fig:fig1}).
Many dimensions of social systems, such as democracy and social complexity, are theoretical constructs that cannot be observed directly but are instead represented using multiple empirical indicators~\cite{blalock1974measurement, treier2008democracy, turchin2018quantitative}.
Treating all indicators as separate state coordinates would yield a high-dimensional state space that is difficult to interpret and would require substantially more data to characterize how transitions depend on the current state.
We construct a substantively interpretable state vector $\bm{x}_t \in \mathbb{R}^d$ using dataset-specific combinations of dimensionality reduction, theoretically guided aggregation, and standardization.
This construction yields a coarse-grained state space in which the continuous-time stochastic dynamics are modeled.
Details of the state-space construction for each application are provided in the Methods and Supplementary Information (SI).

\paragraph*{\textbf{SDE modeling.}}
We next treat these state trajectories as individual realizations of a common stochastic process within each application.
Let $i=1,\dots,N$ index the units in each dataset, such as countries, historical regions, or lexical items.
Although empirical observations are discrete in time, we model the state of unit $i$ at continuous time $t$, denoted by $\bm{x}_t^{(i)}$, as evolving according to an It\^o SDE~\cite{gardiner2004handbook}:
\begin{equation}
  d\bm{x}_t^{(i)} = \bm{F}(\bm{x}_t^{(i)}) dt + \sqrt{2 \mathbf{D}(\bm{x}_t^{(i)})}\,d \bm{W}_t^{(i)},
\label{eq:sde}
\end{equation}
where $d\bm{x}_t^{(i)}$ denotes the infinitesimal state increment of unit $i$, and $\{\bm{W}_t^{(i)}\}_{i=1}^N$ are independent standard $d$-dimensional Wiener processes.
This formulation characterizes state changes through two quantities: the drift field $\bm{F}: \mathbb{R}^d \to \mathbb{R}^d$ representing the conditional mean rate of change, and the state-dependent diffusion matrix $\mathbf{D}: \mathbb{R}^d \to \mathbb{R}^{d \times d}$ characterizing the magnitude and covariance structure of stochastic fluctuations.
The stochastic term provides an effective coarse-grained representation of variability not resolved by the constructed state variables, as can arise from the aggregate effect of many rapidly varying contributions~\cite{zwanzig2001nonequilibrium, van2007stochastic, helbing2010quantitative}.
Equation~\eqref{eq:sde} defines the time-homogeneous form used for DIG and Seshat, with explicit calendar-time dependence introduced later in the cultural-analytics application.
For the main analyses, we infer $\bm{F}$ and $\mathbf{D}$ using Langevin Bayesian networks (LBN)~\cite{bae2025inferring}; the fitted SDE then serves as the reference process for the analyses below (see Methods and SI for inference details).
For notational brevity, we hereafter omit the unit index $i$.

\paragraph*{\textbf{Exogenous Perturbations.}}

Observed trajectories can contain transitions that fall in low-probability regions of the corresponding distribution.
These transitions may coincide with major disruptions, such as wars, revolutions, financial crises, or pandemics, but can also reflect changes in the transition dynamics not captured by the fitted model or limitations of the data.
We use the term \textit{exogenous perturbations} operationally for these statistically atypical transitions. 
In this sense, exogeneity is model-relative: it denotes a statistical departure from what is typical under the reference process and does not identify an external causal origin.

To evaluate such departures along individual trajectories, consider three consecutive observations $\bm{x}_{-}\equiv\bm{x}_{t-\Delta t_-}$, $\bm{x}_t$, and $\bm{x}_{+}\equiv\bm{x}_{t+\Delta t_+}$, where the adjacent observation intervals $\Delta t_-$ and $\Delta t_+$ may differ.
We define three tail probabilities from these observations.
The local probability $p^{\rm loc}_t$ evaluates the incoming transition $\bm{x}_{-}\rightarrow\bm{x}_t$ under the one-step transition distribution from $\bm{x}_{-}$.
The bridge probability $p^{\rm brg}_t$ evaluates the observed intermediate state $\bm{x}_t$ under the bridge distribution conditioned on $\bm{x}_{-}$ and $\bm{x}_{+}$.
The endpoint probability $p^{\rm end}_t$ evaluates $\bm{x}_{+}$ under the two-step predictive distribution from $\bm{x}_{-}$ after marginalizing over the intermediate state.
We transform each tail probability into an exogeneity score,
\begin{equation}
\mathcal{E}^{\alpha}_t
\equiv -\log_{10} p^{\alpha}_t,
\qquad
\alpha \in \{\mathrm{loc},\mathrm{brg},\mathrm{end}\},
\end{equation}
where larger values indicate greater statistical atypicality under the reference process.
For visualizations and aggregate summaries, we cap the scores at $6$, using $\min(\mathcal{E}^{\alpha}_t,6)$, to limit the influence of a small number of extremely small tail probabilities.

These scores support a two-stage diagnostic workflow: first detecting exogenous perturbations and then characterizing their short-term patterns.
Detection is based on the local score, with a high $\mathcal{E}^{\rm loc}_t$ indicating an incoming transition that is statistically atypical under the reference process.
For detected transitions, the joint pattern of the bridge and endpoint scores characterizes whether the departure is localized to the intermediate state, recovered over the two-step interval, persists to the next endpoint, or is atypical in both diagnostics.
Their relative values organize these cases into four qualitative categories: localized, recovered, persistent, and off-path, as summarized in Table~\ref{tab:perturbation_category}.
We derive the score geometry in the local Gaussian limit and verify the relationships in a nonlinear stochastic system with controlled perturbations (Sec.~\Romannum{4} in the SI).

\begin{table}[!t]
\vskip -0.05in
\centering
\caption{Qualitative categories of exogenous perturbation based on bridge and endpoint exogeneity scores. Low/moderate and high indicate qualitative score regimes without fixed thresholds.}
\setlength{\tabcolsep}{8pt}
\begin{tabular}{c|c|c}
    \hline\
    Category & Bridge score $\mathcal{E}^{\rm brg}$ & Endpoint score $\mathcal{E}^{\rm end}$ 
    \\ \hline\hline
    Localized & Low/Moderate & Low/Moderate \\
    Recovered  & High & Low \\
    Persistent & Low & High \\
    Off-path & High  & High \\ \hline
\end{tabular}
\vskip -0.1in
\label{tab:perturbation_category}
\end{table}

\begin{figure*}[!ht]
  \vskip -0.1in
  \centering
  \includegraphics[width=\linewidth]{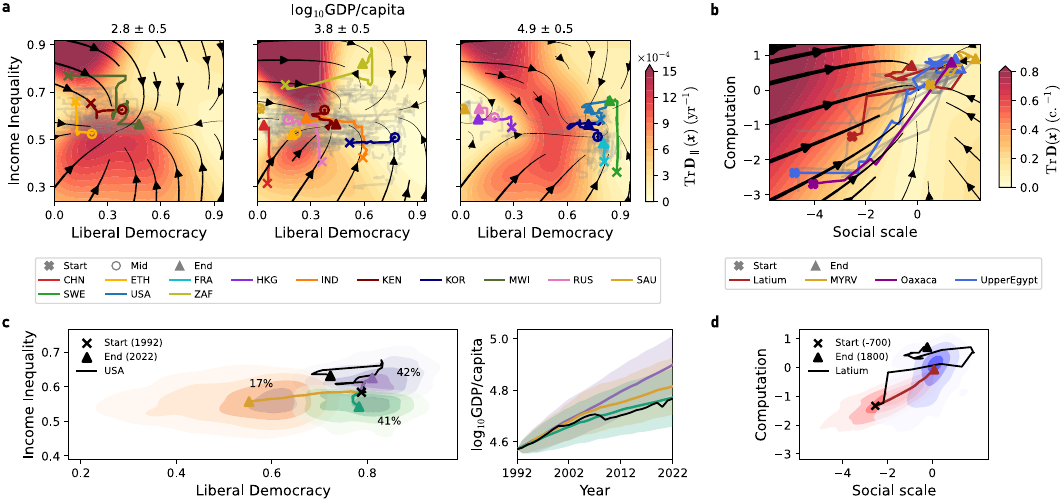}
  \caption{Inferred stochastic dynamics and empirical trajectories in social state spaces.
  \textbf{a} Modern political-economic dynamics on the Democracy--Inequality plane at three levels of $\log_{10}$ GDP/capita; the $\pm 0.5$ windows around these levels are used only to select empirical trajectories. 
  For visualization, PCA coordinates are linearly mapped back to the original indicator scales.
  \textbf{b} Historical dynamics on the Scale--Computation plane.
  In \textbf{a,b}, black streamlines show the projected drift field $\bm{F}_\parallel (\bm{x})$, and background color denotes local projected diffusion magnitude $\mathrm{Tr}\,\mathbf{D}_\parallel(\bm{x})$, where `c.' denotes one century.
  Colored trajectories show selected units' paths, gray trajectories show the remaining paths in the dataset, and markers indicate the initial, intermediate, and final states.
  \textbf{c,d} Stochastic simulations for \textbf{c} the USA and \textbf{d} Latium, initialized at the observed 1992 and 700 BCE states, respectively. 
  For each case, 1,000 trajectories were generated from the inferred SDE over the corresponding empirical time horizon; black lines show the observed trajectories.
  In \textbf{c}, final simulated states are partitioned into three terminal groups for visualization; shaded contours show group-specific kernel-density estimates, percentages indicate group fractions, and colored lines denote group-mean trajectories.
  In \textbf{d}, shaded contours show ensemble snapshots at 500 BCE (red) and 1000 CE (blue), and the colored line denotes the ensemble-mean trajectory.
  }
  \label{fig:fig2}
\end{figure*}

\paragraph*{\textbf{Dynamical irreversibility.}}
Social trajectories often exhibit temporal directionality.
Whereas exogeneity quantifies how atypical an observed transition is, dynamical irreversibility asks whether the observed temporal ordering of a trajectory is more or less probable than its time reversal under the reference dynamics.
Such temporal asymmetry may accompany long-run patterns such as economic development or increasing social complexity.
To quantify this temporal asymmetry, we adapt the concept of entropy production from stochastic thermodynamics~\cite{parrondo2009entropy, seifert2012stochastic}.
For an observed trajectory $\Gamma \equiv \{ \bm{x}_{t_0}, \dots, \bm{x}_{t_L} \}$, we define the dynamical irreversibility $\Sigma[\Gamma] \in \mathbb{R}$ as the log-ratio between the probability of the observed path and that of its time-reversed counterpart $\tilde{\Gamma} = \{ \bm{x}_{t_L}, \dots, \bm{x}_{t_0} \}$:
\begin{equation}
\begin{aligned}
  \Sigma\left[ \Gamma \right] &\equiv \ln \frac{\mathcal{P}\left[ \Gamma \mid \bm{x}_{t_0} \right]}{\mathcal{P} [ \tilde{\Gamma} \mid \bm{x}_{t_L}]} \equiv \sum_{k=0}^{L-1} \sigma_{t_{k+1}},
\label{eq:cum_irreversibility}
\end{aligned}
\end{equation}
where $\mathcal{P}[\Gamma \mid \bm{x}_{t_0}]$ denotes the conditional path density of $\Gamma$, and $\sigma_{t_{k+1}}$ is the local irreversibility associated with the transition $\bm{x}_{t_k}\rightarrow\bm{x}_{t_{k+1}}$.
This log-ratio provides a signed pathwise measure of time-reversal asymmetry: $\Sigma[ \Gamma ] > 0$ indicates that the observed temporal ordering is more probable than its reversal, whereas $\Sigma[ \Gamma ] < 0$ indicates that the reversed ordering is more probable. Its magnitude quantifies the degree of asymmetry between the two orderings.
The partial sum through $t_m$ defines the cumulative irreversibility $\Sigma_{t_m}$, which tracks how this temporal asymmetry accumulates along the trajectory.
A systematic tendency for $\Sigma_{t_m}$ to increase over time across trajectories indicates that the observed temporal ordering is persistently favored over its reversal, providing evidence of a statistical arrow of social change.
The local irreversibility $\sigma_{t_{k+1}}$ resolves the path-level asymmetry into transition-level contributions and inherits the same sign interpretation.
Both $\Sigma$ and $\sigma$ characterize statistical temporal asymmetry only and do not assign normative value to the direction of social change.

\subsection*{Stochastic Dynamics of Social Change}

We first examine the fitted SDEs to characterize the conditional tendencies and stochastic variability that define typical change in each social state space.
We apply the time-homogeneous formulation to two empirical domains.
For modern political economy, we use the DIG dataset, which combines measures of democracy, economic inequality, and GDP per capita for 148 country- or territory-level units during the post-Cold War period (1992--2022).
Hereafter, we use ``country'' as shorthand for country- or territory-level units treated separately in the source datasets.
For quantitative history, we use the Seshat database to construct a two-dimensional state space, with social scale capturing polity size in demographic and territorial terms, and computation representing collective information-processing capacity.
This yields 25 trajectories of Natural Geographic Areas (NGAs), spanning 3800 BCE to 1800 CE.

The inferred DIG dynamics are organized differently across levels of log GDP per capita (Fig.~\ref{fig:fig2}a).
At low and intermediate GDP levels, the projected flow converges toward a region of moderate democracy and relatively high inequality.
Within these slices, the projected dynamics in the high-democracy, high-inequality region favor movement toward lower democracy, consistent with cross-national evidence linking economic inequality to weaker democratic consolidation and democratic erosion~\cite{houle2009inequality, rau2025income}.
At higher GDP levels, however, the organization of the flow changes qualitatively, with locally convergent low- and high-democracy configurations separated by an intermediate-democracy region.
This contrast in dynamical organization qualitatively parallels comparative evidence that semi-democratic regimes are less durable than both democracies and autocracies~\cite{knutsen2015institutional}. 
Beyond these conditional mean tendencies, the diffusion field reveals how stochastic variability is structured across the political-economic state space.
Across GDP levels, the diffusion field is anisotropic, with the major axes of the diffusion ellipses predominantly aligned with the democracy dimension (Fig.~S12).
Its magnitude is also state-dependent, with stochastic variability elevated in the intermediate-democracy region at high GDP (Fig.~\ref{fig:fig2}a).
The combination of elevated variability and comparatively weak drift is reflected in a particularly low drift-to-diffusion ratio, indicating a broad range of possible local changes relative to the average directional tendency (Fig.~S10).

The Seshat dynamics provide a complementary case, connecting broad patterns reported in previous analyses to a local stochastic dynamical description (Fig.~\ref{fig:fig2}b).
In the Scale--Computation plane, the inferred drift shows an overall tendency from low-scale, low-computation states toward greater scale and computation.
This broad direction aligns with the dominant axis of covariation identified in prior Seshat analyses~\cite{turchin2018quantitative}, while the drift field resolves how the conditional direction of change varies locally across the state space.
At the trajectory level, earlier Seshat analyses found that trajectories of individual NGAs become more dispersed in an intermediate region of social complexity and reconverge at higher levels~\cite{shin2020scale}.
The inferred diffusion adds a local dynamical perspective by quantifying stochastic variability conditional on the current state. 
This variability is elevated across much of the low-scale region and is lower toward the high-scale, high-computation region.

To illustrate how the inferred dynamics translate into distributions over trajectories, we simulated stochastic trajectories from the fitted SDEs initialized at observed states (Fig.~\ref{fig:fig2}c,d).
For the United States (USA), simulations starting from the 1992 state generate a broad ensemble of political-economic trajectories through 2022, with terminal states spread across a wide region of the Democracy--Inequality plane.
This diversity is consistent with the differentiated organization of the inferred high-GDP dynamics shown in Fig.~\ref{fig:fig2}a.
For Latium, a Seshat NGA in central Italy, simulations starting from the 700 BCE state show a broad shift toward larger scales and greater computation, with the ensemble mean following the observed long-run direction and individual trajectories retaining substantial stochastic dispersion.
Across both domains, these simulations show how a common reference process gives rise to a distribution of possible trajectories, linking shared dynamical structure to variation among individual paths.

\begin{figure}[!b]
  \centering
  \includegraphics[width=\linewidth]{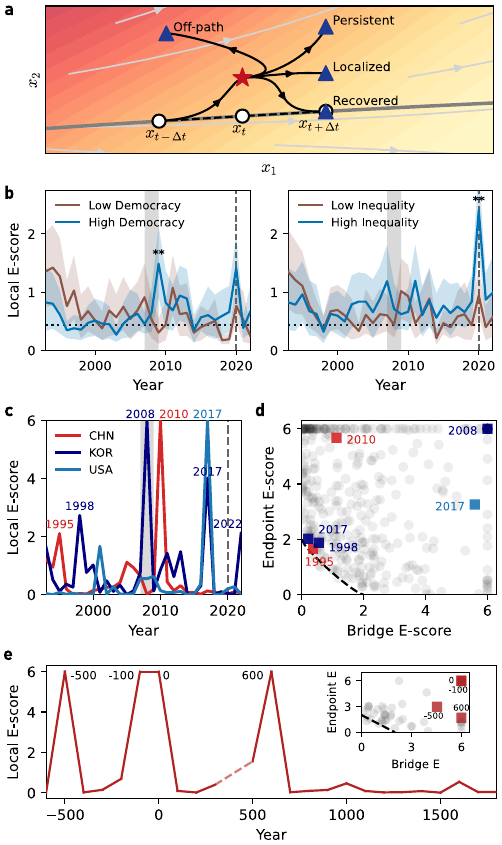}
  \caption{
  \textbf{a} Schematic of exogenous perturbation categories in the bridge--endpoint plane: localized, recovered, persistent, and off-path.
  \textbf{b} Annual local exogeneity $\mathcal{E}^{\rm loc}_t$ in the DIG dataset, stratified by democracy and inequality quartiles.
  Lines show yearly country means for the lowest and highest quartiles, and shaded bands denote pointwise 95\% country-cluster bootstrap confidence intervals.
  Gray shading marks the Great Recession, the dashed vertical line marks the onset of the COVID-19 pandemic, and asterisks indicate Bonferroni-significant quartile differences ($^{**}p_{\rm Bonf}<0.01$).
  The dotted horizontal line denotes the nominal expectation $1/\ln 10$.
  \textbf{c} $\mathcal{E}^{\rm loc}_t$ for CHN, KOR, and USA. 
  \textbf{d} $\mathcal{E}^{\rm brg}$ versus $\mathcal{E}^{\rm end}$ for DIG peak observations, with gray points denoting all country-year peaks and colored squares marking the labeled peaks in \textbf{c}.
  \textbf{e} $\mathcal{E}^{\rm loc}_t$ for Latium in the Seshat dataset. The dashed segment denotes a transition spanning more than a century (300--500 CE).
  The inset shows $\mathcal{E}^{\rm brg}$ and $\mathcal{E}^{\rm end}$ for Seshat peak observations, with Latium peaks highlighted.
  Peaks in \textbf{c}--\textbf{e} are defined by $\mathcal{E}^{\rm loc}\geq2$ for visualization. The 2022 KOR peak in \textbf{c} is omitted from \textbf{d} because no subsequent observation is available.
  The dashed black curves in \textbf{d} and \textbf{e} denote the lower bound implied by the local-score threshold of $2$.
  }
  \label{fig:fig3}
\end{figure}

\subsection*{Exogeneity and Perturbation Patterns}

Having characterized the reference dynamics, we next examine where observed trajectories depart from them.
The local exogeneity score $\mathcal{E}^{\rm loc}_t$ identifies statistically atypical transitions without pre-specifying event dates.
For such transitions, the bridge and endpoint scores, $\mathcal{E}^{\rm brg}_t$ and $\mathcal{E}^{\rm end}_t$, reveal their patterns across successive observations, which we describe using four qualitative categories: localized, recovered, persistent, and off-path (Fig.~\ref{fig:fig3}a).

Applying these diagnostics to the DIG trajectories reveals that local exogeneity is temporally episodic and varies across political-economic strata (Fig.~\ref{fig:fig3}b).
Stratum-specific annual means of $\mathcal{E}^{\rm loc}$ show pronounced elevations around major global disruptions, including the Great Recession (2007--2009) and the onset of the COVID-19 pandemic (2020).
Notably, local exogeneity is significantly higher among high-democracy than low-democracy countries in 2009 and among high-inequality than low-inequality countries in 2020; both contrasts remain significant with $p_{\rm Bonf}<0.01$. 
A pronounced elevation is also observed in the high-democracy stratum in 2020.
Bridge--endpoint profiles further distinguish the temporal signatures of these departures.
In 2009, the high-democracy stratum had mean $(\mathcal{E}^{\rm brg},\mathcal{E}^{\rm end})=(1.05, 1.36)$, with atypicality remaining elevated at the subsequent endpoint.
In 2020, the high-democracy and high-inequality strata showed different aggregate bridge--endpoint profiles despite elevated local exogeneity in both.
The high-democracy stratum showed $(1.29,0.78)$, with lower atypicality at the subsequent endpoint than at the intermediate state.
By contrast, the high-inequality stratum showed $(1.78, 1.73)$, with atypicality remaining elevated in both diagnostics.
These results show that major disruptions can differ both in the strata where atypicality is most pronounced and in the immediate temporal profiles of those departures.

At the country level, the same diagnostics reveal country-specific episodes of atypical change that can be obscured in aggregate summaries (Fig.~\ref{fig:fig3}c,d). 
Across China (CHN), Korea (KOR), and the USA, local exogeneity shows pronounced peaks at different times. 
For example, the 1998 Korean peak coincides with the Asian financial crisis and the first opposition transfer of presidential power; the 2010 Chinese peak corresponds to a transition toward higher inequality alongside rising GDP per capita; and the 2017 USA peak corresponds to a transition marked by a decline along the democracy dimension (Fig.~S15).
These peaks occupy different regions of the bridge--endpoint score plane (Fig.~\ref{fig:fig3}d). 
The 1995 Chinese peak and the 1998 and 2017 Korean peaks show relatively localized profiles, whereas the 2010 Chinese peak is persistent-like and the 2008 Korean and 2017 USA peaks are off-path-like.
These contrasts highlight why local detection alone is insufficient to distinguish different forms of atypical change.

We apply the same diagnostics to sparse, long-run historical trajectories in Seshat (Fig.~\ref{fig:fig3}e). 
We illustrate this with Latium, where $\mathcal{E}^{\rm loc}$ identifies several episodes of perturbation over more than two millennia of historical change.
Despite their elevated local exogeneity, these transitions exhibit different bridge--endpoint profiles and coordinate-wise change.
The consecutive peaks around 100 BCE and the turn of the Common Era illustrate this contrast.
Both exhibit off-path-like profiles, but involve distinct directions of atypical displacement in the state space: the first shows a rapid increase in social scale, whereas the second is dominated by a sharp increase in computation with a comparatively smaller increase in social scale (Fig.~S19).
These two atypical transitions form a two-stage sequence in the state space, with rapid scale expansion followed by a marked increase in information-processing capacity during the transition from the late Republican to early imperial period~\cite{bowman1996provincial}.
By contrast, the 600 CE peak shows a recovery-like pattern, with high $\mathcal{E}^{\rm brg}$ but low $\mathcal{E}^{\rm end}$.
This recovery-like episode around 600 CE overlaps with the disruptions associated with Lombard expansion and the accompanying political fragmentation and reconfiguration of Italy~\cite{humphries2001italy}.

\begin{figure*}[!ht]
  \centering
  \includegraphics[width=\linewidth]{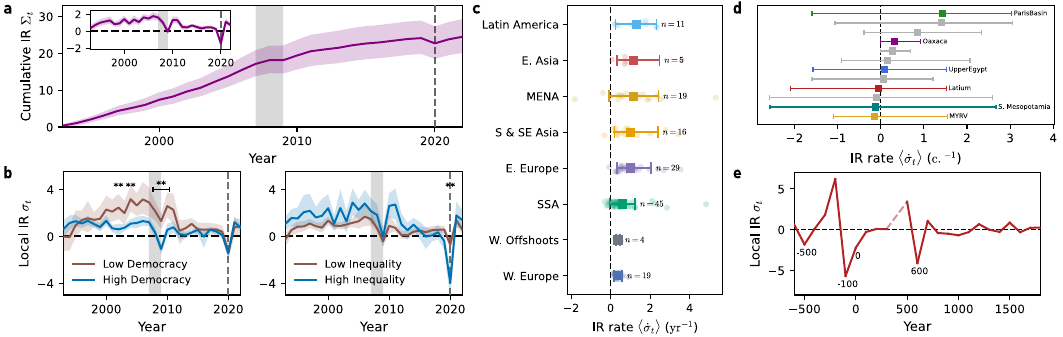}
  \caption{
    Irreversibility in \textbf{a--c} the DIG and \textbf{d,e} Seshat datasets.
    \textbf{a} Cumulative irreversibility (IR) $\Sigma_t$, with the annual local IR $\sigma_t$ shown in the inset.
    \textbf{b} Annual $\sigma_t$ stratified by democracy and inequality quartiles.
    Lines denote country means and shaded bands indicate pointwise 95\% country-cluster bootstrap confidence intervals.
    Gray shading marks the Great Recession, the dashed vertical line marks the COVID-19 pandemic, and asterisks denote Bonferroni-significant quartile differences ($^{**}p_{\rm Bonf}<0.01$).
    \textbf{c} Country-level mean IR rates grouped by world region.
    Pale circles denote individual countries, and $n$ gives the number of countries in each region.
    \textbf{d} NGA-level IR rates.
    Colored symbols highlight selected NGAs, and gray symbols denote the remaining NGAs.
    In \textbf{c,d}, squares denote mean rates and horizontal bars span the 10th--90th percentiles, computed across countries within each region in \textbf{c} and across transition-level rates within each NGA in \textbf{d}.
    Only NGAs with at least ten retained transitions are shown, using transitions with observation gaps of at most two centuries.
    \textbf{e} $\sigma_t$ along the Latium trajectory, with the dashed segment denoting the transition spanning 300--500 CE.
    Black dashed lines mark zero irreversibility.
}
\label{fig:fig4}
\end{figure*}

\subsection*{Statistical Arrow of Social Change}

We shift from transition atypicality to temporal asymmetry, asking whether observed social trajectories are more probable in their observed temporal order than when traversed in reverse under the reference dynamics.
In the DIG application, the cross-country mean of irreversibility $\Sigma_t$ increases overall over the observation period, indicating a sustained preference for the observed temporal ordering of political-economic trajectories (Fig.~\ref{fig:fig4}a). 
The cross-country mean of local irreversibility $\sigma_t$ remains predominantly positive but weakens around the Great Recession and turns negative at the onset of the COVID-19 pandemic, leading to a decline in the mean $\Sigma_t$. 
The subsequent return of mean $\sigma_t$ to positive values indicates that these negative contributions were temporary at the aggregate level. 
Thus, the longer-term statistical arrow remains evident despite temporary reversals in local temporal asymmetry.

Disaggregating this aggregate pattern by political-economic strata reveals substantial heterogeneity (Fig.~\ref{fig:fig4}b).
When stratified by democracy, the low-democracy quartile exhibits higher group-averaged $\sigma_t$ during much of the 2000s, whereas the high-democracy quartile shows a sharper negative excursion around the Great Recession. 
Inequality stratification reveals a different pattern: the two quartiles follow broadly similar trajectories over most years but diverge markedly in 2020, when $\sigma_t$ becomes substantially more negative among high-inequality countries.
These negative excursions parallel the exogeneity results (Fig.~\ref{fig:fig3}b), which showed higher local exogeneity among high-democracy countries during the Great Recession in 2009, and among high-inequality countries in 2020.
At the group level, these episodes therefore mark periods when observed changes were both atypical under the reference dynamics and, on average, ran counter to the locally favored temporal direction.

Geographic disaggregation of the DIG data further reveals significant variation in country-level mean irreversibility rates $\langle \dot{\sigma} \rangle$ across world regions (Welch's one-way ANOVA, $p<0.001$; Fig.~\ref{fig:fig4}c; see SI).
Region-level means are positive across all eight regions, indicating that the statistical arrow is not confined to a particular part of the world, although its strength varies across regions.
Simulations initialized at each country's observed starting state under the common inferred SDE yielded irreversibility rates positively associated with the observed rates across countries (Spearman's $\rho=0.54$) and retained a similar broad regional ordering after aggregation ($\rho=0.71$; Fig.~S16 in the SI).
Neither country nor region is used to condition the fitted dynamics. 
The correspondence therefore suggests that part of the geographic differentiation can emerge from different state-space positions under the same reference dynamics.
Substantial country-level variation within regions shows that regional differences coexist with pronounced heterogeneity among countries within the same region.

The Seshat trajectories show substantial heterogeneity in temporal directionality across the NGAs retained for comparative analysis (Fig.~\ref{fig:fig4}d).
NGA-level mean irreversibility rates $\langle \dot{\sigma} \rangle$ tend to be positive, although several NGAs lie near or below zero (Fig.~\ref{fig:fig4}d). 
For many NGAs, the within-NGA 10th--90th percentile ranges span both positive and negative values. 
Thus, a single long-run mean can mask substantial local temporal asymmetries of opposite sign.
Latium illustrates this distinction particularly clearly: its long-run mean rate lies near zero even though local irreversibility $\sigma_t$ takes pronounced positive and negative values (Fig.~\ref{fig:fig4}e).
Several pronounced irreversibility excursions coincide with the exogeneity peaks discussed above, but excursions around 200 BCE and 500 CE occur without similarly elevated local exogeneity (Figs.~\ref{fig:fig3}e and \ref{fig:fig4}e).
A decomposition by local exogeneity shows that in Latium, transitions with lower local exogeneity scores contribute positively to the long-run rate, whereas those with elevated scores contribute negatively and nearly offset this positive contribution.
Extending the same decomposition to the 12 retained NGAs, lower-exogeneity transitions contribute positively to the mean irreversibility rate in 11 cases.
By contrast, contributions from elevated-exogeneity transitions vary markedly in sign and magnitude, reinforcing the positive contribution in some trajectories and offsetting it in others (Fig.~S20 in the SI).
These results clarify the distinction between the two diagnostics, showing that elevated transition atypicality does not imply a particular direction or strength of local temporal asymmetry.

\begin{figure*}[!ht]
  \centering
  \includegraphics[width=\linewidth]{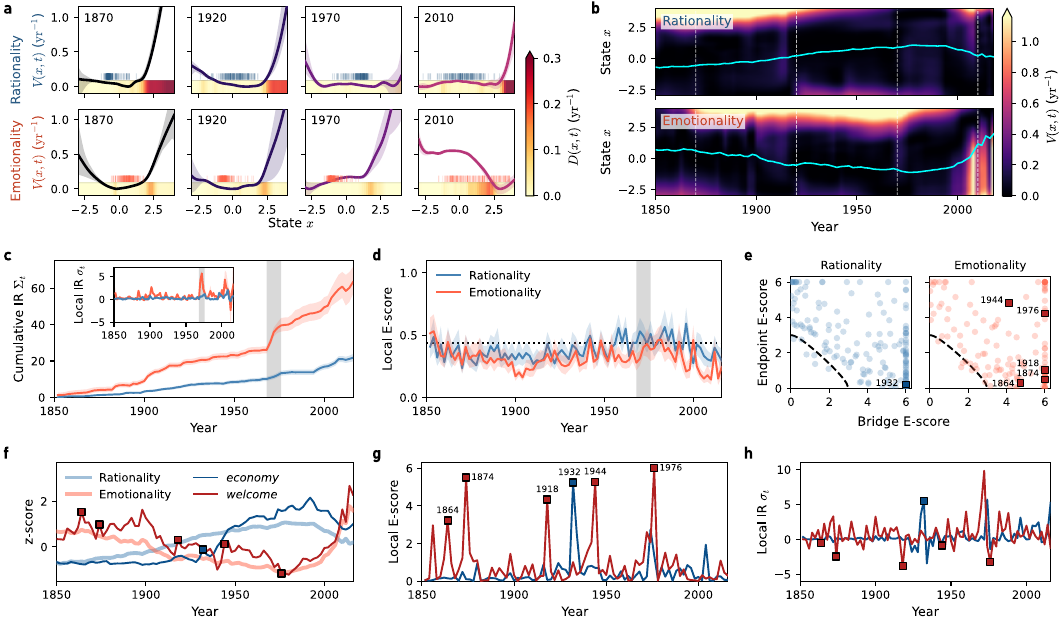}
  \vskip -0.1in
  \caption{
    Time-varying stochastic dynamics of lexical change.
    \textbf{a} Instantaneous drift potentials $V(x,t)$ for the Rationality (top) and Emotionality (bottom) lexicons at selected years.
    Curves and shading denote means and pointwise 95\% ensemble intervals, rugs indicate observed word states, and the lower strips show the inferred diffusion $D(x,t)$.
    \textbf{b} Time-varying potential landscapes with lexicon-mean trajectories (solid cyan); dashed vertical lines mark the snapshots in \textbf{a}.
    In \textbf{a,b}, the potentials are independently shifted at each time so that $\min_x V(x,t)=0$.
    \textbf{c} Lexicon-mean cumulative irreversibility $\Sigma_t$, with lexicon-mean local irreversibility $\sigma_t$ shown in the inset.
    \textbf{d} Lexicon-mean local exogeneity scores; the dotted horizontal line denotes the nominal expectation $1/\ln 10$.
    Shaded bands in \textbf{c,d} denote pointwise 95\% word-cluster bootstrap confidence intervals, and gray shading marks the pronounced increase in cumulative irreversibility during 1968--1976.
    \textbf{e} Bridge and endpoint exogeneity scores for peak transitions across all words; for visual clarity, only peaks with $\mathcal{E}^{\rm loc}_t \geq 3$ are shown.
    The dashed curve denotes the lower bound implied by the local-score threshold $3$.
    Squares mark highlighted transitions of \textit{economy} and \textit{welcome}.
    \textbf{f--h} Trajectories and diagnostics for the selected words: \textbf{f} standardized word-frequency trajectories with their respective lexicon means shown as pale curves, \textbf{g} local exogeneity scores, and \textbf{h} local irreversibility contributions.
    Squares identify the corresponding transitions across \textbf{e--h}.
}
\label{fig:fig5}
\end{figure*}

\subsection*{Time-varying Stochastic Dynamics of Lexical Change}

The preceding analyses used time-homogeneous reference dynamics.
More generally, however, the reference dynamics may vary with calendar time, so that the conditional tendencies and variability associated with a given state need not remain fixed.
We examine this possibility using lexical trajectories from the Google Books Ngram corpus spanning 1850--2019~\cite{michel2011quantitative}, for which dense temporal sampling allows explicit calendar-time dependence to be incorporated into the reference dynamics.
We infer an effective time-inhomogeneous SDE:
\begin{equation}
\begin{aligned}
    d\bm{x}_t = \bm{F}(\bm{x}_t,t)dt + \sqrt{2\mathbf{D}(\bm{x}_t,t)}\,d\bm{W}_t.
\end{aligned}
\end{equation}
Drawing on Refs.~\cite{hamilton2016diachronic,scheffer2021rise}, we analyze two balanced, embedding-derived lexicons, Rationality and Emotionality, each comprising 182 words.
Each word is represented by a one-dimensional z-scored frequency trajectory, so that $\bm{x}_t$ reduces to a scalar state $x_t$ in this application. 
The inferred dynamics vary explicitly with calendar time and are additionally conditioned on lexicon identity (Methods).

We visualize the inferred drift using an instantaneous potential $V(x,t)$, defined by $F(x,t)=-\partial_x V(x,t)$ (Fig.~\ref{fig:fig5}a,b).
The corresponding state- and time-dependent diffusion profiles are shown alongside the potentials in Fig.~\ref{fig:fig5}a.
Consistent with previous findings~\cite{scheffer2021rise}, the lexicon-mean trajectories evolve gradually until around 1980, with Rationality frequencies increasing and Emotionality frequencies decreasing; these trends reverse thereafter.
The inferred potential landscape changes systematically over calendar time, with earlier landscapes broadly favoring pre-1980 tendencies and later landscapes favoring lower-frequency states for Rationality words and higher-frequency states for Emotionality words.
Thus, the lexical reversal coincides with a temporal reorganization of the reference dynamics.
Ensemble simulations from the inferred time-varying dynamics reproduce these contrasting pre- and post-1980 trends, consistent with the fitted dynamics providing an effective description of lexical change (Fig.~S6). 
The diagnostics below therefore evaluate each transition relative to the contemporaneous reference dynamics.

We next apply the irreversibility and exogeneity diagnostics to these time-varying lexical dynamics.
Both lexicons exhibit increasing cumulative irreversibility, revealing sustained directional asymmetry in their long-run trajectories, with stronger accumulation for the Emotionality lexicon (Fig.~\ref{fig:fig5}c).
Cumulative irreversibility increases particularly rapidly during 1968--1976, near the broader period over which the lexicon-mean trajectories reverse.
Moreover, mean irreversibility rates during 1990--2016 are substantially higher than during 1900--1960, increasing from 0.277 to 0.694 per year for the Emotionality lexicon and from 0.103 to 0.307 per year for the Rationality lexicon (Fig.~S22).
This indicates that the increase in directional asymmetry is not confined to the reversal period but extends into the later dynamics.
By contrast, lexicon-mean local exogeneity remains close to its nominal expectation throughout the observation period, with no sustained upward trend or pronounced increase around the lexical reversal (Fig.~\ref{fig:fig5}d).
Thus, stronger temporal asymmetry does not coincide with a corresponding rise in lexicon-wide transition atypicality relative to the contemporaneous reference dynamics.
At the word level, however, individual words exhibit pronounced episodic peaks in local exogeneity.
Across these peak transitions, we observe diverse bridge--endpoint score profiles (Fig.~\ref{fig:fig5}e).

To illustrate these word-level departures, we focus on \textit{economy} and \textit{welcome}, which broadly follow the mean trajectories of their respective lexicons but exhibit local deviations (Fig.~\ref{fig:fig5}f).
Local exogeneity for \textit{welcome} exhibits five pronounced peaks.
Four of these, around 1864, 1918, 1944, and 1976, fall near the closing phases or immediate aftermaths of the U.S. Civil War, the First and Second World Wars, and the Vietnam War, respectively. 
\textit{Economy} shows a distinct peak around 1932 during the Great Depression.
The 1932 transition in \textit{economy} and the 1864, 1874, and 1918 transitions in \textit{welcome} exhibit recovery-like profiles, with elevated bridge scores but comparatively low endpoint scores, while the 1944 and 1976 transitions in \textit{welcome} show elevated bridge and endpoint scores, consistent with the off-path-like category (Fig.~\ref{fig:fig5}e).
Local irreversibility adds a distinct dimension (Fig.~\ref{fig:fig5}h). 
For the atypical transitions in \textit{welcome}, local irreversibility is negative, indicating that the reverse temporal ordering is more probable than the observed ordering under the contemporaneous reference dynamics. 
The sign is reversed for the 1932 transition in \textit{economy}, which has positive local irreversibility, indicating that the observed ordering is more probable than its reversal.
Viewed jointly, these diagnostics show that atypical lexical transitions can either follow or oppose the locally favored temporal direction.

\section*{Discussion}
\label{sec:discussion}

We have developed a stochastic framework that treats observed social trajectories as realizations of a stochastic process. 
This stochastic description captures both systematic tendencies and state-dependent variability, allowing us to separate three aspects of social evolution that are often conflated: the dynamics of typical change, statistically atypical transitions and their short-term patterns, and the directionality of realized histories.
Across three empirical domains, we find that these aspects need not move together and can vary across states, trajectories, and time, with similar disruptions leaving different degrees of atypicality, distinct patterns across successive observations, and different contributions to irreversibility. 
From this perspective, social change is represented not as a single expected path but as a distribution of possible histories within which atypical departures and a statistical arrow of change can be quantified.

Using the inferred SDE as a reference process raises a particularly salient issue for social data. 
Because each country, historical region, or lexical item is observed along only one realized trajectory, pooling such nonreplicable histories requires assumptions about which units and periods are sufficiently comparable for the intended inference~\cite{gelman2013bayesian}. 
The inferred dynamics should thus be interpreted as a coarse-grained description of ensemble-level regularities.
These regularities can provide empirical constraints on candidate mechanisms, although establishing a particular causal explanation requires further evidence~\cite{pearl2009causality}.
Within this model-relative view, exogeneity scores can flag candidate events for further investigation but do not by themselves identify either an external causal origin or a critical juncture, while irreversibility quantifies temporal directionality without implying either normative value or historical inevitability.
Their interpretation therefore depends on both the fitted reference process and the chosen state representation.

Our approach sits at the interface between empirically inferred regularities and mechanistic modeling in quantitative social dynamics.
Mechanistic approaches, including agent-based and statistical-physics models, have examined how phenomena such as social norms~\cite{koster2022spurious}, opinion polarization~\cite{axelrod2021preventing}, and cultural and linguistic dynamics emerge from specified rules of interaction~\cite{castellano2009statistical}.
Complementary empirical approaches have used observed social data to infer dynamical relationships between macro-level variables, identify structure in historical trajectories, and fit formal statistical models of collective behavior~\cite{ranganathan2014bayesian, shin2020scale, lee2025collective}.
Our framework extends this empirical direction by turning inferred dynamics into trajectory-level diagnostics.
This, in turn, provides a route for linking candidate mechanisms to dynamical features of social trajectories that are not captured by aggregate trends, including state-dependent variability, atypical-transition patterns, and temporal asymmetry.

Several limitations of our study point to directions for future work.
Finite and uneven sampling limits the resolution of the inferred dynamics, especially where observations are sparse.
Our uncertainty quantification helps identify where the inferred fields are weakly constrained, while improving inference in such data-limited settings remains an important methodological challenge (Fig.~S4 and Sec.~\Romannum{7} in the SI).
The SDE formulation also assumes that the constructed state is approximately Markovian at the observation scale.
Although we assess this approximation empirically using temporal-correlation and simulation-based analyses across the applications (Figs.~S6 and S7), systems with substantial memory or unresolved slow variables may require an expanded state representation or a non-Markovian formulation.
Finally, the bridge and endpoint diagnostics characterize only the short-term pattern of an atypical transition and do not capture longer-horizon departures that accumulate through individually plausible steps.
Extending these diagnostics to longer-horizon path structure would broaden their temporal scope.
Beyond these methodological extensions, the framework could be applied to other longitudinal social systems, including artistic change~\cite{lee2020dissecting}, cross-national cultural values~\cite{jackson2024worldwide}, public opinion~\cite{claassen2019estimating}, and scientific communities~\cite{fortunato2018science}.
Applying the same probabilistic diagnostics across such domains could reveal recurring dynamical patterns among forms of social change that are usually studied separately, for example in how trajectories depart from their reference dynamics and how those departures unfold over time.
A key direction for future research is to place diverse social trajectories on a common probabilistic footing without presuming a universal dynamical law.

\clearpage

\section*{Methods}

\subsection*{Data and construction of state spaces}

\paragraph*{\textbf{DIG -- Modern political economy.}}
For the modern political-economy analysis, each country- or territory-year observation was represented by a three-dimensional state vector $\bm{x}_t \in \mathbb{R}^3$ comprising political democracy, economic inequality, and economic development.
The state variables were constructed from the \textit{Varieties of Democracy} (V-Dem) project~\cite{coppedge2025vdem, pemstein2025measurement}, the \textit{World Inequality Database} (WID)~\cite{alvaredo2025wid}, and the \textit{Maddison Project Database} (MPD)~\cite{bolt2024maddison}.
The democracy and inequality coordinates were defined as the first principal components of standardized V-Dem and WID indicators, respectively, with their signs oriented so that larger values indicate stronger democratic institutions and greater economic inequality.
Economic development was defined as the standardized logarithm of real GDP per capita.
The resulting dataset is a nearly balanced annual panel of 148 countries from 1992 to 2022, covering the post-Cold War period as a comparatively coherent era of global political-economic transformation.
For regional comparisons, countries were grouped according to the MPD classification into South and Southeast Asia (S~\&~SE Asia), Sub-Saharan Africa (SSA), Eastern Europe (E.~Europe), the Middle East and North Africa (MENA), Latin America, Western Offshoots (W.~Offshoots), Western Europe (W.~Europe), and East Asia (E.~Asia).

\paragraph*{\textbf{Seshat -- Quantitative history.}}
We used the \textit{Seshat: Global History Databank}, which records the sociopolitical characteristics of polities occupying Natural Geographic Areas (NGAs) across broad regions and historical periods~\cite{turchin2015seshat}.
To study long-run regional dynamics, we organized these records into NGA-level trajectories sampled on a century-scale grid, averaging the variables when multiple polities were associated with the same NGA at a given grid year.
Polity-level variables were matched to grid years using a half-century dating tolerance to accommodate minor differences between coded historical intervals and the sampling grid.
Following prior studies~\cite{turchin2018quantitative, turchin2018fitting, shin2020scale}, we used 47 variables to construct a two-dimensional state $\bm{x}_t \in \mathbb{R}^2$.
The first dimension, social scale, was defined as the first principal component of variables measuring polity magnitude, including polity population, territorial extent, and the population of the largest settlement.
The second dimension, computation, was constructed as a composite of four institutional domains---information systems, infrastructure, government, and monetary systems---and represents capacity for information processing and administrative control.
Restricting the analysis to the preindustrial period yielded a sparse, century-scale panel of 25 NGAs spanning 3800 BCE to 1800 CE.
For display, Middle Yellow River Valley, Valley of Oaxaca, and Southern Mesopotamia are abbreviated as MYRV, Oaxaca, and S.~Mesopotamia, respectively.

\paragraph*{\textbf{Google Books Ngram -- Cultural analytics.}}
For the analysis of long-run lexical change, we used the processed dataset of Ref.~\cite{scheffer2021rise}, derived from the Google Books Ngram corpus described in Ref.~\cite{michel2011quantitative}.
Starting from the 5,000 most frequent English words over 1800--2019, preprocessing yielded 4,995 word-frequency trajectories, which we restricted to 1850--2019 for analysis.
Following Ref.~\cite{scheffer2021rise}, each word's annual frequency was normalized by the contemporaneous frequency of the reference word \textit{an} and then standardized over time.
We combined these trajectories with diachronic word embeddings from \textit{HistWords}~\cite{hamilton2016diachronic} and defined an emotionality--rationality axis as the unit-normalized difference between the mean embeddings of prespecified emotionality and rationality seed words, respectively.
Words consistently falling in the lower and upper tails of the projection-score distribution were selected to form balanced Emotionality and Rationality lexicons, respectively, with 182 words in each group.
These labels denote the relative positions of the two lexicons along the embedding axis.
Adjacent annual observations were averaged over nonoverlapping two-year windows, yielding 364 one-dimensional trajectories with 85 observations each, spanning the 1850--1851 through 2018--2019 windows.
All observation windows were retained for SDE fitting, whereas transitions ending in 2018--2019 were excluded from downstream diagnostics because terminal-time dynamics are weakly constrained in the absence of subsequent transitions.

The temporal specification was adapted to the scope and sampling structure of each dataset.
For DIG, we focus on the relatively short 1992--2022 post-Cold War period and use a time-homogeneous reference process to characterize state-dependent dynamics over this interval.
For Seshat, the sparse and irregular observations make explicit calendar-time dependence difficult to distinguish reliably from state dependence.
We thus model DIG and Seshat using time-homogeneous reference dynamics, interpreted as effective conditional dynamics over their respective observation periods.
By contrast, the densely sampled one-dimensional lexical trajectories allow calendar time to be included explicitly in the inferred dynamics.

\subsection*{SDE formulation}

We interpret the empirical records as discrete samples of underlying continuous-time stochastic trajectories.
For an individual unit, these samples form a sequence $\{\bm{x}_{t_k}\}_{k=0}^{L}$, with each state increment $\Delta\bm{x}_{t_k}\equiv\bm{x}_{t_{k+1}}-\bm{x}_{t_k}$ corresponding to an observation interval $\Delta t_k\equiv t_{k+1}-t_k$, which may vary across transitions, units, and datasets.
Each transition is treated as a realization from the finite-time transition density $P_{\Delta t_k}(\bm{x}_{t_{k+1}}\mid\bm{x}_{t_k})$ induced by the SDE in Eq.~\eqref{eq:sde}, with dependence on any conditioning variables left implicit.
Transitions are pooled across units within each application to infer common effective drift and diffusion fields.

The formulation assumes that the constructed state, together with any explicit conditioning variables, provides an approximately Markovian description at the observation scale.
We assess the empirical adequacy of this approximation using temporal-correlation and simulation-based diagnostics (Figs.~S6 and S7).
Within this effective description, the drift $\bm{F}(\bm{x}_t)$ specifies the conditional mean rate of change, and the stochastic term $\sqrt{2\mathbf{D}(\bm{x}_t)}\,d\bm{W}_t$ provides a coarse-grained representation of variability unresolved by the constructed state.
The diffusion matrix $\mathbf{D}$ determines the state-dependent magnitude and covariance structure of these fluctuations, and $d\bm{W}_t$ denotes the increment of a standard Wiener process.
This Gaussian white-noise representation is a standard coarse-grained approximation to the aggregate effect of many weakly dependent, rapidly varying contributions~\cite{zwanzig2001nonequilibrium,van2007stochastic,helbing2010quantitative}.

\subsection*{SDE inference method}

Because the downstream diagnostics require only estimates of the drift field $\bm{F}$ and diffusion matrix $\mathbf{D}$, our framework is modular with respect to the SDE inference method, which may be selected according to the temporal resolution, sample size, and dimensionality of the data.
For the main analyses, we used a Kramers--Moyal-based neural-network estimator adapted from the Langevin Bayesian Network (LBN)~\cite{bae2025inferring}, together with Stochastic Weight Averaging-Gaussian (SWAG) to quantify uncertainty in the inferred fields~\cite{maddox2019simple}.
To demonstrate that the framework is not tied to a specific estimator, we also applied Stochastic Force Inference~\cite{frishman2020learning} in additional analyses presented in the SI.

The LBN represents the drift $\bm{F}$ and diffusion $\mathbf{D}$ using separate neural networks trained to estimate the first two conditional moments of observed increments, following the Kramers--Moyal formulation for drift and diffusion inference~\cite{risken1996fokker, lade2009finite, friedrich2011approaching}.
For a displacement $\Delta\bm{x}_t \equiv \bm{x}_{t+\Delta t}-\bm{x}_t$ over an observation interval $\Delta t$, the estimators are
\begin{equation}
\begin{aligned}
\widehat{\bm{F}}_{\bm{\theta}_F}(\bm{x},\bm{c})
&=
\left\langle \left.
\frac{\Delta\bm{x}_t}{\Delta t} \right| \bm{x}_t=\bm{x},\,\bm{c}_t=\bm{c} \right\rangle,
\\
\widehat{\mathbf{D}}_{\bm{\theta}_D}(\bm{x},\bm{c})
&=
\left\langle \left. \frac{\Delta\bm{x}_t\Delta\bm{x}_t^\top}{2\Delta t}
\right| \bm{x}_t=\bm{x},\,\bm{c}_t=\bm{c} \right\rangle,
\end{aligned}
\end{equation}
where $\bm{\theta}_F$ and $\bm{\theta}_D$ denote the network parameters and $\bm{c}_t$ denotes optional conditioning variables supplied as additional inputs.
For the time-homogeneous DIG and Seshat analyses, no conditioning variables were used.
For the lexical analysis, $\bm{c}_t$ comprised calendar time and the lexical-group label, allowing the inferred fields to vary explicitly with time and between lexicons.

To quantify uncertainty in the inferred fields, we combined SWAG with a 10-fold ensemble of the drift and diffusion networks~\cite{maddox2019simple}.
Data were partitioned into ten folds and a separate SWAG model was trained for each fold-specific partition.
The Monte Carlo predictions were then pooled across all fold-specific models and SWAG weight samples.
The ensemble-averaged predictions across fold-specific models and SWAG weight samples were used as the inferred fields, while their pointwise dispersion quantified model-estimation uncertainty.
This uncertainty is different from the stochastic variability represented by the diffusion matrix.
Details of network architecture, training procedure, finite-interval robustness, and uncertainty quantification are provided in Sec.~\Romannum{2} of the SI.

\subsection*{Dynamical diagnostics}

In empirical applications, we evaluate the dynamical diagnostics using the inferred $\bm{F}$ and $\mathbf{D}$.
The three exogeneity scores share a common probabilistic construction: each is derived from a transition or conditional predictive density and converted into a tail probability, placing local, bridge, and endpoint deviations on a common dimensionless scale.
We validated these scores in synthetic systems by analytically deriving their reference values under controlled perturbations and confirming that the scores computed from the inferred SDEs closely reproduced them (Sec.~\Romannum{4} in the SI).
For brevity, the dependence of $\bm{F}$, $\mathbf{D}$ and the associated transition distributions on $\bm{c}_t$ is left implicit throughout this subsection.

\paragraph*{\textbf{Local exogeneity score.}}

The local exogeneity score quantifies how atypical a single observed transition is under the reference dynamics.
Consider the transition from $\bm{x}_{t-\Delta t_-}$ to $\bm{x}_t$, with incoming displacement
\begin{equation}
\Delta\bm{x}^{\mathrm{in}}_t
\equiv
\bm{x}_t-\bm{x}_{t-\Delta t_-}.
\end{equation}
Under a local Gaussian approximation to the SDE transition density,
\begin{equation}
\Delta\bm{x}^{\mathrm{in}}_t
\mid
\bm{x}_{t-\Delta t_-}
\sim
\mathcal{N}\!\left(
\bm{F}_{-}\Delta t_-,
2\mathbf{D}_{-}\Delta t_-
\right),
\label{eq:transition}
\end{equation}
where $\bm{F}_{-}\equiv\bm{F}(\bm{x}_{t-\Delta t_-})$ and $\mathbf{D}_{-}\equiv\mathbf{D}(\bm{x}_{t-\Delta t_-})$.
The discrepancy between the observed and locally expected displacements is quantified by the squared Mahalanobis distance
\begin{equation}
Q^{\mathrm{loc}}_t
\equiv
\left(
\Delta\bm{x}^{\mathrm{in}}_t-\bm{F}_{-}\Delta t_{-}
\right)^{\top}
\left(
\mathbf{C}^{\rm loc}_t
\right)^{-1}
\left(
\Delta\bm{x}^{\mathrm{in}}_t-\bm{F}_{-}\Delta t_-
\right),
\end{equation}
with $\mathbf{C}^{\rm loc}_t \equiv 2\mathbf{D}_{-}\Delta t_-$.
Under this Gaussian approximation, $Q^{\mathrm{loc}}_t$ follows a chi-squared distribution with $d$ degrees of freedom.
The local tail probability and the local exogeneity score are
\begin{equation}
p^{\mathrm{loc}}_{t}
\equiv
\Pr\!\left(
\chi_d^2 \geq Q^{\mathrm{loc}}_t
\right), \quad 
\mathcal{E}^{\mathrm{loc}}_{t}
\equiv
-\log_{10}p^{\mathrm{loc}}_{t}.
\end{equation}
A large $\mathcal{E}^{\mathrm{loc}}_{t}$ indicates that the observed displacement deviates strongly from its locally expected value after accounting for the magnitude and anisotropy of the stochastic fluctuations.
For example, $\mathcal{E}^{\mathrm{loc}}_t=3$ corresponds to a tail probability of $10^{-3}$ under the reference dynamics.
The score flags statistically atypical incoming transitions as exogenous perturbations under the reference process.
The quadratic form $Q_t^{\rm loc}$ is closely related to classical likelihood-ratio statistics for anomaly detection in multivariate Gaussian time-series models, extended here to nonlinear continuous-time dynamics with state-dependent diffusion~\cite{fox1972outliers, tsay2000outliers, chandola2009anomaly}.
Under the local Gaussian reference process, $p_t^{\mathrm{loc}}$ is uniformly distributed on $[0,1]$, giving a nominal mean local exogeneity score of $1/\ln 10$ (Sec.~\Romannum{4} in the SI).

\paragraph*{\textbf{Two-step exogeneity scores.}}
Because the local exogeneity score uses only the incoming transition, it cannot distinguish a transient deviation from one that persists to the next observation.
To distinguish these possibilities, we consider three consecutive observations,
$\bm{x}_{-}\equiv\bm{x}_{t-\Delta t_{-}}$, $\bm{x}_{t}$, and
$\bm{x}_{+}\equiv\bm{x}_{t+\Delta t_{+}}$, where the adjacent observation intervals $\Delta t_{-}$ and $\Delta t_{+}$ may differ.
Using the same local Gaussian transition approximation, we define the common two-step kernel
\begin{equation}
K_t(\bm{z},\bm{y})
\equiv
P_{\Delta t_{-}}(\bm{z}\mid\bm{x}_{-})
P_{\Delta t_{+}}(\bm{y}\mid\bm{z}).
\label{eq:two_step_kernel}
\end{equation}
This kernel gives the joint predictive density of an intermediate state $\bm{z}$ and a subsequent endpoint $\bm{y}$ conditional on $\bm{x}_{-}$.
The bridge distribution is obtained by conditioning the kernel on the observed endpoint $\bm{x}_{+}$, whereas the endpoint distribution is obtained by marginalizing over the intermediate state.

\paragraph*{Bridge score.}
The bridge score evaluates whether the observed intermediate state $\bm{x}_{t}$ is atypical among trajectories conditioned to connect its two adjacent observations.
By the Markov property, the corresponding conditional bridge density is
\begin{equation}
\pi_t^{\mathrm{brg}}(\bm{z})
\equiv
P(\bm{z}\mid\bm{x}_{-},\bm{x}_{+})
=
\frac{
K_t(\bm{z},\bm{x}_{+})
}{
\displaystyle
\int
K_t(\bm{z}',\bm{x}_{+})\,d\bm{z}'
}.
\label{eq:bridge_density}
\end{equation}
To identify the most probable intermediate state connecting the two adjacent observations, we define the negative log density $\Phi_t^{\mathrm{brg}}(\bm{z}) \equiv -\ln \pi_t^{\mathrm{brg}}(\bm{z})$.
The bridge mode and local curvature are then
\begin{equation}
\begin{aligned}
\bm{\mu}_{t}^{\mathrm{brg}}
&\equiv
\arg\min_{\bm{z}}
\Phi_t^{\mathrm{brg}}(\bm{z}),
\\
\mathbf{H}_{t}^{\mathrm{brg}}
&\equiv
\left.
\nabla_{\bm{z}}^{2}
\Phi_t^{\mathrm{brg}}(\bm{z})
\right|_{\bm{z}=\bm{\mu}_{t}^{\mathrm{brg}}}.
\end{aligned}
\label{eq:bridge_mode_hessian}
\end{equation}
A Laplace approximation around the bridge mode gives
$\pi_t^{\mathrm{brg}}(\bm{z})
\approx
\mathcal{N}
\left(
\bm{\mu}_{t}^{\mathrm{brg}},
[\mathbf{H}_{t}^{\mathrm{brg}}]^{-1}
\right)$.
The bridge discrepancy is 
\begin{equation}
Q_t^{\mathrm{brg}}
\equiv
(\bm{x}_{t}-\bm{\mu}_{t}^{\mathrm{brg}})^{\top}
\mathbf{H}_{t}^{\mathrm{brg}}
(\bm{x}_{t}-\bm{\mu}_{t}^{\mathrm{brg}}).
\label{eq:bridge_Q}
\end{equation}

\paragraph*{Endpoint score.}
The endpoint score evaluates whether the observed endpoint $\bm{x}_{+}$ is atypical under the two-step predictive distribution from $\bm{x}_{-}$ after marginalizing over the intermediate state.
The corresponding marginal predictive density is
\begin{equation}
\pi_t^{\mathrm{end}}(\bm{y})
\equiv
\int K_t(\bm{z}',\bm{y})\,d\bm{z}'.
\label{eq:endpoint_density}
\end{equation}
We approximate the endpoint density by a Gaussian matched to its first two moments.
Let $\bm{Z}
\sim
P_{\Delta t_{-}}(\cdot\mid\bm{x}_{-})$,
and define the conditional second-step mean and covariance as
\begin{equation}
\bm{m}_{+}(\bm{Z})
\equiv
\bm{Z}
+
\bm{F}(\bm{Z})\Delta t_{+},
\quad
\mathbf{C}_{+}(\bm{Z})
\equiv
2\mathbf{D}(\bm{Z})\Delta t_{+}.
\label{eq:endpoint_conditional_moments}
\end{equation}
The resulting endpoint mean and covariance are
\begin{equation}
\begin{aligned}
\bm{\mu}_{t}^{\mathrm{end}}
&\equiv
\mathbb{E}
\left[
\bm{m}_{+}(\bm{Z})
\right],
\\
\mathbf{C}_{t}^{\mathrm{end}}
&\equiv
\operatorname{Cov}
\left[
\bm{m}_{+}(\bm{Z})
\right]
+
\mathbb{E}
\left[
\mathbf{C}_{+}(\bm{Z})
\right].
\end{aligned}
\label{eq:endpoint_covariance}
\end{equation}
These moments were estimated by Monte Carlo sampling of the first-step state $\bm{Z}$.
We then use the moment-matched approximation
\begin{equation}
\pi_t^{\mathrm{end}}(\bm{y})
\approx
\mathcal{N}
\left( \bm{y} \mid
\bm{\mu}_{t}^{\mathrm{end}},
\mathbf{C}_{t}^{\mathrm{end}}
\right),
\label{eq:endpoint_gaussian}
\end{equation}
and define
\begin{equation}
Q_t^{\mathrm{end}}
\equiv
(\bm{x}_{+}-\bm{\mu}_{t}^{\mathrm{end}})^{\top}
\left(
\mathbf{C}_{t}^{\mathrm{end}}
\right)^{-1}
(\bm{x}_{+}-\bm{\mu}_{t}^{\mathrm{end}}).
\label{eq:endpoint_Q}
\end{equation}

Under the respective approximations for the bridge and endpoint distributions, the two discrepancies are mapped to a common tail-probability scale using a chi-squared distribution with $d$ degrees of freedom:
\begin{equation}
\begin{aligned}
p_t^{\alpha}
\equiv
\Pr
\left(
\chi_d^2
\ge
Q_t^{\alpha}
\right), \qquad
\mathcal{E}_t^{\alpha}
\equiv
-\log_{10}
p_t^{\alpha},
\end{aligned}
\label{eq:two_step_exogeneity}
\end{equation}
with $\alpha \in \{ \mathrm{brg}, \mathrm{end} \}.$
Larger values of $\mathcal{E}_t^{\alpha}$ indicate stronger atypicality under the corresponding bridge or endpoint distribution.
For exogenous perturbations identified by the local score, the relative magnitudes of the bridge and endpoint scores characterize localized, recovered, persistent, and off-path categories (Table~\ref{tab:perturbation_category}).
In the local Gaussian limit, we derive the inequality among the three discrepancy measures $Q_t^\alpha$ (Sec.~\Romannum{4} in the SI):
\begin{equation}
\begin{aligned}
Q_t^{\mathrm{brg}} + Q_t^{\mathrm{end}} \geq Q_t^{\mathrm{loc}}.
\end{aligned}
\label{eq:q_score_bound}
\end{equation}
Each $\mathcal{E}_t^\alpha$ is a monotonic transformation of $Q_t^\alpha$ so that this inequality defines a corresponding lower bound in the bridge--endpoint plane for a specified local-score threshold. 
The dashed curves in Figs.~\ref{fig:fig3}d,e and \ref{fig:fig5}e show this bound for the thresholds used to display peak transitions.

\paragraph*{\textbf{Dynamical irreversibility.}}
Dynamical irreversibility quantifies the statistical distinguishability of an observed trajectory from the same sequence of states traversed in reverse.
Consider a trajectory
$\Gamma\equiv\{\bm{x}_{t_0},\bm{x}_{t_1},\ldots,\bm{x}_{t_L}\}$
and its time-reversed counterpart
$\tilde{\Gamma}\equiv\{\bm{x}_{t_L},\bm{x}_{t_{L-1}},\ldots,\bm{x}_{t_0}\}$.
We define the trajectory-level irreversibility as the log ratio of their conditional path densities:
\begin{equation}
\begin{aligned}
\Sigma[\Gamma]
&\equiv
\ln
\frac{
\mathcal{P}_{\mathrm{F}}[\Gamma\mid\bm{x}_{t_0}]
}{
\mathcal{P}_{\mathrm{R}}[\tilde{\Gamma}\mid\bm{x}_{t_L}]
}
\\
&=
\sum_{k=0}^{L-1}
\ln
\frac{
P_{\mathrm{F},\Delta t_k}
\left(
\bm{x}_{t_{k+1}}\mid\bm{x}_{t_k}
\right)
}{
P_{\mathrm{R},\Delta t_k}
\left(
\bm{x}_{t_k}\mid\bm{x}_{t_{k+1}}
\right)
}
\equiv
\sum_{k=0}^{L-1}\sigma_{t_{k+1}},
\end{aligned}
\end{equation}
where $\Delta t_k\equiv t_{k+1}-t_k$.
Here, the subscripts $\mathrm{F}$ and $\mathrm{R}$ denote the forward and reverse protocols, and $\mathcal{P}$ and $P_{\Delta t}$ denote the corresponding conditional path and finite-time transition densities, respectively.
For externally conditioned dynamics, the reverse protocol traverses any time-varying conditioning sequence in reverse together with the trajectory, while time-independent conditioning labels remain unchanged. 
For dynamics without time-varying conditioning, the forward and reverse protocols use the same transition kernel, evaluated along oppositely directed transitions.
The protocol labels are omitted in the Main Text for notational simplicity.
Each local contribution $\sigma_{t_{k+1}}$ is indexed by its arrival time $t_{k+1}$, consistent with the local exogeneity score.

For the It\^o SDE considered here, the local log ratio can be written in Stratonovich form and approximated over a finite transition using midpoint discretization~\cite{spinney2012entropy}:
\begin{equation}
\sigma_{t+\Delta t}
\approx
\left[
\bm{F}(\bar{\bm{x}}_t)
-
\nabla\!\cdot\mathbf{D}(\bar{\bm{x}}_t)
\right]^{\top}
\mathbf{D}(\bar{\bm{x}}_t)^{-1}
\Delta\bm{x}_t,
\end{equation}
where $\bar{\bm{x}}_t \equiv (\bm{x}_t+\bm{x}_{t+\Delta t})/2$, and, for explicitly time-dependent dynamics, the fields are also evaluated at calendar time $t+\Delta t/2$.
The diffusion divergence is defined as
$\left[
\nabla\!\cdot\mathbf{D}
\right]_i
\equiv
\sum_{j=1}^{d}
({\partial D_{ij}}/{\partial x_j})$.
Positive values of $\Sigma[\Gamma]$ and $\sigma_{t_{k+1}}$ indicate that the observed trajectory and transition, respectively, are more probable in the forward direction than their reversed counterparts; negative values indicate the opposite directional asymmetry.

We define the mean irreversibility rate as the cumulative irreversibility per unit elapsed time over the retained transitions,
\begin{equation}
\langle \dot{\sigma} \rangle
\equiv
\frac{\sum_{k\in\mathcal{K}_{\Gamma}} \sigma_{t_{k+1}}}
{\sum_{k\in\mathcal{K}_{\Gamma}} \Delta t_k},
\end{equation}
where $\mathcal{K}_{\Gamma}$ denotes the set of transitions retained for the corresponding analysis according to its inclusion criteria. Thus, the rate measures the accumulated irreversibility per unit elapsed time represented by the retained transitions. 
For the Seshat analysis, $\mathcal{K}_{\Gamma}$ was restricted to transitions with observation intervals $\Delta t_k \leq 200$ years.
As log-density ratios, $\sigma$ and $\Sigma$ are dimensionless, whereas $\langle \dot{\sigma} \rangle$ has units of inverse time.

\subsection*{Statistical analysis}

All analyses involving exogeneity used the capped scores
$\min(\mathcal{E}^{\alpha}_t,6)$,
where $\alpha\in\{\mathrm{loc},\mathrm{brg},\mathrm{end}\}$.
The cap was applied at the observation level before aggregation and resampling, so that tail probabilities at or below $10^{-6}$ received the common value 6.
This limits the influence of a small number of extreme values on aggregate summaries and avoids interpreting differences among tail probabilities below $10^{-6}$ as reliably resolved by the fitted SDE.
Sensitivity analyses across different local-score caps are reported in the SI.

Pointwise 95\% bootstrap confidence intervals for time-resolved aggregate quantities were obtained using a nonparametric cluster bootstrap with 50,000 resamples.
Countries or words were resampled as observational units, depending on the dataset, while retaining all longitudinal observations associated with each resampled unit.
Intervals were defined by the 2.5th and 97.5th percentiles of the bootstrap distributions.
For annual high-versus-low quartile contrasts in Figs.~\ref{fig:fig3} and \ref{fig:fig4}, we computed two-sided bootstrap $p$ values from the proportions of bootstrap differences falling on either side of zero and Bonferroni-adjusted these $p$ values across valid years within each stratification.

\subsection*{Code and data availability}

All code and aggregated datasets are available in the following GitHub repository: \url{https://github.com/qodudrud/SocialSDE}.

\begin{acknowledgments}
The authors thank Peter Turchin and Matilda Peruzzo for sharing the Seshat database. 
YB acknowledges support from the Basic Science Research Program through the National Research Foundation of Korea (NRF), funded by the Ministry of Education (Grant No. RS-2025-02653944). SH acknowledges generous grants from the Siegel Family Endowment and Omidyar Network for supporting this research. SLY acknowledges support from the Cooper Memorial Faculty Fellowship at the Pennsylvania State University Brandywine Campus.

\end{acknowledgments}

\bibliography{main}

\end{document}


\title{{\Large Supplementary Information} 
\\
Inferring Stochastic Dynamics Reveals Atypicality and Irreversibility in Social Change}

\author{Youngkyoung Bae}

\author{Hajime Shimao}

\author{Seungwoong Ha}

\author{Shuo (Luna) Yang}

\author{David Wolpert}

\maketitle


\tableofcontents

\clearpage

\section{Data and Construction of State Spaces}
\label{sec:data}

The empirical analyses draw on three datasets spanning modern political economy, quantitative history, and cultural analytics. 
Because the observed variables differ substantially in their measurement scales, dimensionality, and temporal resolution, the construction of the dynamical state space was tailored to each empirical domain. 
In each case, the observed variables were mapped to a low-dimensional continuous representation $\bm{x}_t \in \mathbb{R}^d$ that retains substantively relevant variation while providing a tractable state space for stochastic differential equation (SDE) inference.
Across all datasets, we did not interpolate or impute missing observations when constructing the empirical trajectories, thereby avoiding artificial trends or smoothing artifacts.
The analyses used only observed states and the transitions connecting available observations, retaining the actual observation intervals when these varied across transitions. 
The following subsections describe the data sources, variable transformations, dimensionality-reduction procedures, and construction of the state coordinates for each application. 
For compactness, supplementary descriptive information on the datasets is consolidated in the final \textit{Dataset Summary} section (Sec.~\ref{sec:data_summary}).

\subsection{Democracy--Inequality--Growth (DIG) Dataset -- Modern Political Economy}

\begin{figure*}[!t]
  \centering
  \includegraphics[width=\linewidth]{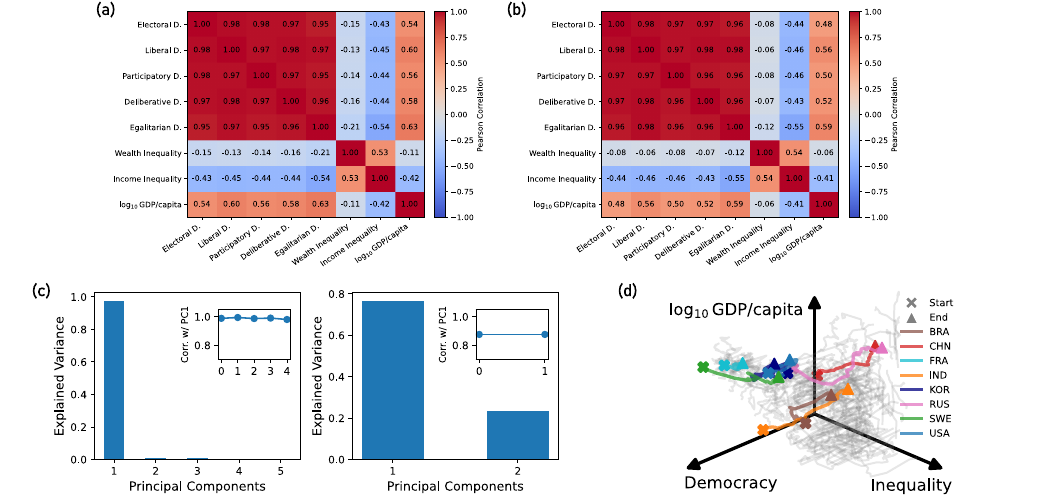}
  \caption{(a, b) Pearson correlation matrices of the original indicators used to construct the political-economy state space, shown for (a) the full complete-case construction sample (1900--2022) and (b) the modern political-economy analysis sample (1992--2022), respectively.
  (c) PCA results for the democracy indicators (left) and the inequality indicators (right). For democracy and inequality, PC1 explains 97.3\% and 76.6\% of the total variance, respectively. Insets show the correlations of the original indicators with their respective PC1 scores. 
  (d) Historical trajectories of representative countries in the resulting three-dimensional state space spanned by democracy, inequality, and standardized $\log_{10}$ GDP/capita. Colored lines denote selected countries, gray lines show the remaining country trajectories, crosses mark the starting year, and triangles mark the final year.
  }
  \label{fig:dig_dataset}
\end{figure*}

To represent modern political-economic dynamics, we construct a three-dimensional state space spanning political democracy, economic inequality, and economic development.
We integrate three primary data sources: the \textit{Varieties of Democracy} (V-Dem) dataset~\cite{coppedge2025vdem, pemstein2025measurement}, the \textit{World Inequality Database} (WID)~\cite{alvaredo2025wid}, and the \textit{Maddison Project Database} (MPD)~\cite{bolt2024maddison}. 
The dataset is organized as a country- or territory-year panel with annual resolution.
For brevity, and consistent with the main text, we use the term ``country'' to refer to country- or territory-level units treated separately in the source datasets.
We define the state of each country at time $t$ as a vector $\bm{x}_t \equiv [x^{\rm D}_{t}, x^{\rm I}_{t}, x^{\rm G}_{t}]^\top \in \mathbb{R}^3$ where $x^{\rm D}_{t}$ denotes democracy, $x^{\rm I}_{t}$ denotes inequality, and $x^{\rm G}_{t}$ denotes the level of economic development.
To synthesize multidimensional social indicators into low-dimensional dynamical coordinates, we employ Principal Component Analysis (PCA) for democracy and inequality, defining $x^{\rm D}_t$ and $x^{\rm I}_t$ as the first-principal-component (PC1) scores of their respective underlying variables. 
Each component is defined as follows:

\begin{itemize}
    \item \textbf{Democracy ($x^{\rm D}_t$):} 
    We utilize the V-Dem dataset to quantify the level of democratic institutions.
    V-Dem provides country-year measures of political institutions based on a large set of expert-coded indicators and measurement models designed for cross-national and cross-temporal comparison.
    Specifically, we apply PCA to the five high-level V-Dem democracy indices: electoral (\texttt{v2x\_polyarchy}), liberal (\texttt{v2x\_libdem}), participatory (\texttt{v2x\_partipdem}), deliberative (\texttt{v2x\_delibdem}), and egalitarian (\texttt{v2x\_egaldem}) democracy, each ranging from 0 to 1 [see Fig.~\ref{fig:dig_dataset}(a, b) for the correlation structure].
    PC1 explains 97.3\% of the total variance, indicating that these indices are highly correlated and can be represented by a single dominant axis of democratic governance [Fig.~\ref{fig:dig_dataset}(c)]. We orient this component so that larger $x^{\rm D}_t$ indicates stronger democratic institutions.
    
    \item \textbf{Economic inequality ($x^{\rm I}_t$):} We obtain the inequality measures from WID, which provides country-year estimates of income and wealth distributions designed for cross-national comparison. 
    We construct $x^{\rm I}_t$ using PC1 of two complementary Gini coefficients: the Gini coefficient of pre-tax national income (\texttt{gptinc992j}) and the Gini coefficient of household wealth (\texttt{ghweal992j}), both computed for the full population percentile group [see Fig.~\ref{fig:dig_dataset}(a, b) for the correlation structure].
    PC1 explains 76.6\% of the total variance and is used as a single coordinate capturing the shared variation in income and wealth inequality [Fig.~\ref{fig:dig_dataset}(c)]. 
    We orient this component so that larger $x^{\rm I}_t$ corresponds to greater economic inequality.
    
    \item \textbf{Economic development ($x^{\rm G}_t$):} We use real GDP per capita (\texttt{gdppc}) from MPD, which provides a historical database of cross-country economic output and population estimates. 
    Because GDP per capita varies multiplicatively across countries and spans several orders of magnitude, we define $x^{\rm G}_t$ as the standardized value of $\log_{10} \mathrm{GDP/capita}$, using the mean and standard deviation computed over the same complete-case construction sample. 
    Larger values of $x^{\rm G}_t$ correspond to higher levels of aggregate economic development.
\end{itemize}

We implement a complete-case filtering protocol before constructing the effective state space. First, we retain only country-year observations for which all variables required to construct $x^{\rm D}_t$, $x^{\rm I}_t$, and $x^{\rm G}_t$ are available. 
Second, for WID, we exclude source-provided estimates assigned the lowest quality category to reduce reliance on weakly constrained inequality estimates.
The PCA transformations for democracy and inequality are fitted on the full complete-case construction sample covering 1900--2022, after standardizing each input indicator to zero mean and unit variance within this sample.
This full construction sample contains 6,834 country-year observations across 148 countries.
After constructing the coordinates, we restrict the dynamical analysis to the post-Cold War period, 1992--2022, to reduce long-run temporal heterogeneity while retaining broad cross-national coverage.
The resulting analysis sample is nearly balanced: 144 countries have complete annual coverage from 1992 to 2022, whereas four countries enter later because of source-specific data availability, namely the United Arab Emirates (1993--2022), Bahrain (2002--2022), Montenegro (1998--2022), and Slovakia (1993--2022).
The final dataset therefore consists of 4,570 country-year observations across 148 countries.
For downstream regional comparisons, countries are grouped according to the MPD regional classification into South and Southeast Asia (S~\&~SE Asia), Sub-Saharan Africa (SSA), Eastern Europe (E.~Europe), the Middle East and North Africa (MENA), Latin America, Western Offshoots (W.~Offshoots), Western Europe (W.~Europe), and East Asia (E.~Asia).

Figure~\ref{fig:dig_dataset} summarizes the construction of the DIG state space.
The correlation structure of the raw indicators is highly similar between the full construction sample and the post-Cold War analysis sample.
Consistently, PCA repeated using only observations from 1992 onward yielded PC1 scores that were nearly identical to those obtained using the full 1900--2022 construction sample when evaluated over the common post-1992 observations ($r=0.999999$ for democracy and $r=0.999871$ for inequality), indicating that the constructed state coordinates are robust to the choice of PCA fitting period.
The resulting three-dimensional coordinates provide an interpretable state representation of national trajectories in democracy, inequality, and economic development, as illustrated in Fig.~\ref{fig:dig_dataset}(d).

\subsection{Seshat: Global History Databank -- Quantitative History}

For the quantitative-history analysis, we use data from the \textit{Seshat: Global History Databank}~\cite{turchin2015seshat}, which records the sociopolitical characteristics of historical polities across a broad range of regions and time periods.
In Seshat, a polity denotes an independent political unit, ranging in scale from relatively small local communities to territorially expansive and multiethnic empires~\cite{turchin2018quantitative,turchin2018fitting}.
Because the identity and territorial extent of individual polities change through time, we organize polity-level records into trajectories associated with fixed Natural Geographic Areas (NGAs).
An NGA is a geographically fixed reference region used by Seshat to organize long-run historical observations, with boundaries that remain fixed through time and a typical spatial scale of approximately $100\,\mathrm{km}\times100\,\mathrm{km}$ (within roughly $\pm 50\%$).
Successive polities associated with the same NGA can therefore contribute to a common NGA trajectory despite changes in polity identity over time.

Polity-level records were processed using the official Seshat Python package \texttt{SeshatDatasetAnalysis}, while NGA--polity relation records were retrieved through the Seshat API.
Polity records were organized on a 100-year temporal grid, and coded variables were sampled at the corresponding polity-grid years using the package's zero-order sampling rule.
For numerical records reported as ranges, we used the midpoint of the coded lower and upper bounds.
For short-lived polities whose coded lifespans contained no century-grid point, a fallback grid point was assigned to the immediately preceding century mark and retained when it lay within 50 years of the coded polity start.
We used the resulting non-imputed social-complexity variables (SCVs) throughout the analysis.

To construct NGA-level trajectories, the temporally resolved NGA--polity relation records were matched to the same 100-year grid used for the polity-level SCVs.
At each grid year $t$, a polity was associated with an NGA when its coded NGA--polity relation interval overlapped the century-scale window $[t-50,t+50]$.
The resulting NGA--year--polity records were then matched exactly by polity and year to the non-imputed SCV records described above.
When multiple polities were associated with the same NGA at a given grid year, their variables were averaged across polities to obtain a single NGA-level observation.
This procedure yields longitudinal NGA trajectories in which successive observations may correspond to different historical polities while referring to the same fixed geographic region.

To represent these dynamics in a continuous state space, we define a two-dimensional state vector
$\bm{x}_t \equiv [x_t^{\rm S},x_t^{\rm C}]^\top \in \mathbb{R}^2$,
where $x_t^{\rm S}$ represents \textit{social scale} and $x_t^{\rm C}$ represents \textit{computation}, or collective information-processing capacity.
Following previous studies~\cite{turchin2018quantitative,turchin2018fitting,shin2020scale}, the construction uses 47 SCVs: three variables contributing to social scale and 44 variables contributing to computation across four domains---government (11 variables), infrastructure (12), information (15), and monetary systems (6).
The complete set of variables and their domain assignments is summarized in Table~\ref{tab:seshat_variables} in Sec.~\ref{sec:data_summary}.
The two coordinates summarize complementary aspects of sociopolitical complexity:

\begin{figure*}[!t]
  \centering
  \includegraphics[width=\linewidth]{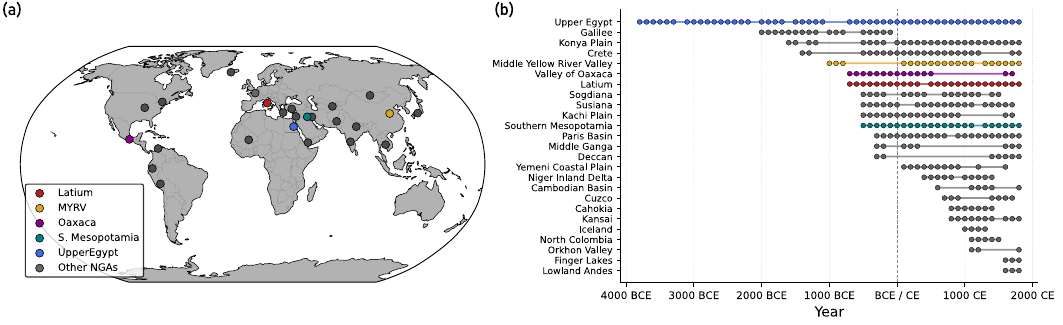}
  \caption{(a) Geographic locations of the Natural Geographic Areas (NGAs). The five NGAs are marked in color, while all other NGAs are shown in gray. MYRV denotes the Middle Yellow River Valley, S. Mesopotamia denotes Southern Mesopotamia, and Oaxaca denotes the Valley of Oaxaca. 
  (b) Temporal coverage of each NGA, ordered by the earliest year represented in the dataset. Points indicate years for which data are available, and horizontal lines span the period between the earliest and latest observations for each NGA. The five NGAs are shown using the same colors as in panel (a), with all other NGAs shown in gray. 
  The dashed vertical line marks the BCE/CE boundary.
  }
  \label{fig:seshat_dataset}
\end{figure*}

\begin{itemize}
    \item \textbf{Social scale ($x_t^{\rm S}$):}
    The social-scale coordinate captures the demographic and territorial magnitude of a polity. We use three quantitative indicators: polity population, polity territory, and the population of the largest settlement. Each variable is first transformed using $\log_{10}$ and then standardized to zero mean and unit variance before applying PCA. 
    PC1 explains 90.6\% of the total variance and is used as $x_t^{\rm S}$, providing a common coordinate that summarizes variation in the demographic, territorial, and urban scale of historical polities.

    \item \textbf{Computation ($x_t^{\rm C}$):}
    The computation coordinate characterizes the collective capacity of a polity to store, transmit, and process information and to coordinate activity across large social scales. 
    It is constructed from 44 variables spanning four institutional domains: government, infrastructure, information, and monetary systems. 
    These domains capture complementary aspects of administrative organization, infrastructure supporting large-scale coordination, information storage and transmission, and monetary systems.
    The constituent SCVs within each domain were aggregated using the domain-specific rules implemented in \texttt{SeshatDatasetAnalysis}.
    We then combine these four Seshat-derived domain scores into a single computation coordinate using a weighted mean, with weights proportional to the number of constituent SCVs in each domain (11 for government, 12 for infrastructure, 15 for information, and 6 for monetary systems).
    The resulting composite is standardized to zero mean and unit variance to define the computation coordinate $x_t^{\rm C}$.
    Larger values of $x_t^{\rm C}$ correspond to greater collective information-processing and coordination capacity.
    As a robustness check, we also constructed an alternative computation coordinate by applying PCA to the four standardized domain scores, with PC1 explaining 77.7\% of their total variance.
    The weighted-mean and PCA-based coordinates are strongly correlated ($r \simeq 0.995$), showing that the two aggregation procedures capture nearly the same underlying dimension of variation.
\end{itemize}

We retained records with complete values for the three social-scale variables and the four derived computation-domain scores.
We restrict the dynamical analysis to the preindustrial period, retaining observations through 1800 CE.
Across the retained NGAs, observations span 3800 BCE to 1800 CE.
Because the variables required for state construction are not available at every century-grid point, individual NGA trajectories are sparse and irregular, with gaps occasionally exceeding one century.
NGAs represented by only a single retained observation were excluded because they do not define a trajectory, leaving 25 NGA trajectories and 358 NGA-year states for dynamical analysis.
Their geographic distribution and temporal coverage are shown in Fig.~\ref{fig:seshat_dataset}.

\subsection{Google Books Ngram -- Cultural Analytics}

For the cultural-analytics analysis, we draw on the processed Google Books Ngram dataset introduced in Ref.~\cite{scheffer2021rise}, which is based on the \textit{Google Books corpus} documented in Ref.~\cite{michel2011quantitative}. 
The source dataset was constructed from the 5,000 most frequent English words over the period 1800--2019, with individual words treated as the observational units. 
Using the preprocessing pipeline provided with Ref.~\cite{scheffer2021rise}, matching the 5,000-word list to the available Ngram trajectory table and applying the prescribed filtering and normalization steps yields 4,995 word-frequency trajectories.
We use the provided z-score trajectories, obtained by normalizing each word's annual frequency by the contemporaneous frequency of the reference word `an' and standardizing the resulting relative-frequency trajectory over 1800--2019.
Here, we restrict the analysis to 1850--2019, consistent with the temporal scope of Ref.~\cite{scheffer2021rise}.
We represent each word at time $t$ by the resulting one-dimensional state $x_t\in\mathbb{R}$, corresponding to its standardized relative frequency. 
Because lexical usage is represented by a single scalar coordinate, no further dimensionality reduction is required. 
This construction places words with substantially different absolute frequencies on a common standardized scale while retaining their longitudinal variation in relative usage.

To identify lexical groups along a Rationality--Emotionality dimension, we use diachronic word embeddings from \textit{HistWords}~\cite{hamilton2016diachronic}. 
HistWords provides time-resolved 300-dimensional vector representations of words derived from historical corpora, allowing semantic relationships among words to vary across historical periods spanning 1800--1990.
Rather than defining the lexical groups directly from principal components of their temporal frequency trajectories, whose semantic interpretation has been debated in relation to other lexical contrasts~\cite{sun2022colloquialization,scheffer2022reply}, we construct the Rationality--Emotionality dimension from predefined semantic seed sets.
We use 29 Rationality seed words (e.g., analysis, chemical, philosophy, and science) and 26 Emotionality seed words (e.g., anger, disgusting, friendly, and happy), listed in Tables~\ref{tab:rationality_lexicon} and \ref{tab:emotionality_lexicon}, respectively.
To account for historical changes in semantic relationships, we construct the embedding axis separately at four snapshots, $t_{\rm HW}\in\{1850,1900,1950,1990\}$. At each snapshot, we compute the mean embedding vectors of the Rationality and Emotionality seed sets, denoted by $\bar{\bm{v}}_{R,t_{\rm HW}}$ and $\bar{\bm{v}}_{E,t_{\rm HW}}$, and define the unit-normalized Emotionality-to-Rationality axis as
\begin{equation}
\bm{a}_{t_{\rm HW}}
=
\frac{\bar{\bm{v}}_{R,t_{\rm HW}}-\bar{\bm{v}}_{E,t_{\rm HW}}}
{\left\|\bar{\bm{v}}_{R,t_{\rm HW}}-\bar{\bm{v}}_{E,t_{\rm HW}}\right\|},
\end{equation}
where $\|\cdot\|$ denotes the Euclidean norm.
For each word $w$ represented in the HistWords vocabulary at snapshot $t_{\rm HW}$, we then calculate its projection score $s_{w,t_{\rm HW}}=\bm{v}_{w,t_{\rm HW}}^{\top}\bm{a}_{t_{\rm HW}}$. 
By construction, larger projection scores correspond to positions toward the Rationality pole, whereas smaller scores correspond to positions toward the Emotionality pole.
We note that these labels are operational and denote relative positions along the constructed axis rather than unique semantic classifications of individual words.

We next use the projection scores to identify words that consistently occupy the two extremes of the constructed semantic axis across historical periods.
Of the 4,995 processed Ngram words, 4,961 are represented in HistWords at all four snapshots and can therefore be assigned projection scores.
At each HistWords snapshot, we define the upper 5\% and lower 5\% of the projection-score distribution over the full HistWords vocabulary as the Rationality and Emotionality tails, respectively.
Among the 4,961 eligible Ngram words, we then retain only words that remain in the same semantic tail across all four snapshots, thereby selecting words whose relative positions toward the same semantic pole persist across the four embedding periods.
From these persistent candidates, we further retain alphabetic English 1-grams containing more than one character and remove standard English stop words together with selected function-word and part-of-speech categories, including pronouns, determiners, prepositions, auxiliary or modal terms, and proper nouns.

\begin{figure*}[!t]
  \centering
  \includegraphics[width=\linewidth]{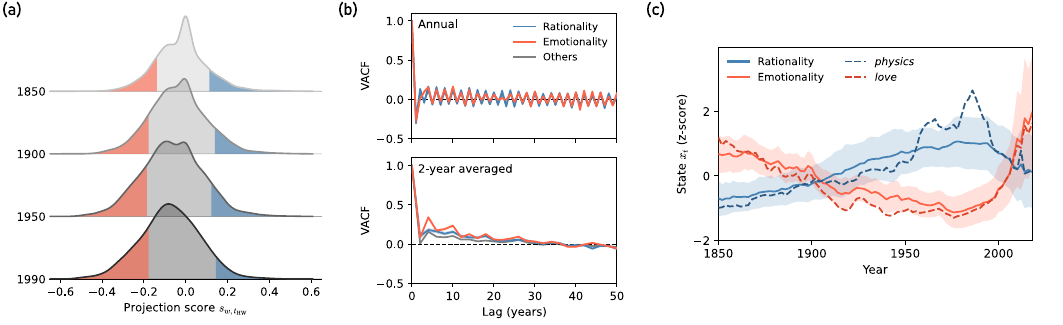}
  \caption{
  (a) Distributions of Rationality--Emotionality projection scores for the 4,961 Ngram words represented in HistWords at the four historical embedding snapshots (1850, 1900, 1950, and 1990).
  Red and blue shading mark the projection-score ranges spanned by the final Emotionality and Rationality lexicons, respectively.
  (b) Velocity autocorrelation functions (VACFs) calculated from the annual trajectories (top) and after averaging adjacent observations over nonoverlapping two-year windows (bottom), shown separately for the Emotionality, Rationality, and remaining words.
  The pronounced alternating short-lag structure at annual resolution is strongly suppressed after two-year temporal coarse-graining.
  (c) Standardized relative-frequency (z-score) trajectories of the final Emotionality and Rationality lexicons after two-year averaging.
  Solid lines show the group means, and shaded bands indicate $\pm 1$ standard deviation across words.
  Dashed lines show two representative individual trajectories, \textit{love} and \textit{physics}, from the Emotionality and Rationality lexicons, respectively.
}
  \label{fig:ngram_dataset}
\end{figure*}

We additionally impose a frequency-coverage criterion to ensure that the selected words are sufficiently represented throughout the full analysis period. 
For each candidate word $w$, we define its annual relative frequency as
\begin{equation}
r_w(t)
=
\frac{f_w(t)}{f_{\textit{an}}(t)},
\end{equation}
where $f_w(t)$ and $f_{\textit{an}}(t)$ denote the frequencies of word $w$ and the reference word \textit{an}, respectively. 
We retain only words satisfying $r_w(t)>5\times10^{-4}$ for every year from 1850 to 2019. 
After this filtering, 182 Rationality candidates and 409 Emotionality candidates remain. 
To construct balanced lexical groups, we retain all 182 Rationality words and select 182 Emotionality words with the largest minimum relative frequency, $\min_t r_w(t)$, over the analysis period. 
This procedure yields two balanced lexicons containing 182 words each, for a total of 364 lexical trajectories; the resulting Emotionality and Rationality lexicons are listed in Tables~\ref{tab:emotionality_lexicon} and \ref{tab:rationality_lexicon}, respectively.
The distributions of projection scores and the positions of the resulting lexical groups along the historical semantic axes are illustrated in Fig.~\ref{fig:ngram_dataset} (a).

Before fitting the stochastic dynamics, we coarse-grain the annual trajectories in time. 
At annual resolution, the velocity autocorrelation function exhibits a pronounced alternating pattern across successive lags for the Rationality, Emotionality, and remaining-word groups, indicating a common short-timescale component across lexical categories.
Because this structure is shared across both selected lexicons and the remaining vocabulary, it is consistent with a common annual-scale component rather than a lexicon-specific feature.
We therefore average adjacent annual observations over nonoverlapping two-year windows.
This choice provides a minimal temporal coarse-graining that suppresses the alternating annual-scale component while preserving longer-term variation in lexical usage [Fig.~\ref{fig:ngram_dataset}(b)].
The resulting dataset consists of 364 one-dimensional lexical trajectories, comprising 182 Rationality words and 182 Emotionality words, with 85 observations per trajectory spanning the 1850--1851 through 2018--2019 windows. 
Thus, each word contributes 84 adjacent transitions, yielding 30,576 transitions in total.
All 85 observation windows are retained for SDE fitting, with calendar time and lexicon identity supplied as conditioning variables in the time-varying lexical model.
Transitions terminating in the final 2018--2019 window are excluded from downstream dynamical diagnostics because the terminal-time dynamics are weakly constrained in the absence of subsequent observations.
The resulting coarse-grained lexical trajectories are illustrated in Fig.~\ref{fig:ngram_dataset}(c).

\clearpage

\section{SDE Inference Method}
\label{sec:method}

Having constructed the empirical state trajectories in Sec.~\ref{sec:data}, we next infer the effective stochastic dynamics underlying their evolution.
Specifically, the observed finite-time transitions are used to estimate the drift field $\bm{F}$ and diffusion matrix $\mathbf{D}$ of the effective SDE (It\^o's convention),
\begin{equation}
\begin{aligned}
    d\bm{x}_t = \bm{F}(\bm{x}_t, \bm{c}_t)dt + \sqrt{2\mathbf{D}(\bm{x}_t, \bm{c}_t)}\,d\bm{W}_t,
\end{aligned}
\label{eq:ito_sde}
\end{equation}
where $\bm{x}_t\in\mathbb{R}^d$ denotes the state vector in the constructed state space, $\bm{F}$ describes the conditional mean tendency of change, and $\mathbf{D}$ characterizes the magnitude and covariance structure of stochastic fluctuations.
Here, $d\bm{W}_t$ denotes an increment of a standard $d$-dimensional Wiener process, and $\bm{c}_t$ represents optional conditioning variables included where appropriate.
For the main analyses, we use a Kramers--Moyal-based neural-network estimator adapted from the Langevin Bayesian Network (LBN) framework~\cite{bae2025inferring}, together with Stochastic Weight Averaging-Gaussian (SWAG) to quantify estimation uncertainty in the inferred fields~\cite{maddox2019simple}.
The downstream diagnostics depend only on the inferred drift and diffusion fields and can therefore be combined with alternative inference methods, as demonstrated using Stochastic Force Inference (SFI)~\cite{frishman2020learning} in Sec.~\ref{sec:other_sde}.


\begin{figure*}[!b]
  \centering
  \includegraphics[width=.95\linewidth]{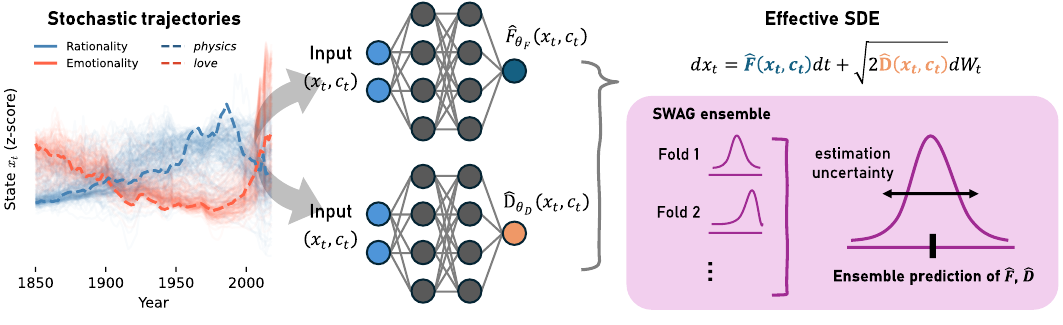}
  \caption{
  Schematic of the LBN-based SDE inference framework. 
  Finite-time transitions extracted from the observed stochastic trajectories are used to train separate neural networks for the drift $\hat{\bm F}_{\theta_F}(\bm x_t,\bm c_t)$ and diffusion $\hat{\mathbf D}_{\theta_D}(\bm x_t,\bm c_t)$, where $\bm{x}_t$ denotes the state vector in the constructed state space and $\bm{c}_t$ denotes optional conditioning variables.
  The lexical trajectories are shown here as an illustrative example; in the time-varying lexical model, $\bm{c}_t$ includes calendar time and lexicon identity. 
  For each cross-validation fold, SWAG provides an approximate distribution over network parameters. 
  Predictions sampled from the fold-specific SWAG posteriors are pooled to obtain ensemble estimates of the drift and diffusion together with their estimation uncertainty.
  The ensemble-averaged fields define the fitted effective SDE.
}
  \label{fig:method_LBN}
\end{figure*}

The LBN framework estimates the drift and diffusion of an SDE from observed stochastic trajectories by learning the Kramers--Moyal coefficients with neural networks~\cite{bae2025inferring}. 
In this approach, the drift field and diffusion matrix are represented by separate neural networks whose outputs approximate the first and second conditional moments of the observed state increments, respectively.
The framework has been shown to recover stochastic dynamics across a range of nonlinear and non-stationary systems with diverse drift and diffusion structures~\cite{bae2025inferring}.
Here, we adapt this basic inference scheme to the empirical trajectories considered in this study and combine it with a cross-validation and Stochastic Weight Averaging-Gaussian (SWAG) ensemble procedure to quantify estimation uncertainty in the inferred fields.
The overall inference procedure is illustrated in Fig.~\ref{fig:method_LBN}.

\subsection{Inference Procedure}
\label{subsec:inference_procedure}

While the original LBN implementation employs variational inference, here we construct the neural-network ensemble using k-fold cross-validation together with SWAG~\cite{maddox2019simple}.
The use of multiple train--validation partitions reduces sensitivity to any single data split while retaining held-out data for monitoring model training, which is particularly useful for the finite empirical datasets considered here.
For each fold, SWAG provides an approximate distribution over network parameters around the region explored during training, from which multiple parameter samples are drawn to generate drift and diffusion predictions (right panel in Fig.~\ref{fig:method_LBN}).
Predictions are then pooled across folds and SWAG samples. 
Their ensemble means define the estimated drift $\widehat{\bm{F}}$ and diffusion $\widehat{\mathbf{D}}$ of the fitted reference SDE, while their dispersion provides an ensemble-based estimate of uncertainty in these inferred fields.
This estimation uncertainty incorporates variation across both train--validation partitions and SWAG parameter samples, and is distinct from the stochastic variability represented by the diffusion matrix $\mathbf{D}$ itself.
Application-specific uncertainty estimates for the inferred fields are presented in Sec.~\ref{sec:additional_results}.

The complete training and ensemble-construction procedure is summarized in Algorithm~\ref{alg:swag_lbn}. The procedure is applied separately to the drift and diffusion networks, with the targets defined below.
\begin{algorithm}[H]
\caption{Training and ensemble construction for LBN with SWAG}
\label{alg:swag_lbn}
\begin{algorithmic}[1]
    \Require
    Network architecture, optimizer, and dataset
    $\mathcal{D}\equiv\{(\bm{q}_t,\bm{y}_t)\}$,
    where $\bm{q}_t\equiv(\bm{x}_t,\bm{c}_t)$ denotes the network input
    and $\bm{y}_t$ denotes the target associated with the quantity to infer.

    \State Split $\mathcal{D}$ into $k$ approximately equally sized folds
    $\{\mathcal{D}_1,\dots,\mathcal{D}_k\}$.

    \For{$i=1,\dots,k$}
        \State Set
        $\mathcal{D}_{\rm val}^{(i)}\gets\mathcal{D}_i$
        and
        $\mathcal{D}_{\rm tr}^{(i)}
        \gets\mathcal{D}\setminus\mathcal{D}_i$.

        \State Initialize network parameters $\bm{\theta}_i$.

        \For{epoch $=1,\dots,N_{\rm epoch}$}
            \State Compute the training loss
            \[
            \mathcal{L}
            \bigl(
            \bm{\theta}_i,
            \mathcal{D}_{\rm tr}^{(i)}
            \bigr)
            =
            \frac{1}{2}
            \left\langle
            \left\|
            \bm{y}
            -
            \bm{y}_{\bm{\theta}_i}(\bm{q})
            \right\|^2
            \right\rangle_{\mathcal{D}_{\rm tr}^{(i)}}.
            \]
            \State Perform an optimization step on
            $\mathcal{L}(\bm{\theta}_i,\mathcal{D}_{\rm tr}^{(i)})$.
            \State Evaluate the validation loss
            $\mathcal{L}(\bm{\theta}_i,\mathcal{D}_{\rm val}^{(i)})$.
            \If{the validation loss improves}
                \State Save the current parameters as the best-validation checkpoint
                $\bm{\theta}_{i}^{\ast}$.
            \EndIf
        \EndFor

        \State Reload the best-validation checkpoint
        $\bm{\theta}_i \gets \bm{\theta}_{i}^{\ast}$.

        \State Continue training from $\bm{\theta}_{i}^{\ast}$ for the SWAG collection phase.
        \For{epoch $=1,\dots,N_{\rm epoch}^{\rm SWAG}$}
            \State Update $\bm{\theta}_i$ using
            $\mathcal{D}_{\rm tr}^{(i)}$.
            \If{the SWAG collection interval is reached}
                \State Record the current parameter snapshot and update the SWAG statistics.
            \EndIf
        \EndFor

        \State Construct the fold-specific SWAG approximation
        $\mathcal{P}_i(\bm{\theta})
        \approx
        \mathcal{N}
        \left(
        \bar{\bm{\theta}}_i,
        \mathbf{C}_{\theta,i}
        \right)$
        from the collected parameter snapshots.
    \EndFor

    \State Draw $N_{\rm ens}/k$ parameter samples from each
    $\mathcal{P}_i(\bm{\theta})$
    to form the ensemble
    $\{\bm{\theta}^{(j)}\}_{j=1}^{N_{\rm ens}}$.

    \State For a given input $\bm{q}=(\bm{x},\bm{c})$, evaluate
    $\bm{y}_{\bm{\theta}^{(j)}}(\bm{q})$
    for all ensemble members.

    \State Compute the ensemble mean
    $\hat{\bm{y}}(\bm{q})$
    and covariance
    $\hat{\mathbf{C}}_{\bm{y}}(\bm{q})$,
    representing the ensemble estimate and its associated estimation covariance, respectively.
\end{algorithmic}
\end{algorithm}
\noindent
To connect the generic learning procedure in Algorithm~\ref{alg:swag_lbn} to SDE inference, we construct transition-level targets from the observed state increments.
For a transition $\bm{x}_t \rightarrow \bm{x}_{t+\Delta t}$, the drift and diffusion targets are defined as
\begin{equation}
\bm{y}^{(F)}_t
\equiv
\frac{\Delta \bm{x}_t}{\Delta t},
\qquad
\mathbf{Y}^{(D)}_t
\equiv
\frac{\Delta\bm{x}_t\Delta\bm{x}_t^\top}{2\Delta t},
\end{equation}
where $\Delta \bm{x}_t \equiv \bm{x}_{t+\Delta t}-\bm{x}_t$.
Accordingly, the generic target $\bm{y}_t$ in Algorithm~\ref{alg:swag_lbn} is given by
$\bm{y}^{(F)}_t$ for the drift network and by the independent components of
$\mathbf{Y}^{(D)}_t$ for the diffusion network.
When observation intervals vary across transitions, each target is constructed using each transition's observation interval $\Delta t$.
Under the squared-error loss used in Algorithm~\ref{alg:swag_lbn}, the network outputs estimate the conditional means of these targets, yielding
\begin{equation}
\begin{aligned}
\widehat{\bm{F}}_{\bm{\theta}_F}(\bm{x},\bm{c})
&=
\left\langle
\left.
\frac{\Delta\bm{x}_t}{\Delta t}
\right|
\bm{x}_t=\bm{x},\,\bm{c}_t=\bm{c}
\right\rangle,
\qquad
\widehat{\mathbf{D}}_{\bm{\theta}_D}(\bm{x},\bm{c})
=
\left\langle
\left.
\frac{\Delta\bm{x}_t\Delta\bm{x}_t^\top}{2\Delta t}
\right|
\bm{x}_t=\bm{x},\,\bm{c}_t=\bm{c}
\right\rangle.
\end{aligned}
\label{eq:OLE_estimators}
\end{equation}
These conditional moments are finite-interval approximations to the first two Kramers--Moyal coefficients and are widely used to infer drift and diffusion from observed stochastic trajectories~\cite{risken1996fokker, friedrich2011approaching, frishman2020learning, gao2024learning, bae2025inferring}.

\subsection{Network Architecture and Training Configuration}

The drift and diffusion estimators are implemented as separate feed-forward neural networks with the same hidden-layer architecture, as summarized in Table~\ref{table:LBN_architecture}. 
Each network consists of four hidden fully connected (FC) layers with $H=256$ units, followed by a task-specific output layer. An ELU activation is applied after each hidden layer. 
For a network input $\bm{q}\equiv(\bm{x},\bm{c})$, the drift network produces a $d$-dimensional output representing $\widehat{\bm{F}}_{\bm{\theta}_F}$, whereas the diffusion network produces $d(d+1)/2$ outputs that are subsequently mapped to the symmetric diffusion matrix $\widehat{\mathbf{D}}_{\bm{\theta}_D}$.
Hidden-layer weights were initialized using Kaiming-normal initialization, whereas the output-layer weights were initialized using Xavier-normal initialization; all biases were initialized to zero.

\begin{table}[!t]
\caption{
Network architectures of the drift (left) and diffusion (right) estimators.
Both networks consist of four hidden fully connected (FC) layers with
$H=256$ hidden units and ELU activations, followed by an output layer.
Here, $d$ denotes the dimension of the state space and $d_{\rm feat}$ denotes the dimension of $\bm{q}_{\rm feat}\equiv(\bm{x},\bm{c}_{\rm cont})$ supplied to the first hidden layer.
For the diffusion estimator, the $d(d+1)/2$ output components are mapped to
a symmetric positive-definite $d\times d$ matrix as described in the text.
}
\centering
\begin{minipage}[t]{0.49\textwidth}
\centering
\textbf{Drift estimator}\\[0.5em]
\begin{tabular}{c|c|c|c}
\hline
Layer & Input dim. & Output dim. & Activation \\
\hline
FC layer 1 & $d_{\rm feat}$ & $H$ & ELU \\
FC layer 2 & $H$ & $H$ & ELU \\
FC layer 3 & $H$ & $H$ & ELU \\
FC layer 4 & $H$ & $H$ & ELU \\
Output layer & $H$ & $d$ & None \\
\hline
\end{tabular}
\end{minipage}
\hfill
\begin{minipage}[t]{0.49\textwidth}
\centering
\textbf{Diffusion estimator}\\[0.5em]
\begin{tabular}{c|c|c|c}
\hline
Layer & Input dim. & Output dim. & Activation \\
\hline
FC layer 1 & $d_{\rm feat}$ & $H$ & ELU \\
FC layer 2 & $H$ & $H$ & ELU \\
FC layer 3 & $H$ & $H$ & ELU \\
FC layer 4 & $H$ & $H$ & ELU \\
Output layer & $H$ & $d(d+1)/2$ & None \\
Symmetric map & $d(d+1)/2$ & $d\times d$ & Softplus eig. \\
\hline
\end{tabular}
\end{minipage}
\label{table:LBN_architecture}
\end{table}

Continuous and categorical conditioning variables are incorporated differently in the network architecture.
We decompose the conditioning variables as
$\bm{c}\equiv(\bm{c}_{\rm cont},\bm{c}_{\rm cat})$,
where $\bm{c}_{\rm cont} \in \mathbb{R}^{d_{\rm cont}}$ denotes continuous conditioning variables and
$\bm{c}_{\rm cat}\in\mathbb{R}^{d_{\rm cat}}$ denotes the one-hot representation of categorical conditioning variables.
Continuous conditioning variables are treated as ordinary input features and concatenated with the state vector, so that the feature input to the first hidden layer is
$\bm{q}_{\rm feat}\equiv(\bm{x},\bm{c}_{\rm cont})\in\mathbb{R}^{d_{\rm feat}}$ with $d_{\rm feat}=d + d_{\rm cont}$.
Categorical conditioning variables instead modulate the hidden representations through a FiLM-style feature-wise affine transformation~\cite{perez2018film}.
Specifically, for each hidden layer $\ell=1,\ldots,4$, we apply
\begin{equation}
\bm{h}_{\ell}
\mapsto
\left[
\bm{1}+\bm{\gamma}_{\ell}(\bm{c}_{\rm cat})
\right]
\odot
\bm{h}_{\ell}
+
\bm{\beta}_{\ell}(\bm{c}_{\rm cat}),
\end{equation}
before the ELU activation, where $\bm{1}$ denotes the vector of ones, $\bm{h}_{\ell}\in\mathbb{R}^{H}$ denotes the pre-activation hidden representation,
$\bm{\gamma}_{\ell},\bm{\beta}_{\ell}:\mathbb{R}^{d_{\rm cat}}\rightarrow\mathbb{R}^{H}$
are learned linear mappings specific to layer $\ell$, and $\odot$ denotes element-wise multiplication.
The weights and biases of these modulation mappings are initialized to zero, so that the affine modulation initially reduces to the identity transformation.
No categorical modulation is applied to the output layer.
When no categorical conditioning variables are present, this modulation is omitted and the hidden layers reduce to the standard FC layers described above.
In the Google Books Ngram application, calendar time was rescaled to 0.01 time units per year and mean-centered before being supplied as a continuous input feature, whereas lexicon identity was incorporated as a categorical condition.

To satisfy the symmetry and positive-semidefiniteness requirements of the diffusion matrix~\cite{risken1996fokker}, we parameterize its numerical estimate to be positive definite.
The $d(d+1)/2$ outputs of the diffusion network are first arranged as the independent elements of a symmetric matrix $\mathbf{S}$.
We then perform the eigendecomposition
$\mathbf{S}=\mathbf{U}\operatorname{diag}(\bm{\lambda})\mathbf{U}^{\top}$
and transform the eigenvalues according to
\begin{equation}
\widehat{\mathbf{D}}_{\bm{\theta}_D}
=
\mathbf{U}
\operatorname{diag}
\left[
\operatorname{Softplus}(\bm{\lambda})+\epsilon
\right]
\mathbf{U}^{\top},
\end{equation}
where $\operatorname{Softplus}(u)=\log(1+e^u)$ is applied element-wise and $\epsilon=10^{-9}$ is included for numerical stability.
This construction preserves symmetry while ensuring strictly positive eigenvalues of the estimated diffusion matrix.

Across all three empirical applications, the drift and diffusion estimators were trained independently using the same overall optimization procedure and identical 10-fold cross-validation partitions. The transitions within each trajectory were randomly divided into ten approximately equal subsets. In each fold, one subset from each trajectory was used for validation and the remaining transitions were used for training.
The initial learning rate and minibatch size were tuned separately for each estimator and application based on the validation loss averaged across the ten folds, with a single selected configuration held fixed across all fold-specific models. The resulting learning rate was the same for both estimators within each application: $5\times10^{-6}$ for DIG, $10^{-5}$ for Seshat, and $5\times10^{-4}$ for the Google Books Ngram analysis. The corresponding minibatch sizes for the drift and diffusion estimators were $(512,256)$, $(128,64)$, and $(512,512)$, respectively.
During the initial optimization phase, network parameters were optimized for 1,000 epochs using AdamW with a weight-decay coefficient of $10^{-3}$. With no learning-rate warm-up, the learning rate was cosine-annealed from its selected initial value toward a minimum value of $10^{-9}$. For each fold, the checkpoint attaining the lowest validation loss was retained.

Starting from the retained best-validation checkpoint, training was continued for the SWAG collection phase. The SWAG learning rate was set to $10^{-3}$ for DIG and Seshat and to $10^{-4}$ for the Google Books Ngram analysis. We collected 50 parameter snapshots at three-epoch intervals and used them to construct a fold-specific SWAG approximation characterized by the parameter mean, diagonal variance, and a low-rank covariance component of rank 20. 
For prediction, we drew 20 parameter samples from each fold-specific SWAG approximation and pooled them across folds as described in Sec.~\ref{subsec:inference_procedure}, yielding 200 ensemble members.

\subsection{Robustness to Finite Observation Intervals}
\label{subsec:LBN_finite_interval}

The conditional moments in Eq.~\eqref{eq:OLE_estimators} converge to the infinitesimal Kramers--Moyal coefficients in the short-time limit $\Delta t \rightarrow 0$ and generally contain finite-time corrections at finite observation intervals~\cite{lade2009finite}.
Importantly, however, whether a given observation interval is effectively ``short'' is system-dependent and must be judged in relation to how rapidly the constructed state evolves, rather than by its absolute duration in calendar time.
Thus, annual sampling in DIG, two-year sampling in the Google Books Ngram trajectories, or century-scale sampling in Seshat cannot be regarded as intrinsically fine or coarse based on their calendar durations alone.

\begin{figure*}[!t]
  \centering
  \includegraphics[width=.9\linewidth]{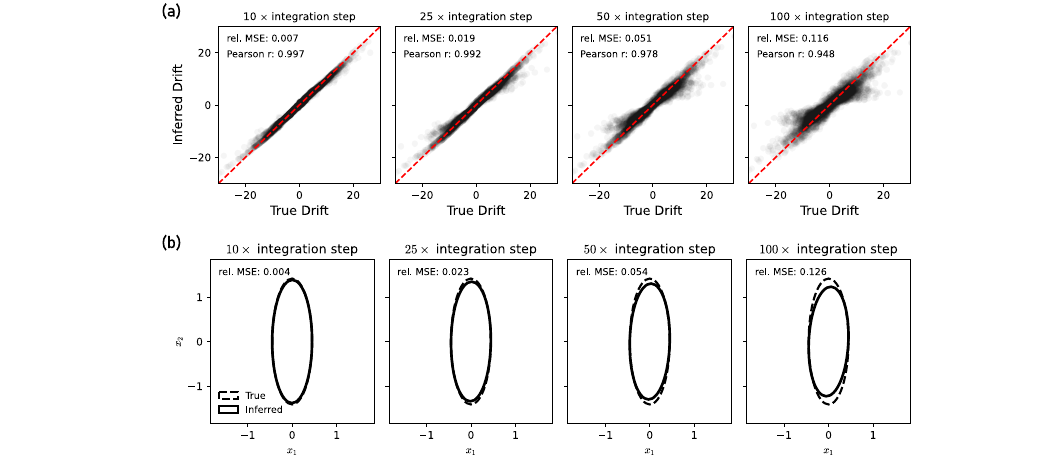}
  \caption{
  LBN inference under temporal subsampling of a stochastic van der Pol system. Columns correspond to trajectories subsampled every 10, 25, 50, and 100 numerical integration steps, respectively. (a) Comparison between the true and inferred drift components. Gray points denote individual evaluations, and the red dashed line indicates equality between the inferred and true values. Insets report the relative mean-squared error (rel. MSE) and Pearson correlation coefficient. (b) Comparison of the true (dashed) and inferred (solid) diffusion ellipses for these sampling intervals. Inferred ellipses correspond to the diffusion matrices averaged over the evaluation states. Insets report the relative mean-squared error.
}
\label{fig:LBN_finite_interval}
\end{figure*}

To assess the practical robustness of LBN inference to finite observation intervals, we performed a controlled subsampling experiment using a stochastic van der Pol system:
\begin{equation}
\begin{aligned}
F_1(\bm{x}) &= x_2, \qquad
F_2(\bm{x}) = \mu(1-x_1^2)x_2-x_1,
\end{aligned}
\end{equation}
with $\mu=1$ and $\mathbf{D}=\operatorname{diag}(0.2,\,2.0)$.
Original trajectories were generated using a numerical integration step $\delta t_{\rm int}=10^{-3}$ and then subsampled every 10, 25, 50, and 100 integration steps, corresponding to observation intervals $\Delta t=0.01$, $0.025$, $0.05$, and $0.1$, respectively. 
The same LBN inference procedure was applied independently to each subsampled dataset. 
Here, we quantify estimation error using the relative mean-squared error,
$\mathrm{rel.\ MSE}=\|\widehat{Y}-Y\|^2/\|Y\|^2$,
where $Y$ and $\widehat{Y}$ denote the true and inferred values of either the drift or diffusion, evaluated over the common test states.
As shown in Fig.~\ref{fig:LBN_finite_interval}, quantitative deviations increase progressively with temporal coarsening, but the inferred drift remains closely aligned with the true field and the inferred diffusion retains the principal magnitude and anisotropy of the true diffusion across the range examined, including at $\Delta t=100\,\delta t_{\rm int}$.
These results therefore demonstrate that the principal structure of the underlying drift and diffusion can remain recoverable from observations sampled at finite temporal intervals.

\section{Empirical Adequacy of the Inferred Dynamics}
\label{sec:empirical_adequacy}

For empirical systems, no ground-truth drift and diffusion fields are available against which the inferred fields can be compared directly.
We therefore assess the empirical adequacy of the fitted SDEs using complementary simulation-based and temporal-correlation diagnostics. 
Specifically, we examine whether trajectories generated from the inferred dynamics reproduce the broad ensemble-level evolution, dispersion, and temporal persistence of the empirical trajectories, and whether substantial temporal dependence remains after conditioning on the inferred local dynamics. 
These diagnostics are not intended to establish that the fitted SDEs provide exact descriptions of the empirical processes, but to assess whether they capture the relevant dynamical structure sufficiently well to serve as probabilistic reference dynamics for the subsequent analyses.

\begin{figure*}[!t]
  \centering
  \includegraphics[width=\linewidth]{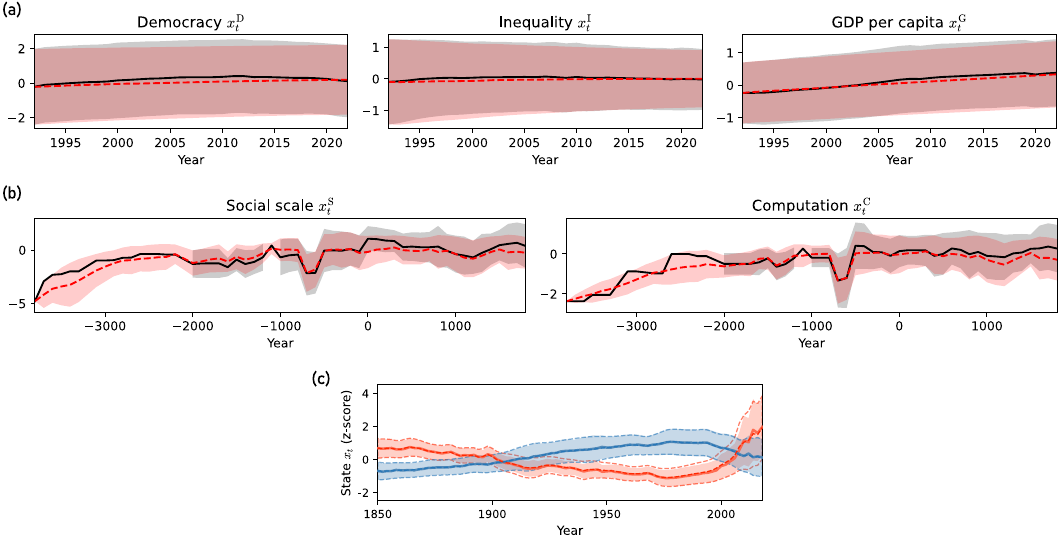}
  \vskip -0.1in
  \caption{Stochastic trajectories generated from the inferred SDEs are compared with the empirical trajectories for (a) the Democracy–Inequality–Growth (DIG) dataset, (b) the Seshat dataset, and (c) the Google Books Ngram dataset. 
  For each empirical trajectory, simulations were initialized at the observed initial state and time, and propagated over the same observation period. 
  In (a,b), black solid lines and gray shaded bands show the empirical mean and $\pm 1$ standard deviation across trajectories represented at each observation time, respectively, while red dashed lines and red shaded bands show the simulated mean and $\pm 1$ standard deviation. 
  In (c), blue and red denote the Rationality and Emotionality lexicons, respectively; solid lines and shaded bands show the empirical mean and $\pm 1$ standard deviation across words, while dashed lines show the mean and $\pm 1$ standard-deviation envelope for simulated trajectories.}
  \label{fig:mean_trajs}
\vskip -0.1in
\end{figure*}

We assessed the generative behavior of the inferred SDEs by initializing stochastic simulations for each empirical trajectory at its observed initial state and time, propagating them over the same time horizon, and evaluating them at the observation times.
All stochastic trajectories were generated using the Euler–Maruyama scheme with an integration step of $\delta t_{\rm int}=10^{-2}$, with the inferred fields evaluated at each integration step.
Because the fitted SDE defines a distribution over possible trajectories, we assess agreement at the ensemble level, focusing on the statistics of the simulated and observed trajectory distributions instead of pathwise agreement between individual simulations and observations.
Figure~\ref{fig:mean_trajs} compares the resulting simulated and empirical ensembles in terms of their time-dependent means and dispersions.
For both DIG and Seshat, the simulated ensembles capture the broad temporal evolution of the observed states together with comparable levels of dispersion. 
For the Google Books Ngram data, the simulated ensembles closely reproduce the contrasting long-run evolution of the Rationality and Emotionality lexicons, including their reversal around 1980.

\begin{figure*}[!t]
  \centering
  \includegraphics[width=\linewidth]{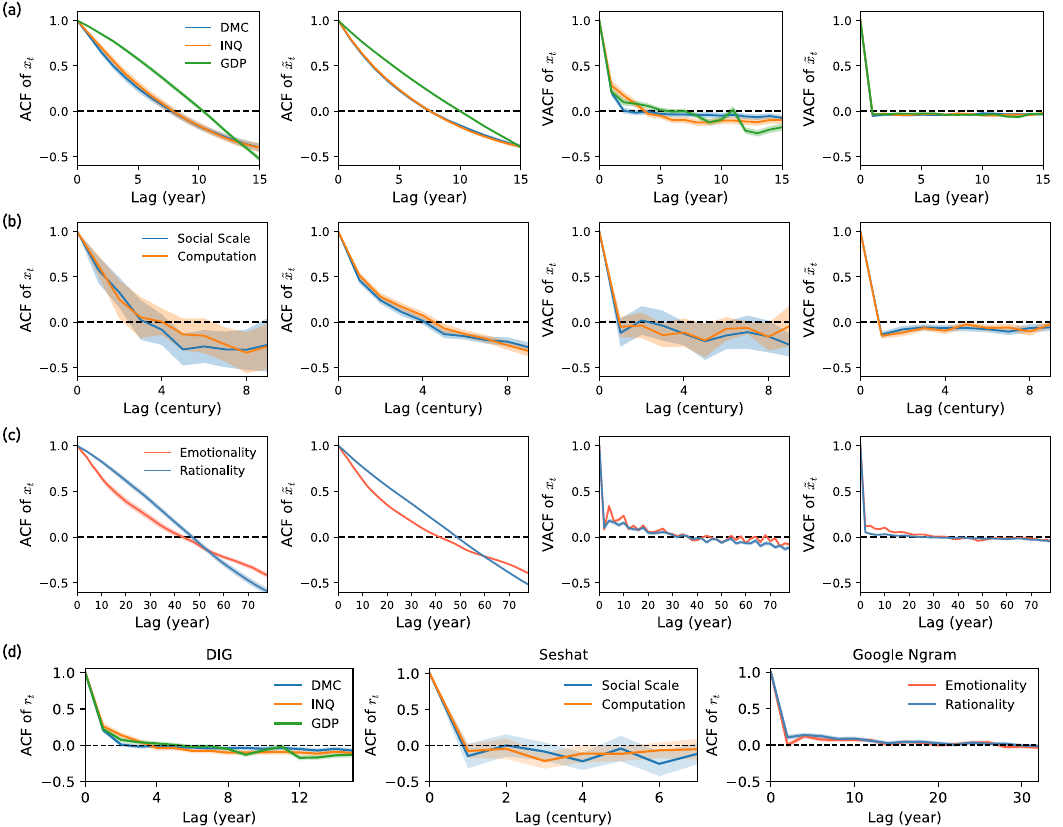}
  \vskip -0.1in
  \caption{Autocorrelation functions (ACFs) and velocity autocorrelation functions (VACFs) are shown for (a) the Democracy–Inequality–Growth (DIG) dataset, (b) the Seshat dataset, and (c) the Google Books Ngram dataset. From left to right, columns show the ACFs of the observed states $\bm{x}_t$, the ACFs of states $\tilde{\bm{x}}_t$ generated from the inferred SDE, the VACFs of the observed trajectories, and the VACFs of the generated trajectories. 
  For each trajectory, VACFs were computed from finite-difference velocities $(\bm{x}_{t+\Delta t} - \bm{x}_t)/\Delta t$, and ACFs and VACFs were evaluated separately before averaging the resulting correlation functions across trajectories.
  (d) ACFs of the standardized residual increments $\bm{r}_t$ (see Eq.~\eqref{eq:residual}) for the DIG, Seshat, and Google Books Ngram datasets.
  Colors denote the state variables or lexical groups identified in the legends. Shaded bands denote pointwise 95\% confidence intervals estimated from trajectory-to-trajectory variation in the correlation functions, and horizontal dashed lines mark zero correlation.}
  \label{fig:ACFs}
  \vskip -0.1in
\end{figure*}

We next examined the temporal correlation structure of the empirical and simulated trajectories by comparing their autocorrelation functions (ACFs) and velocity autocorrelation functions (VACFs) [Fig.~\ref{fig:ACFs}(a--c)].
Here, ACFs were computed separately for each trajectory after centering each state component by its trajectory-specific temporal mean and normalizing the lagged covariance by its trajectory-level variance. 
VACFs were computed analogously from the finite-difference velocities $(\bm{x}_{t+\Delta t}-\bm{x}_t)/\Delta t$.
For irregularly sampled trajectories, observation pairs were matched according to their elapsed-time lags, and the resulting trajectory-level correlation functions were averaged with equal weight across trajectories.
Across all three domains, the generated trajectories reproduce the broad decay of the empirical state ACFs, including correlations that persist across multiple successive observations and differences in decay rates among state variables or lexical groups.
Notably, the empirical state correlations remain appreciable beyond the first observation lag, indicating temporal persistence that extends across multiple successive observations.
In contrast to the persistent state correlations, the empirical VACFs decay sharply after zero lag, with only limited correlations remaining at subsequent lags. The simulated trajectories exhibit similarly rapid relaxation.
This separation between persistent state correlations and rapidly relaxing velocity correlations is consistent with an effective overdamped Langevin description at the temporal resolution of the observations.

To examine more directly whether persistent temporal dependence remains after accounting for the inferred local dynamics, we analyzed the ACFs of the standardized residuals [Fig.~\ref{fig:ACFs}(d)]. For each observed transition, we defined the residual as 
\begin{equation}
\bm{r}_t \equiv \left[2\hat{\mathbf{D}}(\bm{x}_t, \bm{c}_t) \Delta t \right]^{-1/2} \left( \Delta \bm{x}_t - \hat{\bm{F}}(\bm{x}_t, \bm{c}_t)\Delta t \right),
\label{eq:residual}
\end{equation}
which removes the inferred conditional mean increment and rescales the fluctuation by its local covariance. 
Across all three domains, the residual correlations drop sharply at the first lag and remain small at subsequent lags. Thus, little persistent temporal correlation remains after conditioning on the inferred local drift and diffusion.
Together with the simulation-based comparisons and the rapid relaxation of the VACFs, this result provides empirical support for an effective overdamped SDE with approximate Markovian closure at the temporal resolution of the observations.

\section{Exogeneity Diagnostics}
\label{sec:exo_scores}

Having established the empirical adequacy of the inferred SDEs in Sec.~\ref{sec:empirical_adequacy}, we use the fitted SDE as a probabilistic reference for evaluating individual observed transitions.
For each observed transition, the reference process defines a finite-time predictive distribution against which the realized state change can be evaluated.
We therefore define \textit{exogeneity} operationally as statistical atypicality relative to the fitted reference dynamics.
A transition assigned low probability by the inferred SDE is interpreted as an exogenous perturbation in this model-relative sense, without implying that its underlying cause is necessarily external to the social system.

To quantify such departures, we introduce three complementary exogeneity scores based on local, bridge, and endpoint probabilities. 
Although these probabilities evaluate different aspects of an observed trajectory, they share a common construction: each maps a standardized discrepancy under its diagnostic distribution to an upper-tail probability. 
We transform these probabilities into exogeneity scores as
\begin{equation}
\mathcal{E}^{\alpha}_t=-\log_{10} p^{\alpha}_t,\qquad \alpha\in\{\mathrm{loc},\mathrm{brg},\mathrm{end}\}.
\end{equation}
Larger values of $\mathcal{E}^{\alpha}_t$ therefore indicate greater atypicality under the relevant diagnostic distribution.
We describe the local score first, followed by the two-step bridge and endpoint scores and their joint interpretation.

For the derivations below, we take the reference SDE as given, with drift field $\bm{F}$ and diffusion matrix $\mathbf{D}$.
For notational simplicity, we suppress the dependence of $\bm{F}$ and $\mathbf{D}$ on $\bm{c}_t$ throughout this section, with the understanding that they are evaluated at the relevant conditioning values where applicable.
In the empirical applications, the exogeneity scores are evaluated using the drift and diffusion fields inferred from the observed trajectories.

\subsection{Local Exogeneity Score}

The local exogeneity score quantifies how atypical a single observed transition is relative to the expected transition distribution under the reference dynamics.
Consider the incoming transition from $\bm{x}_{-}\equiv\bm{x}_{t-\Delta t_-}$ to $\bm{x}_t$, with displacement $\Delta\bm{x}^{\mathrm{in}}_t \equiv \bm{x}_t-\bm{x}_{-}$.
Here, $\Delta t_-$ denotes the observation interval between $\bm{x}_{-}$ and $\bm{x}_t$, which may vary across transitions.
Conditional on $\bm{x}_{-}$, the reference SDE defines an ensemble of possible one-step increments over $\Delta t_-$.
For the local score, we approximate the distribution of these increments by
\begin{equation}
\Delta\bm{x}^{\mathrm{in}}_t
\mid
\bm{x}_{-}
\sim
\mathcal{N}\!\left(
\bm{F}_{-}\Delta t_-,
2\mathbf{D}_{-}\Delta t_{-}
\right),
\label{eq:transition}
\end{equation}
where $\bm{F}_{-}\equiv\bm{F}(\bm{x}_{-})$ and $\mathbf{D}_{-}\equiv\mathbf{D}(\bm{x}_{-})$.
To evaluate the observed displacement relative to the local stochastic scale, we define its whitened residual
\begin{equation} 
\bm{r}^{\mathrm{in}}_t \equiv 
\left( 2\mathbf{D}_{-} \Delta t_- \right)^{-1/2} \left( \Delta\bm{x}^{\mathrm{in}}_t - \bm{F}_{-}\Delta t_- \right). 
\end{equation}
For increments drawn from the local Gaussian approximation in Eq.~\eqref{eq:transition}, $\bm{r}^{\mathrm{in}}_t$ follows a standard $d$-dimensional Gaussian distribution.
The squared norm of $\bm{r}^{\rm in}_t$ gives the local Mahalanobis statistic
\begin{equation} 
Q^{\mathrm{loc}}_t \equiv 
\left\| \bm{r}^{\mathrm{in}}_t \right\|^2 
= \left( \Delta\bm{x}^{\mathrm{in}}_t - \bm{F}_{-}\Delta t_- \right)^{\top} \left( 2\mathbf{D}_{-}\Delta t_- \right)^{-1} \left( \Delta\bm{x}^{\mathrm{in}}_t - \bm{F}_{-}\Delta t_- \right), 
\end{equation}
which follows a chi-squared distribution with $d$ degrees of freedom.
The covariance normalization in $Q^{\mathrm{loc}}_t$ accounts for both the scale and anisotropy of the local stochastic fluctuations, so that an equally large displacement is less atypical along directions of high intrinsic variability than along directions of low intrinsic variability.
Consequently, we define the local tail probability and exogeneity score as \begin{equation} 
p^{\mathrm{loc}}_{t} \equiv \Pr\!\left( \chi_d^2 \geq Q^{\mathrm{loc}}_t \right), 
\qquad 
\mathcal{E}^{\mathrm{loc}}_{t} \equiv -\log_{10}p^{\mathrm{loc}}_{t}. 
\end{equation}
A large $\mathcal{E}^{\mathrm{loc}}_t$ indicates that the observed transition lies in a low-probability tail of the transition distribution expected under the reference dynamics and is thus poorly accounted for by the reference SDE alone.
Such transitions are flagged as exogenous perturbations in the model-relative sense defined above.

The local score has a natural baseline under the reference dynamics.
Let $F_{\chi_d^2}$ denote the cumulative distribution function of a chi-squared random variable with $d$ degrees of freedom.
For transitions generated by the reference dynamics, the local Gaussian approximation gives $Q^{\mathrm{loc}}\sim\chi_d^2$.
Hence, for any $u\in[0,1]$,
\begin{equation}
\begin{aligned}
\Pr\!\left(p^{\mathrm{loc}}\leq u\right)
=
\Pr\!\left(1-F_{\chi_d^2}(Q^{\mathrm{loc}})\leq u\right)
=
\Pr\!\left(Q^{\mathrm{loc}}\geq F_{\chi_d^2}^{-1}(1-u)\right)
=
u,
\end{aligned}
\end{equation}
where $F_{\chi_d^2}^{-1}(1-u)$ is the $(1-u)$-quantile, leaving the probability $u$ in the upper tail.
Thus, $p^{\mathrm{loc}}\sim U(0,1)$ for such transitions, which gives
\begin{equation}
\Pr\!\left(\mathcal{E}^{\mathrm{loc}}\geq s\right)
=
\Pr\!\left(p^{\mathrm{loc}}\leq 10^{-s}\right)
=
10^{-s},
\qquad s\geq0,
\end{equation}
where $U$ denotes the uniform distribution.
A score threshold $s$ is exceeded by a fraction $10^{-s}$ of transitions generated by the reference dynamics alone.
The uniformity of $p^{\mathrm{loc}}$ also yields a nominal expectation for the local exogeneity score.
Averaging $-\log_{10}p^{\mathrm{loc}}$ over $p^{\mathrm{loc}}\sim  U(0,1)$ gives
\begin{equation}
\langle \mathcal{E}^{\rm loc} \rangle 
=
\int_0^1 -\log_{10}p\,dp
=
\frac{1}{\ln 10}
\simeq 0.434.
\label{eq:nominal_expectation}
\end{equation}
This nominal expectation $1/\ln 10$ is shown as the dashed horizontal line in Fig.~5(d) of the main text, providing a baseline against which the lexicon-mean local exogeneity scores are compared.

\subsection{Two-step Exogeneity Scores}

The local score determines whether the incoming transition is atypical, but does not characterize how the departure unfolds over the next transition. 
To examine this two-step pattern, we consider three consecutive observations, $\bm{x}_{-} \equiv \bm{x}_{t-\Delta t_{-}}$, $\bm{x}_{t}$, and $\bm{x}_{+} \equiv \bm{x}_{t+\Delta t_{+}}$, where the adjacent observation intervals $\Delta t_{-}$ and $\Delta t_{+}$ may differ across transitions.
The two-step dynamics can be viewed from two complementary perspectives: whether the observed intermediate state $\bm{x}_{t}$ is atypical given its neighboring observations, and whether the subsequent endpoint $\bm{x}_{+}$ is atypical after accounting for the range of possible intermediate states. 
These perspectives motivate the bridge and endpoint diagnostics introduced below.

For a candidate intermediate state $\bm{z}$ and a candidate endpoint $\bm{y}$, the local Gaussian transition approximation gives
\begin{equation}
\begin{aligned}
P_{\Delta t_{-}}\!(\bm{z}\mid\bm{x}_{-}) \approx \mathcal{N}\!\left( \bm{z}\mid \bm{m}_{-}, \mathbf{C}_{-} \right), 
\qquad
P_{\Delta t_{+}}\!(\bm{y}\mid\bm{z}) \approx \mathcal{N} \left( \bm{y}\mid \bm{m}_{+}(\bm{z}), \mathbf{C}_{+}(\bm{z}) \right),
\end{aligned}
\label{eq:two_step_transitions}
\end{equation}
where
\begin{equation}
\begin{aligned}
\bm{m}_{-}
&\equiv
\bm{x}_{-}
+
\bm{F}(\bm{x}_{-})\Delta t_{-},
&
\mathbf{C}_{-}
&\equiv
2\mathbf{D}(\bm{x}_{-})\Delta t_{-},\\
\bm{m}_{+}(\bm{z})
&\equiv
\bm{z}
+
\bm{F}(\bm{z})\Delta t_{+},
&
\mathbf{C}_{+}(\bm{z})
&\equiv
2\mathbf{D}(\bm{z})\Delta t_{+}.
\end{aligned}
\end{equation}
These transition densities define the common two-step predictive kernel
\begin{equation}
K_t(\bm{z},\bm{y}) \equiv P_{\Delta t_{-}}\!(\bm{z}\mid\bm{x}_{-}) P_{\Delta t_{+}}\!(\bm{y}\mid\bm{z}).
\label{eq:two_step_kernel}
\end{equation}
This kernel $K_t(\bm{z},\bm{y})$ represents the joint predictive density of the intermediate state $\bm{z}$ and the endpoint $\bm{y}$, conditional on $\bm{x}_{-}$.
The bridge and endpoint diagnostics are obtained from this kernel by conditioning on the observed endpoint and marginalizing over the intermediate state, respectively.
As in the local diagnostic, each is summarized by a Mahalanobis-type discrepancy under an appropriate Gaussian approximation.
We then convert this discrepancy to an upper-tail probability and define the exogeneity score.

\paragraph*{\textbf{Bridge score.}}
The bridge score evaluates whether the observed intermediate state $\bm{x}_{t}$ is atypical conditional on its two neighboring observations $\bm{x}_{-}$ and $\bm{x}_{+}$.
By the Markov property, the conditional bridge density is
\begin{equation}
\pi_t^{\mathrm{brg}}(\bm{z}) \equiv P(\bm{z}\mid\bm{x}_{-},\bm{x}_{+}) = 
\frac{K_t(\bm{z},\bm{x}_{+})}{\displaystyle \int K_t(\bm{z}',\bm{x}_{+})\,d\bm{z}' }.
\label{eq:bridge_density}
\end{equation}
Using Eqs.~\eqref{eq:two_step_transitions} and~\eqref{eq:two_step_kernel}, the negative log of the bridge density is expressed as
\begin{equation}
\begin{aligned}
\Phi_t^{\mathrm{brg}}(\bm{z})
&\equiv\; -\ln \pi_t^{\mathrm{brg}}(\bm{z})
\\
&= \frac{1}{2}
(\bm{z}-\bm{m}_{-})^{\top}
\mathbf{C}_{-}^{-1}
(\bm{z}-\bm{m}_{-})
+
\frac{1}{2}
\left( \bm{x}_{+} - \bm{m}_{+}(\bm{z}) \right)^{\top}
\mathbf{C}_{+}(\bm{z})^{-1}
\left( \bm{x}_{+} - \bm{m}_{+}(\bm{z}) \right)
+
\frac{1}{2}
\ln \det\mathbf{C}_{+}(\bm{z})
+ \mathrm{const.}
\end{aligned}
\label{eq:bridge_potential}
\end{equation}
The first two quadratic terms are one-half the squared Mahalanobis discrepancies under
$P_{\Delta t_{-}}(\bm{z}\mid\bm{x}_{-})$ and
$P_{\Delta t_{+}}(\bm{x}_{+}\mid\bm{z})$, respectively.
The log-determinant term accounts for the state dependence of $\mathbf{C}_{+}(\bm{z})$. 
Because $\mathbf{C}_{-}$ is independent of $\bm{z}$, the term involving $\mathbf{C}_{-}$ is absorbed into the constant.

From Eq.~\eqref{eq:bridge_potential}, we identify the most probable intermediate state $\bm{\mu}_{t}^{\mathrm{brg}}$, which we refer to as the bridge mode, and characterize the local curvature of $\Phi_t^{\mathrm{brg}}$ around this mode by
\begin{equation}
\bm{\mu}_{t}^{\mathrm{brg}}
\equiv
\arg\min_{\bm{z}}
\Phi_t^{\mathrm{brg}}(\bm{z}), \qquad \mathbf{H}_{t}^{\mathrm{brg}}
\equiv
\left.
\nabla_{\bm{z}}^{2}
\Phi_t^{\mathrm{brg}}(\bm{z})
\right|_{\bm{z}=\bm{\mu}_{t}^{\mathrm{brg}}}.
\label{eq:bridge_mode}
\end{equation}
A Laplace approximation around the bridge mode gives
\begin{equation}
\pi_t^{\mathrm{brg}}(\bm{z})
\approx
\mathcal{N}
\!\left( \bm{z} \mid
\bm{\mu}_{t}^{\mathrm{brg}},
[\mathbf{H}_{t}^{\mathrm{brg}}]^{-1}
\right).
\label{eq:bridge_laplace}
\end{equation}
Based on Eq.~\eqref{eq:bridge_laplace}, we define the bridge discrepancy as
\begin{equation}
Q_t^{\mathrm{brg}}
\equiv
(\bm{x}_{t}-\bm{\mu}_{t}^{\mathrm{brg}})^{\top}
\mathbf{H}_{t}^{\mathrm{brg}}
(\bm{x}_{t}-\bm{\mu}_{t}^{\mathrm{brg}}).
\label{eq:bridge_Q}
\end{equation}
Here, $Q_t^{\mathrm{brg}}$ is the squared Mahalanobis distance of the observed intermediate state $\bm{x}_t$ under the bridge approximation.
A large $Q_t^{\mathrm{brg}}$ indicates that the observed intermediate state is atypical relative to the bridge distribution conditioned on its two neighboring observations.
Numerically, $Q_t^{\mathrm{brg}}$ is evaluated using a Hessian--vector product, without explicitly constructing $\mathbf{H}_{t}^{\mathrm{brg}}$ or its inverse.

\paragraph*{\textbf{Endpoint score.}}
The endpoint score measures the atypicality of the observed endpoint $\bm{x}_{+}$ relative to the two-step predictive distribution from $\bm{x}_{-}$, with the intermediate state marginalized out.
The marginal predictive density is
\begin{equation}
\pi_t^{\mathrm{end}}(\bm{y})
\equiv
P(\bm{y}\mid\bm{x}_{-})
=
\int
K_t(\bm{z}',\bm{y})\,d\bm{z}'.
\label{eq:endpoint_density}
\end{equation}
Because the dependence of the drift and diffusion on the intermediate state generally makes this marginal density analytically intractable, we characterize it through its first two moments.
Let $\bm{Z}$ denote the random intermediate state under the first-step predictive distribution
\begin{equation}
\bm{Z}
\sim
P_{\Delta t_{-}}\!(\cdot\mid\bm{x}_{-})\approx \mathcal{N}\!\left( \bm{m}_{-}, \mathbf{C}_{-} \right).
\label{eq:endpoint_intermediate}
\end{equation}
Conditional on a realization of $\bm{Z}$, the second-step endpoint distribution has mean $\bm{m}_{+}(\bm{Z})$ and covariance $\mathbf{C}_{+}(\bm{Z})$.
Marginalizing over the distribution of $\bm{Z}$ gives the endpoint mean and covariance
\begin{equation}
\begin{aligned}
\bm{\mu}_{t}^{\mathrm{end}}
&\equiv
\mathbb{E}_{\bm{Z}}\!\left[
\bm{m}_{+}(\bm{Z})
\right],
\qquad
\mathbf{C}_{t}^{\mathrm{end}}
\equiv
\operatorname{Cov}_{\bm{Z}}\!\left[
\bm{m}_{+}(\bm{Z})
\right]
+
\mathbb{E}_{\bm{Z}}\!\left[
\mathbf{C}_{+}(\bm{Z})
\right],
\label{eq:endpoint_moments}
\end{aligned}
\end{equation}
where $\mathbb{E}_{\bm{Z}}[\cdot]$ and $\operatorname{Cov}_{\bm{Z}}[\cdot]$ denote expectation and covariance with respect to $\bm{Z}$.
The endpoint covariance $\mathbf{C}_{t}^{\mathrm{end}}$ combines variation in the propagated mean across possible intermediate states, $\operatorname{Cov}_{\bm{Z}}[\bm{m}_{+}(\bm{Z})]$, with the expected stochastic variability of the second transition, $\mathbb{E}_{\bm{Z}}[\mathbf{C}_{+}(\bm{Z})]$.
We then approximate the marginal endpoint density by the moment-matched Gaussian
\begin{equation}
\pi_t^{\mathrm{end}}(\bm{y})
\approx
\mathcal{N}
\left( \bm{y} \mid
\bm{\mu}_{t}^{\mathrm{end}},
\mathbf{C}_{t}^{\mathrm{end}}
\right),
\label{eq:endpoint_gaussian}
\end{equation}
and define the endpoint discrepancy as
\begin{equation}
Q_t^{\mathrm{end}}
\equiv
(\bm{x}_{+}-\bm{\mu}_{t}^{\mathrm{end}})^{\top}
\left(
\mathbf{C}_{t}^{\mathrm{end}}
\right)^{-1}
(\bm{x}_{+}-\bm{\mu}_{t}^{\mathrm{end}}).
\label{eq:endpoint_Q}
\end{equation}
Here, $Q_t^{\mathrm{end}}$ is the squared Mahalanobis distance of the observed endpoint $\bm{x}_+$ under the endpoint approximation.
A large $Q_t^{\mathrm{end}}$ indicates that the observed endpoint is atypical under the two-step predictive distribution after marginalizing over possible intermediate states.
In numerical calculations, these moments are estimated by Monte Carlo sampling from the Gaussian approximation to the first-step predictive distribution in Eq.~\eqref{eq:endpoint_intermediate}.

\paragraph*{\textbf{Common definitions.}}
Under the Gaussian approximations above, both
$Q_t^{\mathrm{brg}}$ and $Q_t^{\mathrm{end}}$ take the form of $d$-dimensional squared Mahalanobis distances.
We convert these discrepancies to upper-tail probabilities using the $\chi_d^2$ distribution and define
\begin{equation}
p_t^{\alpha}
\equiv
\Pr
\left(
\chi_d^2
\ge
Q_t^{\alpha}
\right), \qquad \mathcal{E}_t^{\alpha}
\equiv
-\log_{10}
p_t^{\alpha},
\label{eq:two_step_scores}
\end{equation}
with $\alpha \in \{ \mathrm{brg}, \mathrm{end} \}$.
Larger $\mathcal{E}_t^{\alpha}$ indicates greater atypicality under the evaluated two-step distribution.
We additionally note that the tail probabilities have the same nominal uniform distribution as the local diagnostic, giving the same nominal expectation $1/\ln 10$ in Eq.~\eqref{eq:nominal_expectation}.
Together, $\mathcal{E}_t^{\mathrm{brg}}$ and $\mathcal{E}_t^{\mathrm{end}}$ characterize whether two-step atypicality appears primarily at the intermediate state, at the subsequent endpoint, at both, or at neither.
The numerical procedure for evaluating the bridge and endpoint exogeneity scores is summarized in Algorithm~\ref{alg:two_step_exogeneity}.

\begin{algorithm}[H]
\caption{Numerical evaluation of the two-step exogeneity scores}
\label{alg:two_step_exogeneity}
\begin{algorithmic}[1]

\Require Three consecutive observations
$(\bm{x}_{-},\,\bm{x}_{t},\, \bm{x}_{+})$,
intervals $\Delta t_{-}$ and $\Delta t_{+}$,
drift $\bm{F}$, diffusion $\mathbf{D}$.

\State Compute
$\bm{m}_{-} \gets
\bm{x}_{-}+\bm{F}(\bm{x}_{-})\Delta t_{-}$
and
$\mathbf{C}_{-} \gets
2\mathbf{D}(\bm{x}_{-})\Delta t_{-}$.

\Statex
\Statex \textbf{Bridge diagnostic}

\State Construct $\Phi_t^{\mathrm{brg}}(\bm{z})$ [Eq.~\eqref{eq:bridge_potential}]:
\begin{equation}
\begin{aligned}
\Phi_t^{\mathrm{brg}}(\bm{z}) = \frac{1}{2}
(\bm{z}-\bm{m}_{-})^{\top}
\mathbf{C}_{-}^{-1}
(\bm{z}-\bm{m}_{-})
+
\frac{1}{2}
\left( \bm{x}_{+} - \bm{m}_{+}(\bm{z}) \right)^{\top}
\mathbf{C}_{+}(\bm{z})^{-1}
\left( \bm{x}_{+} - \bm{m}_{+}(\bm{z}) \right)
+
\frac{1}{2}
\ln \det\mathbf{C}_{+}(\bm{z})
\nonumber
\end{aligned}
\end{equation}
where $\bm{m}_{+}(\bm{z}) = \bm{z} + \bm{F}(\bm{z})\Delta t_{+}$ and $\mathbf{C}_{+}(\bm{z}) =
2\mathbf{D}(\bm{z})\Delta t_{+}$.
The additive constant is omitted because it does not affect the optimization or curvature calculation.

\State Find
$\bm{\mu}_{t}^{\mathrm{brg}}
\gets
\arg\min_{\bm{z}}\Phi_t^{\mathrm{brg}}(\bm{z})$ by numerical minimization. 
\State Set
$\bm{\delta}\gets
\bm{x}_{t}-\bm{\mu}_{t}^{\mathrm{brg}}$ and evaluate
$\mathbf{H}_{t}^{\mathrm{brg}}\bm{\delta}$
using a Hessian--vector product.
\State Compute
$Q_t^{\mathrm{brg}}
\gets
\bm{\delta}^{\top}
(\mathbf{H}_{t}^{\mathrm{brg}}\bm{\delta})$ and obtain
$p_t^{\mathrm{brg}}$ and
$\mathcal{E}_t^{\mathrm{brg}}$ from Eq.~\eqref{eq:two_step_scores}.

\Statex
\Statex \textbf{Endpoint diagnostic}

\State Draw
$\bm{z}^{(m)}
\sim
\mathcal{N}(\bm{m}_{-},\mathbf{C}_{-})$ with
$m=1,\ldots,M$, where $M$ is the number of samples.

\For{$m=1,\ldots,M$}
    \State Compute
    $\bm{m}_{+}^{(m)}
    \gets
    \bm{z}^{(m)}
    +\bm{F}(\bm{z}^{(m)})\Delta t_{+}$ and    $\mathbf{C}_{+}^{(m)}
    \gets
    2\mathbf{D}(\bm{z}^{(m)})\Delta t_{+}$.
\EndFor

\State Estimate
$\bm{\mu}_{t}^{\mathrm{end}}$
and
$\mathbf{C}_{t}^{\mathrm{end}}$
from the sample mean and covariance of
$\{\bm{m}_{+}^{(m)}\}_{m=1}^{M}$
and the sample mean of
$\{\mathbf{C}_{+}^{(m)}\}_{m=1}^{M}$,
according to Eq.~\eqref{eq:endpoint_moments}.

\State Compute
$Q_t^{\mathrm{end}}$
using Eq.~\eqref{eq:endpoint_Q}, and obtain
$p_t^{\mathrm{end}}$ and
$\mathcal{E}_t^{\mathrm{end}}$ from Eq.~\eqref{eq:two_step_scores}.
\end{algorithmic}
\end{algorithm}

\subsection{Exogeneity Scores in the Local Gaussian Limit}

To clarify the distinction between the local, bridge, and endpoint diagnostics, we consider three consecutive observed states
$\bm{x}_{-} \rightarrow \bm{x}_t \rightarrow \bm{x}_+$
separated by equal time intervals $\Delta t$.
For analytical clarity, we consider a local Gaussian limit in which coefficient variations over the scale of the deviations are neglected and the two one-step transitions share the covariance
$\mathbf{C}\equiv2\mathbf{D}\Delta t$.
Starting from the observed initial state $\bm{x}_{-}$, we define the unperturbed mean evolution by
\begin{equation}
\bar{\bm{x}}_{t}
=
\bm{x}_{-}+\bm{F}(\bm{x}_{-})\Delta t,
\qquad
\bar{\bm{x}}_{+}
=
\bar{\bm{x}}_{t}+\bm{F}(\bar{\bm{x}}_{t})\Delta t.
\end{equation}
We define the intermediate-state and endpoint deviations from this unperturbed mean evolution as
\begin{equation}
\bm{\delta}
\equiv
\bm{x}_t-\bar{\bm{x}}_t,
\qquad
\bm{\eta}
\equiv
\bm{x}_+-\bar{\bm{x}}_+.
\end{equation}

The local diagnostic evaluates the incoming transition $\bm{x}_{-}\rightarrow\bm{x}_t$.
Because $\bar{\bm{x}}_t$ is the one-step mean state from $\bm{x}_{-}$ and the covariance is $\mathbf{C}$, the local discrepancy reduces to
\begin{equation}
\begin{aligned}
Q_t^{\mathrm{loc}}
&= \left(\Delta \bm{x}^{\rm in}_t - \bm{F}(\bm{x}_{-})\Delta t \right)^{\top}
\mathbf{C}^{-1}
\left(\Delta \bm{x}^{\rm in}_t - \bm{F}(\bm{x}_{-})\Delta t \right)
=
(\bm{x}_t-\bar{\bm{x}}_t)^{\top}
\mathbf{C}^{-1}
(\bm{x}_t-\bar{\bm{x}}_t)
=
\bm{\delta}^{\top}
\mathbf{C}^{-1}
\bm{\delta} \equiv k^2,
\label{eq:local_gaussian_Qloc}
\end{aligned}
\end{equation}
where $k$ measures the magnitude of the local deviation in units of the one-step stochastic fluctuation scale, with $k<1$ indicating a deviation within the unit covariance ellipsoid and larger $k$ indicating deviations that span multiple local fluctuation scales.

The bridge diagnostic evaluates whether the observed intermediate state $\bm{x}_t$ is consistent with the two-step evolution conditional on the observed initial and endpoint states, $\bm{x}_-$ and $\bm{x}_+$.
For a candidate intermediate state $\bm{z}$, define its deviation from the unperturbed intermediate mean as
$\bm{u}\equiv\bm{z}-\bar{\bm{x}}_t$.
In the locally constant-coefficient limit, this deviation is carried unchanged into the mean of the second transition.
Thus,
\begin{equation}
P_{\Delta t}(\bm{z}\mid\bm{x}_{-})
\simeq
\mathcal{N}(\bm{z}\mid\bar{\bm{x}}_t,\mathbf{C}),
\qquad
P_{\Delta t}(\bm{x}_{+}\mid\bm{z})
\simeq
\mathcal{N}
\left(
\bm{x}_{+}\mid
\bar{\bm{x}}_{+}+\bm{u},
\mathbf{C}
\right).
\end{equation}
Using $\bm{x}_{+}=\bar{\bm{x}}_{+}+\bm{\eta}$, the residual of the observed endpoint under the second transition is $\bm{\eta}-\bm{u}$.
The conditional bridge density is therefore proportional to
\begin{equation}
\begin{aligned}
\pi_t^{\mathrm{brg}}(\bm{z})
&\propto
P_{\Delta t}(\bm{z}\mid\bm{x}_{-})
P_{\Delta t}(\bm{x}_{+}\mid\bm{z})
\\
&\propto
\exp\left[
-\frac{1}{2}\bm{u}^{\top}\mathbf{C}^{-1}\bm{u}
-\frac{1}{2}
(\bm{\eta}-\bm{u})^{\top}
\mathbf{C}^{-1}
(\bm{\eta}-\bm{u})
\right].
\end{aligned}
\end{equation}
Completing the square in $\bm{u}$ gives
\begin{equation}
\pi_t^{\mathrm{brg}}(\bm{z})
=
\mathcal{N}
\left(
\bm{z}
\mid
\bar{\bm{x}}_t+\frac{1}{2}\bm{\eta},
\frac{1}{2}\mathbf{C}
\right).
\label{eq:local_gaussian_bridge}
\end{equation}
Thus, conditioning on the observed endpoint shifts the bridge mean from $\bar{\bm{x}}_t$ by one-half of the endpoint deviation $\bm{\eta}$ while reducing the covariance to $\mathbf{C}/2$.
In this Gaussian limit, the Laplace approximation in Eq.~\eqref{eq:bridge_laplace} is exact, with bridge mode and curvature
\begin{equation}
\bm{\mu}_t^{\mathrm{brg}}
=
\bar{\bm{x}}_t+\frac{1}{2}\bm{\eta},
\qquad
\mathbf{H}_t^{\mathrm{brg}}
=
2\mathbf{C}^{-1}.
\end{equation}
Equivalently, conditional on the initial and endpoint states, the most probable intermediate deviation from the unperturbed mean is $\bm{\eta}/2$.
Using $\bm{x}_t=\bar{\bm{x}}_t+\bm{\delta}$, the bridge discrepancy becomes
\begin{equation}
\begin{aligned}
Q_t^{\mathrm{brg}}
&=
(\bm{x}_t-\bm{\mu}_t^{\mathrm{brg}})^{\top}
\mathbf{H}_t^{\mathrm{brg}}
(\bm{x}_t-\bm{\mu}_t^{\mathrm{brg}})
\\
&=
2
\left(
\bm{\delta}-\frac{1}{2}\bm{\eta}
\right)^{\top}
\mathbf{C}^{-1}
\left(
\bm{\delta}-\frac{1}{2}\bm{\eta}
\right).
\end{aligned}
\label{eq:local_gaussian_Qbrg}
\end{equation}
Here, $Q_t^{\mathrm{brg}}$ is the squared Mahalanobis mismatch between the observed intermediate-state deviation $\bm{\delta}$ and the deviation $\bm{\eta}/2$.
We note that $Q_t^{\mathrm{brg}}=0$ when $\bm{\delta}=\bm{\eta}/2$, for which the observed intermediate state coincides with the bridge mode.

The endpoint diagnostic instead evaluates whether the observed endpoint $\bm{x}_+$ is atypical under the two-step predictive distribution from $\bm{x}_{-}$ after marginalizing over the intermediate state.
In the present limit, this marginal distribution can be evaluated analytically:
\begin{equation}
\pi_t^{\mathrm{end}}(\bm{y})
=
\int
P_{\Delta t}(\bm{z}'\mid\bm{x}_{-})
P_{\Delta t}(\bm{y}\mid\bm{z}')
\,d\bm{z}'.
\end{equation}
Using
$\bm{u}'\equiv\bm{z}'-\bar{\bm{x}}_t$,
the first transition gives
$\bm{u}'\sim\mathcal{N}(\bm{0},\mathbf{C})$,
while the second transition satisfies
\begin{equation}
P_{\Delta t}(\bm{y}\mid\bm{z}')
\simeq
\mathcal{N}
\left(
\bm{y}
\mid
\bar{\bm{x}}_+ + \bm{u}',
\mathbf{C}
\right)
\end{equation}
Marginalizing over $\bm{u}'$ gives the Gaussian convolution
\begin{equation}
\begin{aligned}
\pi_t^{\mathrm{end}}(\bm{y})
&= \int \mathcal{N}(\bm{u}' \mid \bm{0}, \mathbf{C}) \mathcal{N} \left( \bm{y} \mid \bar{\bm{x}}_+ + \bm{u}', \mathbf{C} \right) \,d\bm{u}'
\\
&=
\mathcal{N}
\left(
\bm{y}
\mid
\bar{\bm{x}}_+,
2\mathbf{C}
\right).
\label{eq:local_gaussian_endpoint}
\end{aligned}
\end{equation}
In this Gaussian limit, the moment-matched approximation in Eq.~\eqref{eq:endpoint_gaussian} becomes exact, with
\begin{equation}
\bm{\mu}_t^{\mathrm{end}}
=
\bar{\bm{x}}_+,
\qquad
\mathbf{C}_t^{\mathrm{end}}
=
2\mathbf{C}.
\end{equation}
The same endpoint mean and covariance are recovered directly from
Eq.~\eqref{eq:endpoint_moments}. For $\bm{Z}\sim\mathcal{N}(\bar{\bm{x}}_t,\mathbf{C})$, the present limit gives
$\bm{m}_+(\bm{Z})
=
\bar{\bm{x}}_+
+
(\bm{Z}-\bar{\bm{x}}_t)$
and
$\mathbf{C}_+(\bm{Z})=\mathbf{C}$.
Hence,
\begin{equation}
\begin{aligned}
\mathbb{E}_{\bm{Z}}
\left[
\bm{m}_+(\bm{Z})
\right]
&=
\bar{\bm{x}}_+
+
\mathbb{E}_{\bm{Z}}
\left[
\bm{Z}-\bar{\bm{x}}_t
\right]
=
\bar{\bm{x}}_+,
\\
\operatorname{Cov}_{\bm{Z}}
\left[
\bm{m}_+(\bm{Z})
\right]
+
\mathbb{E}_{\bm{Z}}
\left[
\mathbf{C}_+(\bm{Z})
\right]
&=
\operatorname{Cov}_{\bm{Z}}
\left[
\bm{Z}-\bar{\bm{x}}_t
\right]
+
\mathbf{C}
\\
&=
\mathbf{C}+\mathbf{C}
=
2\mathbf{C}.
\end{aligned}
\end{equation}
Using $\bm{x}_+=\bar{\bm{x}}_++\bm{\eta}$, the endpoint discrepancy becomes
\begin{equation}
\begin{aligned}
Q_t^{\mathrm{end}}
&=
(\bm{x}_+-\bm{\mu}_t^{\mathrm{end}})^{\top}
\left(
\mathbf{C}_t^{\mathrm{end}}
\right)^{-1}
(\bm{x}_+-\bm{\mu}_t^{\mathrm{end}})
\\
&=
\bm{\eta}^{\top}
(2\mathbf{C})^{-1}
\bm{\eta}
=
\frac{1}{2}
\bm{\eta}^{\top}
\mathbf{C}^{-1}
\bm{\eta}.
\end{aligned}
\label{eq:local_gaussian_Qend}
\end{equation}
Here, $Q_t^{\mathrm{end}}$ is the squared Mahalanobis deviation of the observed endpoint from the unperturbed two-step mean, with the covariance $2\mathbf C$ accounting for stochastic variability accumulated over both transitions.
We note that $Q_t^{\mathrm{end}}=0$ when $\bm{\eta}=\bm{0}$, for which the observed endpoint coincides with the unperturbed endpoint mean.

Collecting the three results, the local, bridge, and endpoint discrepancies in this Gaussian limit are
\begin{equation}
\boxed{
\begin{aligned}
Q_t^{\mathrm{loc}}
&=
\bm{\delta}^{\top}
\mathbf{C}^{-1}
\bm{\delta}
\equiv
k^2,
\\
Q_t^{\mathrm{brg}}
&=
2
\left(
\bm{\delta}-\frac{1}{2}\bm{\eta}
\right)^{\top}
\mathbf{C}^{-1}
\left(
\bm{\delta}-\frac{1}{2}\bm{\eta}
\right),
\\
Q_t^{\mathrm{end}}
&=
\frac{1}{2}
\bm{\eta}^{\top}
\mathbf{C}^{-1}
\bm{\eta}.
\end{aligned}
}
\label{eq:exogeneity_summary}
\end{equation}
These expressions obey the identity
\begin{equation}
Q_t^{\mathrm{brg}}
+
Q_t^{\mathrm{end}}
=
Q_t^{\mathrm{loc}}
+
(\bm{\eta}-\bm{\delta})^{\top}
\mathbf{C}^{-1}
(\bm{\eta}-\bm{\delta}).
\label{eq:local_gaussian_identity}
\end{equation}
Because $\mathbf{C}^{-1}$ is positive definite, this gives the bound
\begin{equation}
\boxed{
Q_t^{\mathrm{brg}}
+
Q_t^{\mathrm{end}}
\geq
Q_t^{\mathrm{loc}}.
}
\label{eq:local_gaussian_bound}
\end{equation}
The term $(\bm{\eta}-\bm{\delta})^{\top} \mathbf{C}^{-1} (\bm{\eta}-\bm{\delta})$ is the squared Mahalanobis magnitude of the change in deviation from the unperturbed mean evolution over the second transition, since 
$\bm{\eta}-\bm{\delta} = (\bm{x}_+-\bm{x}_t) - (\bar{\bm{x}}_+-\bar{\bm{x}}_t)$.
Thus, the combined bridge and endpoint discrepancies cannot be smaller than the local discrepancy, with equality when $\bm{\eta}=\bm{\delta}$.
In this case, the intermediate deviation is carried forward without further change and the bridge and endpoint discrepancies are equal with $Q_t^{\mathrm{brg}}=Q_t^{\mathrm{end}}=k^2/2$.

This bound also explains the nearly empty lower-left region of the bridge--endpoint score plane for transitions selected above a local-exogeneity threshold.
For a local-score threshold $\mathcal{E}_t^{\mathrm{loc}}\ge s_0$, the local discrepancy satisfies
\begin{equation}
Q_t^{\mathrm{loc}}
\ge
F_{\chi_d^2}^{-1}
\left(
1-10^{-s_0}
\right),
\end{equation}
so that, in the present Gaussian limit,
\begin{equation}
Q_t^{\mathrm{brg}}
+
Q_t^{\mathrm{end}}
\ge
F_{\chi_d^2}^{-1}
\left(
1-10^{-s_0}
\right).
\end{equation}
Because the exogeneity scores are monotonic functions of their respective discrepancies, the boundary in the bridge--endpoint score plane is given by
\begin{equation}
F_{\chi_d^2}^{-1}
\left(
1-10^{-\mathcal{E}_t^{\mathrm{brg}}}
\right)
+
F_{\chi_d^2}^{-1}
\left(
1-10^{-\mathcal{E}_t^{\mathrm{end}}}
\right)
=
F_{\chi_d^2}^{-1}
\left(
1-10^{-s_0}
\right).
\label{eq:local_gaussian_score_bound}
\end{equation}
The shape of this boundary depends on the state-space dimension $d$ through the $\chi_d^2$ transformation.
For $d=2$, the boundary reduces to the straight line $\mathcal{E}_t^{\mathrm{brg}}+\mathcal{E}_t^{\mathrm{end}}=s_0$, whereas for other dimensions it is generally curved.
We use the dimension-specific boundaries and show them in the bridge--endpoint score planes in the main text.
The empirical points lie predominantly on or above these boundaries, providing a geometric explanation for the nearly empty lower-left regions and showing that the observed score geometry remains broadly consistent with the Gaussian-limit prediction.

\subsection{Numerical Validation}

The preceding analysis establishes analytical relationships among the local, bridge, and endpoint discrepancies in the local Gaussian limit. 
To obtain explicit predictions for numerical validation, we consider a controlled one-parameter setup of perturbations.
We prescribe the intermediate displacement by $\bm{\delta}^{\top}\mathbf{C}^{-1}\bm{\delta}=k^2$ and set the endpoint displacement to $\bm{\eta}=\beta\bm{\delta}$, with $\beta \geq 0$.
Here, $k$ controls the Mahalanobis magnitude of the intermediate perturbation, whereas $\beta$ controls the endpoint displacement relative to it.
Substituting this relation into the local-Gaussian expressions in Eq.~\eqref{eq:exogeneity_summary} gives
\begin{equation}
Q_t^{\mathrm{loc}}=k^2,
\qquad
Q_t^{\mathrm{brg}}
=
2\left(1-\frac{\beta}{2}\right)^2k^2,
\qquad
Q_t^{\mathrm{end}}
=
\frac{\beta^2}{2}k^2.
\label{eq:vdp_beta_prediction}
\end{equation}
Thus, $Q_t^{\mathrm{loc}}$ remains fixed as $\beta$ varies, while the bridge and endpoint discrepancies change systematically.
The representative values $\beta=0$, $1$, and $2$ produce discrepancy patterns associated with recovered, localized, and persistent perturbation cases, respectively. For $\beta>2$, both the bridge and endpoint discrepancies increase with $\beta$, yielding an off-path-like pattern for sufficiently large endpoint displacements (Table~\ref{tab:exo_footprint_beta}).
Note that these qualitative categories describe qualitative relationships among the discrepancy measures and do not define categories with fixed numerical thresholds.
Because the exogeneity scores are monotonic transformations of these discrepancies, the same relationships appear in the bridge–endpoint score plane.

\begin{table}[H]
\vskip -0.1in
\caption{Summary of representative exogenous perturbation footprints in the local Gaussian approximation. The recovered, localized, and persistent cases follow the one-parameter setup $\bm{\eta}=\beta\bm{\delta}$, whereas sufficiently large $\beta>2$ produces an off-path-like profile.}
\centering
\begin{tabular}{c|c|c|c}
    Perturbation footprint & Local discrepancy $Q^{\rm loc}$ & Bridge discrepancy $Q^{\rm brg}$ & Endpoint discrepancy $Q^{\rm end}$ \\ \hline\hline
    Recovered ($\beta = 0$)& $k^2$ &  $2k^2$  &  $0$  \\ \hline
    Localized ($\beta = 1$) & $k^2$ & $(1/2)k^2$ & $(1/2)k^2$ \\ \hline
    Persistent ($\beta = 2$)& $k^2$ & $0$ & $2k^2$ \\ \hline
    Off-path-like (sufficiently large $\beta>2$) & $k^2$  & High & High
\end{tabular}
\label{tab:exo_footprint_beta}
\end{table}

\begin{figure*}[!t]
  \centering
  \includegraphics[width=\linewidth]{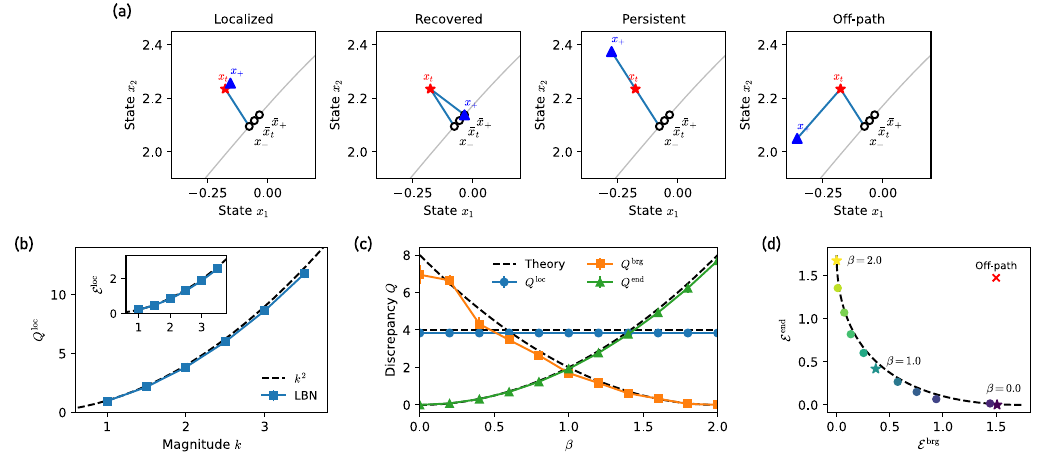}
  \vskip -0.1in
  \caption{
  (a) Representative perturbation patterns. 
  For an unperturbed mean segment $(x_{-}, \bar{x}_{t}, \bar{x}_{+})$ sampled from a drift-only van der Pol trajectory, we impose a controlled displacement at the intermediate state (red star) and vary its expression at the endpoint (blue triangle), generating localized, recovered, persistent, and off-path patterns. Open circles denote the unperturbed segment, and the gray line indicates the deterministic trajectory.
  (b) Local discrepancy $Q^{\rm loc}$ as the perturbation magnitude $k$ varies. The inset shows the local exogeneity score $\mathcal{E}^{\rm loc}$.
  (c) Local, bridge, and endpoint discrepancies, $Q^{\rm loc}$ (blue), $Q^{\rm brg}$ (orange), and $Q^{\rm end}$ (green), as the control parameter $\beta$ varies at fixed $k=2$.
  (d) Bridge--endpoint score plane $(\mathcal{E}^{\rm brg}, \mathcal{E}^{\rm end})$ as $\beta$ varies at fixed $k=2$. The red cross denotes a representative off-path perturbation.
  In (b--d), dashed lines denote the theoretical predictions in Eq.~\eqref{eq:vdp_beta_prediction}, and colored lines and markers denote values obtained from the LBN-inferred SDE trained at the finite observation interval $\Delta t=10 \,\delta t_{\rm int}$.
}
  \label{fig:exo_validation}
\end{figure*}

We test these predictions using the two-dimensional stochastic van der Pol oscillator introduced in Sec.~\ref{subsec:LBN_finite_interval}. 
To define an unperturbed mean sequence, we generate a deterministic trajectory by integrating the drift-only dynamics and select three states separated by the observation interval, $\bm{x}_{-}\rightarrow\bar{\bm{x}}_t\rightarrow\bar{\bm{x}}_+$. 
We then construct the perturbed sequence by setting $\bm{x}_t=\bar{\bm{x}}_t+\bm{\delta}$ and $\bm{x}_+=\bar{\bm{x}}_{+}+\bm{\eta}$.
Here, $\bm{\delta}$ is applied along the local normal to this deterministic trajectory and scaled such that $\bm{\delta}^{\top}\mathbf{C}^{-1}\bm{\delta}=k^2$.
Following the one-parameter setup $\bm{\eta}=\beta\bm{\delta}$, we vary $\beta$ from $0$ to $2$, spanning the recovered, localized, and persistent footprints [Fig.~\ref{fig:exo_validation}(a)].
Although sufficiently large $\beta>2$ can produce an off-path-like footprint, this requires a large collinear endpoint displacement. 
We instead construct a separate off-path perturbation by directing the endpoint from the perturbed intermediate state along the negative local tangent direction.
The resulting diagnostics are evaluated using the LBN-inferred SDE trained on stochastic trajectories sampled at the finite observation interval $\Delta t=10 \, \delta t_{\rm int}$, rather than directly using the known drift and diffusion fields.

As the perturbation magnitude $k$ increases, the local discrepancy evaluated from the inferred SDE closely follows the predicted quadratic scaling $Q_t^{\mathrm{loc}}=k^2$, while the local exogeneity score increases monotonically [Fig.~\ref{fig:exo_validation}(b)]. 
At fixed $k=2$, varying $\beta$ leaves $Q_t^{\mathrm{loc}}$ approximately constant, while $Q_t^{\mathrm{brg}}$ decreases and $Q_t^{\mathrm{end}}$ increases in close agreement with the predictions in Eq.~\eqref{eq:vdp_beta_prediction} [Fig.~\ref{fig:exo_validation}(c)]. 
In the bridge--endpoint score plane, this variation produces a continuous progression from recovered through localized to persistent geometries, while the separately applied off-path perturbation lies in the region where both scores are elevated [Fig.~\ref{fig:exo_validation}(d)]. 
These results show that the discrepancy relationships and associated score geometries predicted by the local Gaussian analysis remain evident when the diagnostics are evaluated from an inferred nonlinear SDE at a finite observation interval.

\section{Dynamical Irreversibility}
\label{sec:irreversibility}

Dynamical irreversibility measures the temporal asymmetry between the path density of an observed trajectory and that of the same sequence of states followed in reverse temporal order.
Consider an observed state trajectory
$\Gamma\equiv\{\bm{x}_{t_0},\bm{x}_{t_1},\ldots,\bm{x}_{t_L}\}$,
together with its conditioning sequence
$\mathcal{C}\equiv\{\bm{c}_{t_0},\bm{c}_{t_1},\ldots,\bm{c}_{t_L}\}$.
Under time reversal, the state trajectory and conditioning sequence are traversed in reverse temporal order, $\tilde{\Gamma} \equiv \{\bm{x}_{t_L},\bm{x}_{t_{L-1}},\ldots,\bm{x}_{t_0}\}$ and $\tilde{\mathcal{C}} \equiv \{\bm{c}_{t_L},\bm{c}_{t_{L-1}},\ldots,\bm{c}_{t_0}\}$.
All components of $\bm{x}_t$ considered here are taken to be even under time reversal, and $\mathcal{C}$ is treated as an input to the state dynamics and is not included in the stochastic state.
We define the trajectory-level irreversibility as the log ratio of the conditional path densities under the forward and reverse protocols:
\begin{equation}
\Sigma[\Gamma]
\equiv
\ln
\frac{
\mathcal{P}_{\mathrm{F}}[\Gamma\mid\bm{x}_{t_0},\mathcal{C}]
}{
\mathcal{P}_{\mathrm{R}}[\tilde{\Gamma}\mid\bm{x}_{t_L},\tilde{\mathcal{C}}]
},
\label{eq:IR_trj-level}
\end{equation}
where $\mathrm{F}$ and $\mathrm{R}$ label the forward and reverse protocols, respectively, and $\mathcal{P}$ denotes the conditional path density.
For simplicity, the dependence of $\Sigma$ on $\mathcal{C}$ is left implicit.
By the Markov property, the conditional path densities factorize over the observed transitions.
Substituting these factorizations into Eq.~\eqref{eq:IR_trj-level} gives
\begin{equation}
\Sigma[\Gamma]
=
\sum_{k=0}^{L-1}
\sigma_{t_{k+1}},
\quad \text{where}
\quad 
\sigma_{t_{k+1}} \equiv
\ln
\frac{
P_{\mathrm{F},\Delta t_k}
(\bm{x}_{t_{k+1}}\mid\bm{x}_{t_k}, \bm{c}_{t_k})
}{
P_{\mathrm{R},\Delta t_k}
(\bm{x}_{t_k}\mid\bm{x}_{t_{k+1}}, \bm{c}_{t_{k+1}})
}.
\end{equation}
Here, $\Delta t_k\equiv t_{k+1}-t_k$, and
$P_{\mathrm{F},\Delta t_k}$ and $P_{\mathrm{R},\Delta t_k}$
denote the finite-time transition densities used for the forward and reversed trajectories, respectively.
For time-homogeneous dynamics, both are evaluated using the same transition kernel.
When the dynamics depend on time-varying conditioning variables, the reversed trajectory is evaluated with the corresponding conditioning sequence traversed in reverse temporal order.

The local irreversibility $\sigma_{t_{k+1}}$ thus measures the temporal asymmetry of the observed transition $\bm{x}_{t_k}\rightarrow\bm{x}_{t_{k+1}}$.
Positive values of $\sigma_{t_{k+1}}$ indicate that the observed transition is more probable in the forward direction than its reversed counterpart, whereas negative values indicate that the reverse ordering is more probable.
For the It\^o diffusion in Eq.~\eqref{eq:ito_sde}, the infinitesimal contribution to the forward-to-reverse log path-density ratio can be written in Stratonovich form~\cite{seifert2012stochastic,spinney2012entropy}:
\begin{equation}
d\Sigma_t
=
\left[
\bm{F}(\bm{x}_t,\bm{c}_t)
-
\nabla\!\cdot\!\mathbf{D}(\bm{x}_t,\bm{c}_t)
\right]^{\top}
\mathbf{D}(\bm{x}_t,\bm{c}_t)^{-1}
\circ d\bm{x}_t ,
\label{eq:IR_stratonovich}
\end{equation}
where
$[\nabla\!\cdot\!\mathbf{D}]_i \equiv
\sum_{j=1}^{d}\partial D_{ij}/\partial x_j$ with the conditioning variables held fixed, and $\circ$ denotes the Stratonovich product.
The Stratonovich product admits a time-symmetric midpoint discretization, in which the integrand is evaluated at the midpoint of each state increment.
For a finite observed transition $\bm{x}_{t_k}\rightarrow\bm{x}_{t_{k+1}}$, we therefore approximate the integral of Eq.~\eqref{eq:IR_stratonovich} over $[t_k,t_{k+1}]$ as
\begin{equation}
\sigma_{t_{k+1}}
\simeq
\left[
\bm{F}(\bar{\bm{x}}_{t_k},\bar{\bm{c}}_{t_k})
-
\nabla\!\cdot\!\mathbf{D}(\bar{\bm{x}}_{t_k},\bar{\bm{c}}_{t_k})
\right]^{\top}
\mathbf{D}(\bar{\bm{x}}_{t_k},\bar{\bm{c}}_{t_k})^{-1}
\Delta\bm{x}_{t_k},
\end{equation}
where $\Delta \bm{x}_{t_k} \equiv \bm{x}_{t_{k+1}} - \bm{x}_{t_k}$, $\bar{\bm{x}}_{t_k} \equiv (\bm{x}_{t_k}+\bm{x}_{t_{k+1}})/{2}$, and $\bar{\bm{c}}_{t_k}$ denotes the conditioning variables evaluated at the transition midpoint.
For explicitly time-dependent dynamics, the calendar-time component is evaluated at $\bar t_k\equiv(t_k+t_{k+1})/2$.
The cumulative irreversibility up to observation time $t_m$ is defined by the partial sum $\Sigma_{t_m} \equiv \sum_{k=0}^{m-1}\sigma_{t_{k+1}}$, which records how the local directional asymmetries accumulate along the observed trajectory.

To compare trajectories spanning different durations or containing irregular observation intervals, we define the mean irreversibility rate as the accumulated irreversibility per unit elapsed time over the retained transitions,
\begin{equation}
\left\langle \dot{\sigma} \right\rangle
\equiv
\frac{\sum_{k\in\mathcal{K}_{\Gamma}}\sigma_{t_{k+1}}
}{\sum_{k\in\mathcal{K}_{\Gamma}}\Delta t_k
},
\label{eq:IR_rate}
\end{equation}
where $\mathcal{K}_{\Gamma}$ denotes the set of transitions retained for the analysis.
For the Seshat analysis in the main text, we restrict this set to transitions with observation intervals of at most 200 years to reduce sensitivity to long observation gaps.
Because $\sigma$ and $\Sigma$ are log-density ratios, they are dimensionless, whereas $\langle\dot{\sigma}\rangle$ has units of inverse time.

In stochastic-thermodynamic terminology, the dynamical irreversibility defined in Eq.~\eqref{eq:IR_trj-level} formally corresponds to the medium entropy-production contribution~\cite{seifert2012stochastic}.
For physical Langevin systems satisfying local detailed balance, this contribution represents entropy transferred to the surrounding medium and can be related to dissipated heat.
The total entropy production additionally includes the change in stochastic system entropy associated with the initial and final state distributions:
\begin{equation}
\Sigma_{\mathrm{tot}}[\Gamma]
=
\Sigma[\Gamma]
+
\Delta s_{\mathrm{sys}},
\qquad
\Delta s_{\mathrm{sys}}
\equiv
\ln
\frac{
\rho_{t_0}(\bm{x}_{t_0})
}{
\rho_{t_L}(\bm{x}_{t_L})
}.
\end{equation}
For trajectories sampled from a stationary state, the ensemble-averaged system entropy change vanishes, $\langle\Delta s_{\mathrm{sys}}\rangle=0$, so that
$\langle\Sigma_{\mathrm{tot}}[\Gamma]\rangle = \langle\Sigma[\Gamma]\rangle$.
Our measure characterizes the directional asymmetry of the conditional transition dynamics without including this boundary contribution.
In this work, we use the term dynamical irreversibility to emphasize this statistical interpretation and do not interpret $\Sigma$ as thermodynamic entropy production of the social system.

\section{Compatibility Across SDE Inference Methods}
\label{sec:other_sde}

The analyses in the main text infer the reference SDE using LBN~\cite{bae2025inferring}. 
However, the downstream diagnostics introduced in Secs.~\ref{sec:exo_scores} and \ref{sec:irreversibility} depend only on the drift and diffusion fields of the fitted SDE, rather than on the particular method used to infer them.
To demonstrate that the diagnostic framework is not tied to LBN, we additionally inferred the reference dynamics using Stochastic Force Inference (SFI)~\cite{frishman2020learning} for both the DIG and Seshat datasets.

SFI provides a conceptually distinct inference strategy from LBN.
Whereas LBN represents the drift and diffusion fields using neural networks trained from local conditional moments, SFI estimates these fields through projection onto a prescribed basis~\cite{frishman2020learning}.
For the SFI fits, we used polynomial drift bases of order two for DIG and order one for Seshat, together with a constant diffusion matrix for both datasets.
Using the same state representations and observed trajectories as in the main analyses, we applied the exogeneity and irreversibility diagnostics without modification.

\begin{figure*}[!h]
  \centering
  \includegraphics[width=\linewidth]{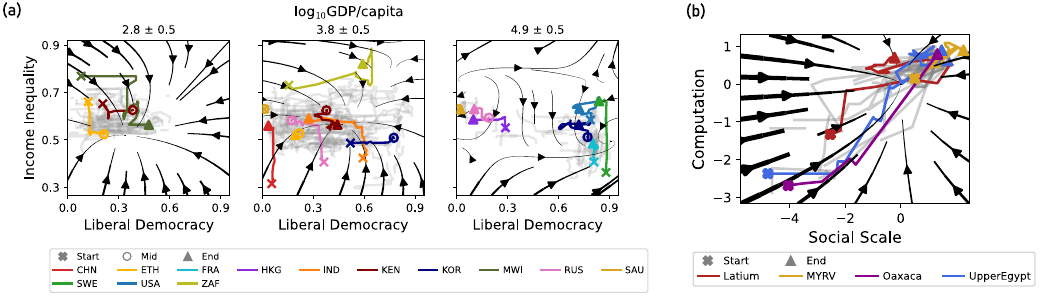}
  \vskip -0.1in
  \caption{Stochastic dynamics inferred using Stochastic Force Inference (SFI).
  (a) Modern political-economic dynamics on the Democracy--Inequality plane at three levels of $\log_{10}$ GDP/capita; the $\pm 0.5$ windows around these levels are used only to select empirical trajectories. 
  (b) Historical dynamics in the Seshat dataset on the Scale--Computation plane.
  In both panels, black streamlines show the drift field $\bm{F}(\bm{x})$ inferred using SFI.
  Colored trajectories show selected units' observed paths, gray trajectories show the remaining observed paths in each dataset, and markers indicate the initial, intermediate, and final observed states.
  For visualization in (a), PCA coordinates are linearly mapped back to the original indicator scales.
}
  \label{fig:sfi_results}
\end{figure*}

The resulting SFI-inferred drift fields are shown in Fig.~\ref{fig:sfi_results}.
Using these independently inferred reference dynamics, we then examined whether the downstream diagnostics yield similar rankings of the observed transitions.
Across both datasets, the LBN- and SFI-based diagnostics showed positive rank correlations, with Spearman correlations ranging from $0.701$ to $0.903$.
These results provide an empirical illustration that the proposed diagnostic framework can be applied to SDEs obtained from a distinct inference procedure while yielding broadly consistent rankings of observed transitions.

\begin{table}[H]
\centering
\caption{
Spearman correlations between transition-level diagnostics obtained from the LBN- and SFI-inferred reference dynamics for the DIG and Seshat datasets.}
\label{tab:sfi_correlation}
\setlength{\tabcolsep}{12pt}
\begin{tabular}{c|cccc}
\hline\hline
Dataset
& Local E-score $\mathcal{E}^{\rm loc}$
& Bridge E-score $\mathcal{E}^{\rm brg}$
& Endpoint E-score $\mathcal{E}^{\rm end}$
& Local irreversibility $\sigma$ \\
\hline
DIG
& 0.892
& 0.903
& 0.872
& 0.701 \\
Seshat
& 0.837
& 0.746
& 0.773
& 0.793 \\
\hline\hline
\end{tabular}
\end{table}

\clearpage

\section{Additional Empirical Results}
\label{sec:additional_results}

This section presents application-specific analyses that complement the results reported in the main text.
For each application, we first examine the inferred reference dynamics together with model-estimation uncertainty in the drift and diffusion fields, and then present additional exogeneity and irreversibility analyses supporting the aggregate and trajectory-level patterns discussed in the main text.
Model-estimation uncertainty is quantified from the dispersion of predictions across cross-validation folds and SWAG samples and is distinct from the stochastic variability represented by the diffusion matrix $\mathbf{D}$ itself (Sec.~\ref{sec:method}).
These uncertainty maps therefore indicate where the inferred reference dynamics are more or less well constrained by the available observations under the adopted inference procedure.

\subsection{Modern Political Economy}

\subsubsection{Inferred Stochastic Dynamics and Uncertainty}

\begin{figure}[!h]
  \centering
  \includegraphics[width=\linewidth]{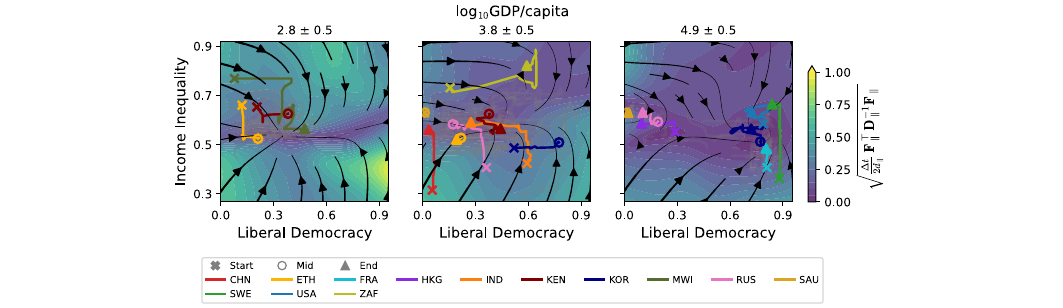}
  \caption{Inferred drift-to-diffusion balance in modern political-economic dynamics. The inferred dynamics are shown on the Democracy--Inequality plane across three levels of $\log_{10}$ GDP per capita. Black streamlines denote the projected drift field $\bm{F}_{\parallel}(\bm{x})$. Background color shows the local drift-to-diffusion ratio with $d_{\parallel}=2$ and $\Delta t$ set to one year. 
  Smaller values indicate weaker drift relative to diffusion, whereas larger values indicate stronger drift relative to diffusion over the interval $\Delta t$.
  Colored trajectories show selected countries, gray trajectories show the remaining countries, and markers indicate initial, intermediate, and final states. For visualization, the Democracy and Inequality coordinates are linearly mapped back to the original indicator scales.
}
\label{fig:drift-diff-ratio}
\end{figure}

\begin{figure}[!t]
  \centering
  \includegraphics[width=\linewidth]{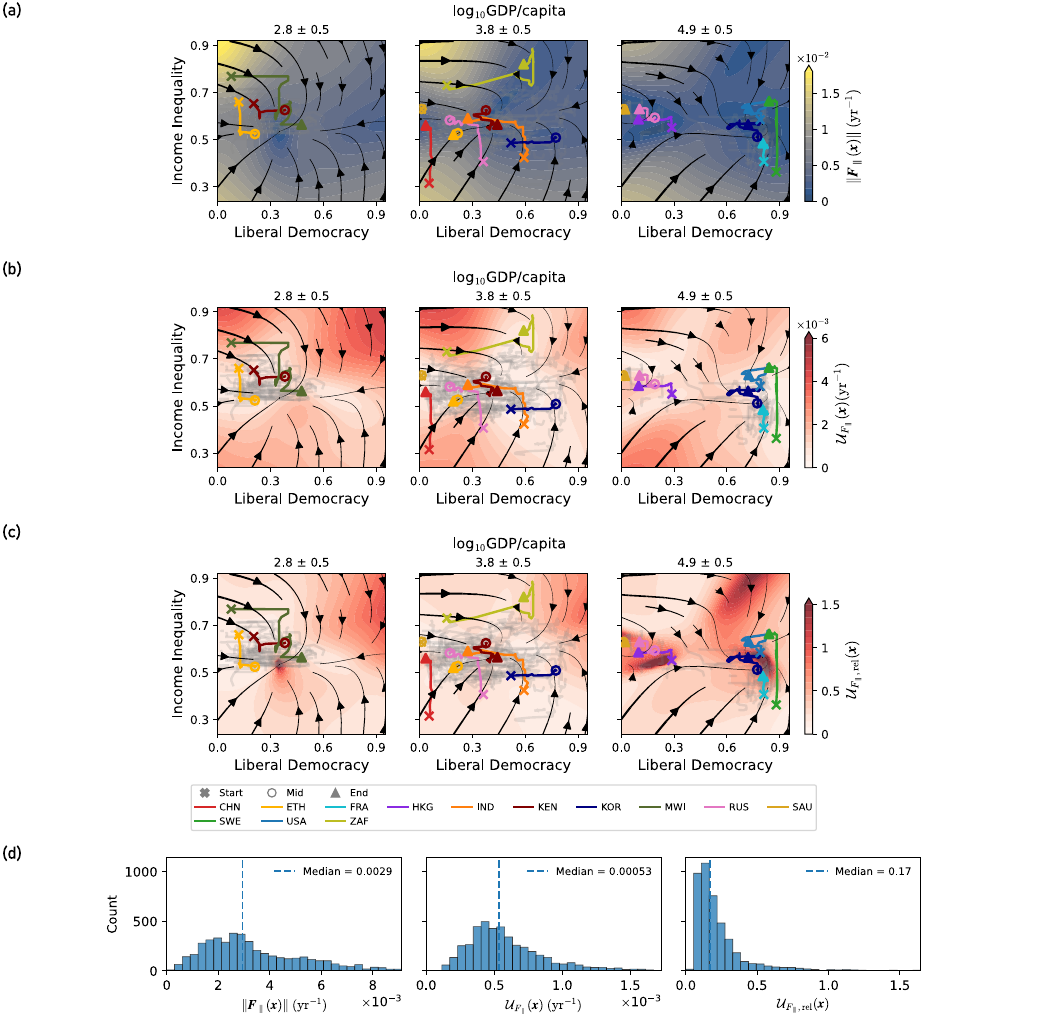}
  \vskip -0.1in
  \caption{
  Projected drift magnitude and model-estimation uncertainty in the modern political-economic dynamics. 
  (a) Magnitude of the drift projected onto the Democracy--Inequality plane, $\|{\mathbf F}_{\parallel}(\mathbf x)\|$, across three levels of $\log_{10}$ GDP/capita; the $\pm 0.5$ windows around these levels are used only to select empirical trajectories. 
  (b) Absolute model-estimation uncertainty of the projected drift, $\mathcal{U}_{F_{\parallel}}(\mathbf x)$, quantified from the pointwise dispersion of drift predictions across the model ensemble.
  (c) Relative projected-drift uncertainty, $\mathcal{U}_{F_{\parallel},\mathrm{rel}}(\bm{x})=\mathcal{U}_{F_{\parallel}}(\bm{x})/\|{\mathbf F}_{\parallel}(\bm{x})\|$. 
  Large values in (c) indicate uncertainty that is large relative to the local mean drift and can arise from either large absolute uncertainty or a weak mean drift.
  (d) Distributions across observed country-year states of $\|{\bm{F}}_{\parallel}(\bm{x})\|$, $\mathcal{U}_{F_{\parallel}}(\bm x)$, and $\mathcal{U}_{F_{\parallel},\mathrm{rel}}(\bm{x})$, from left to right. Dashed vertical lines indicate the median of each distribution.
  In (a--c), black streamlines show the projected drift field, colored trajectories show selected countries, and gray trajectories show the remaining countries in the dataset; markers indicate initial, intermediate, and final states. 
  For visualization, PCA coordinates are linearly mapped back to the original indicator scales.
}
\label{fig:DIG_Drift_w_uct}
\end{figure}

\begin{figure}[!t]
  \centering
  \includegraphics[width=\linewidth]{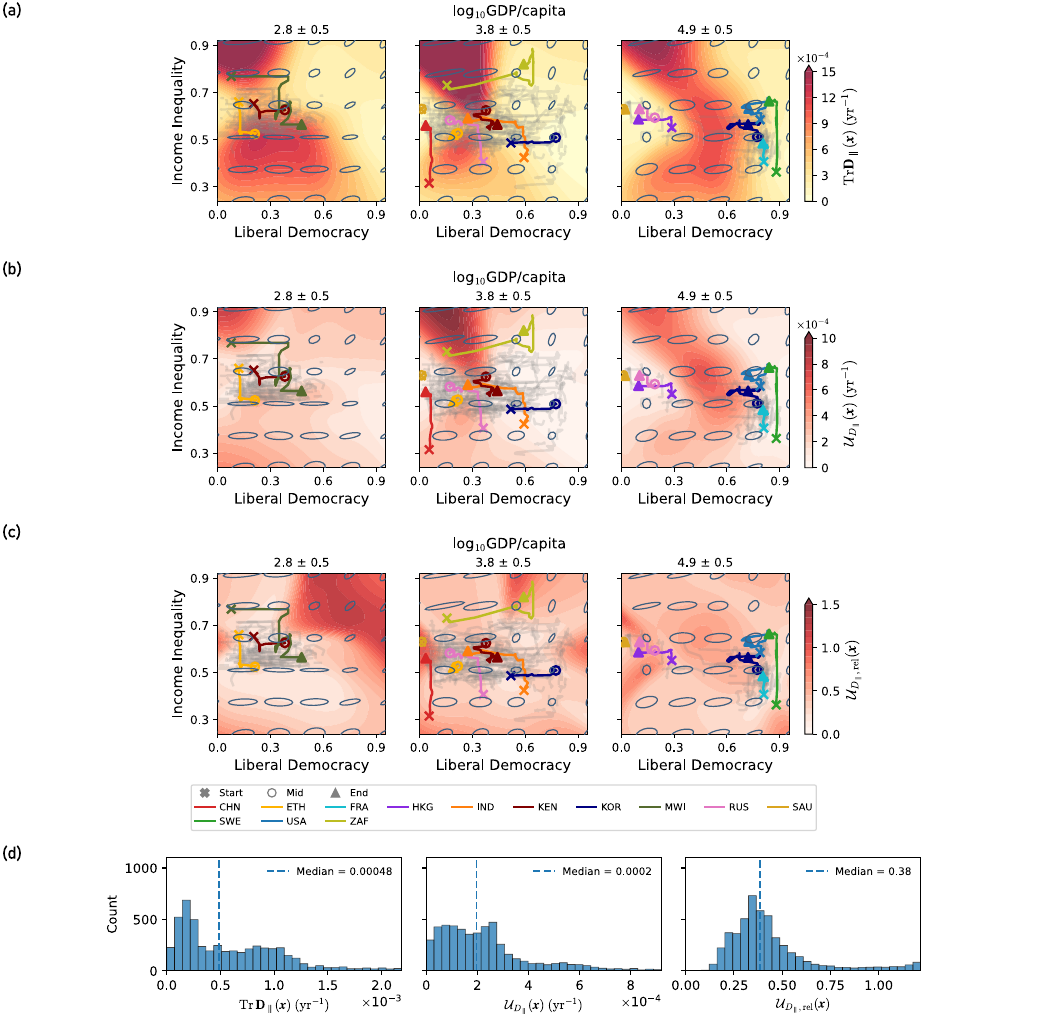}
  \vskip -0.1in
  \caption{
  Projected diffusion magnitude and model-estimation uncertainty in the modern political-economic dynamics.
  (a) Projected diffusion magnitude on the Democracy--Inequality plane, $\mathrm{Tr}\,{\mathbf D}_{\parallel}(\bm{x})$, across three levels of $\log{10}$ GDP/capita; the $\pm 0.5$ windows around these levels are used only to select empirical trajectories. 
  (b) Absolute model-estimation uncertainty of the projected diffusion magnitude, $\mathcal{U}_{D_{\parallel}}(\bm{x})$, quantified as the pointwise standard deviation of $\mathrm{Tr}\,{\mathbf D}_{\parallel}(\bm{x})$ across the model ensemble.
  (c) Relative uncertainty of the projected diffusion magnitude, $\mathcal{U}_{D_{\parallel},\mathrm{rel}}(\bm{x})=\mathcal{U}_{D_{\parallel}}(\bm{x})/\mathrm{Tr}\,{\mathbf D}_{\parallel}(\bm{x})$. 
  Large values in (c) indicate uncertainty that is large relative to the local diffusion strength and can arise from either large absolute uncertainty or weak local diffusion.
  (d) Distributions across observed country-year states of $\mathrm{Tr}\,{\mathbf D}_{\parallel}(\bm{x})$, $\mathcal{U}_{D_{\parallel}}(\bm{x})$, and $\mathcal{U}_{D_{\parallel}, \mathrm{rel}}(\bm{x})$, from left to right. Dashed vertical lines indicate the median of each distribution.
  In (a--c), blue ellipses depict the local anisotropy and principal directions of the projected diffusion tensor, with their axis lengths and orientations determined by its eigenvalues and eigenvectors, respectively.
  Colored trajectories show selected countries, and gray trajectories show the remaining countries in the dataset; markers indicate initial, intermediate, and final states. For visualization, PCA coordinates are linearly mapped back to the original indicator scales.
}
\label{fig:DIG_Diffusion_w_uct}
\end{figure}

Figures~\ref{fig:drift-diff-ratio}, \ref{fig:DIG_Drift_w_uct} and \ref{fig:DIG_Diffusion_w_uct} provide a more detailed view of the political-economic reference dynamics shown in Fig.~2(a) of the main text.
Across the same three GDP-per-capita slices, we examine the magnitudes of the projected drift and diffusion fields, together with their model-estimation uncertainty.

As discussed in the main text, the organization of the projected drift field changes across GDP-per-capita slices, from flow predominantly toward a single convergent region at low and intermediate GDP per capita to two locally convergent configurations at high GDP per capita [Fig.~2(a)].
The drift-magnitude maps provide additional information about the strength of this directional organization [Fig.~\ref{fig:DIG_Drift_w_uct}(a)].
The projected drift $\bm{F}_\parallel$ is generally stronger in lower-democracy regions and weakens overall as GDP per capita increases; at high GDP per capita, it is particularly weak in the intermediate-democracy region separating the two convergent configurations.
The diffusion field provides complementary information about stochastic variability in the same regions [Fig.~\ref{fig:DIG_Diffusion_w_uct}(a)].
The projected diffusion tensors $\mathbf{D}_\parallel$ are anisotropic across much of the state space, with the major axes of the diffusion ellipses predominantly aligned with the democracy dimension.
In the high-GDP slice, diffusion is elevated in the intermediate-democracy region, where its major axis remains predominantly aligned with the democracy direction.
To compare these deterministic and stochastic contributions directly, we additionally evaluate the dimensionless drift-to-diffusion measure $\sqrt{\Delta t\,\bm{F}_\parallel^{\top}\mathbf{D}_\parallel^{-1}\bm{F}_\parallel/(2d_\parallel)}$, which quantifies the local strength of drift relative to diffusion over the observation interval [Fig.~\ref{fig:drift-diff-ratio}].
This measure tends to decrease with increasing GDP per capita and remains particularly small in the intermediate-democracy region at high GDP per capita.
Thus, although the projected drift retains a differentiated two-configuration structure at high GDP per capita, the directional tendency in the intervening region is weak relative to local stochastic variability, whose principal direction is predominantly aligned with the democracy dimension.
This local structure is qualitatively consistent with the USA simulations in Fig.~2(c), where trajectories spread broadly along the democracy dimension and occupy multiple terminal regions rather than remaining tightly concentrated near the high-democracy configuration.

\clearpage
\clearpage

\clearpage

We next assess how stable these inferred structures are across the model ensemble constructed from cross-validation folds and SWAG samples.
For a prediction $A^{(s)}(\bm{x})$ from the $s$th ensemble member, we define the ensemble-mean estimate and its absolute model-estimation uncertainty as
\begin{equation}
\hat{A}(\bm{x}) \equiv \mathbb{E}_{\rm ens}\left[ A^{(s)} (\bm{x}) \right], \qquad
\mathcal{U}_{A}(\bm{x}) \equiv \sqrt{\mathbb{E}_{\rm ens}\left[ \left\| A^{(s)}(\bm{x}) - \hat{A}(\bm{x}) \right\|^2 \right]},
\label{eq:def_uct}
\end{equation}
where $\mathbb{E}_{\rm ens}$ denotes the expectation over the cross-validation and SWAG ensemble.
The relative uncertainty is $\mathcal{U}_{A, {\rm rel}}(\bm{x}) \equiv \mathcal{U}_{A}(\bm{x}) / \| \hat{A}(\bm{x})\|$, which expresses the ensemble dispersion relative to the magnitude of the ensemble-mean estimate.
We take $A=\bm{F}_\parallel$ for the drift and $A=\mathrm{Tr}\,\mathbf{D}_\parallel$ for the diffusion.
Absolute uncertainty varies across the state space for both the projected drift and diffusion fields [Figs.~\ref{fig:DIG_Drift_w_uct}(b) and \ref{fig:DIG_Diffusion_w_uct}(b)].
Because absolute uncertainty can vary with the local magnitude of the inferred quantity, we additionally examine relative uncertainty [Figs.~\ref{fig:DIG_Drift_w_uct}(c) and \ref{fig:DIG_Diffusion_w_uct}(c)].
Across the observed states, the median relative uncertainty is $0.17$ for the projected drift and $0.38$ for the projected diffusion magnitude, with $97.3\%$ and $93.6\%$ of observations, respectively, having relative uncertainty below unity [Figs.~\ref{fig:DIG_Drift_w_uct}(d) and \ref{fig:DIG_Diffusion_w_uct}(d)].
We note that relative uncertainty can nevertheless become large in regions where the corresponding field is weak, even when the absolute ensemble dispersion is modest, so the absolute and relative uncertainty maps should be interpreted jointly.
Together, these results indicate that the broad organization of the inferred stochastic dynamics is comparatively stable across the model ensemble, while fine-scale features in weak-field or high-uncertainty regions should be interpreted more cautiously.

\subsubsection{Exogeneity Diagnostics}

\begin{figure}[!b]
  \centering
  \includegraphics[width=\linewidth]{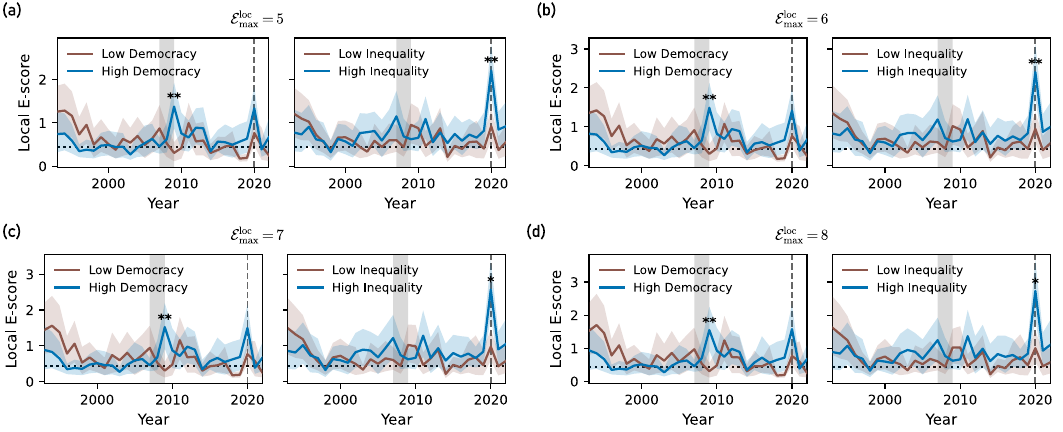}
  \caption{Sensitivity of local exogeneity patterns to the E-score cap. 
  Annual local exogeneity scores $\mathcal{E}^{\rm loc}$ in the modern political-economic dataset, stratified by democracy and inequality quartiles, using score caps (a) $\mathcal{E}^{\rm loc}_{\max}=5$, (b) $6$, (c) $7$, and (d) $8$. 
  Lines show yearly country means for the lowest and highest quartiles, and shaded bands denote pointwise 95\% country-cluster bootstrap confidence intervals. 
  Local E-scores are capped at the indicated value before computing group-level summaries and bootstrap statistics. 
  Gray shading marks the Great Recession (2007–2009), the dashed vertical line marks the onset of the COVID-19 pandemic (2020), and the dotted horizontal line denotes the nominal expectation $1/\ln 10$ under ideal calibration [Eq.~\eqref{eq:nominal_expectation}]. Asterisks indicate Bonferroni-adjusted significance ($^{*}p_{\rm Bonf}<0.05$, $^{**}p_{\rm Bonf}<0.01$).
}
\label{fig:dig_exogeneity_cap}
\end{figure}

We examined the sensitivity of the aggregate local-exogeneity patterns to the cap applied to $\mathcal{E}^{\rm loc}$.
We denote the cap by $\mathcal{E}_{\max}^{\rm loc}$, such that each score is replaced by $\min(\mathcal{E}^{\rm loc},\mathcal{E}_{\max}^{\rm loc})$ before aggregation and resampling; the main-text analyses use $\mathcal{E}_{\max}^{\rm loc}=6$.
The cap limits the influence of a small number of extremely large scores on aggregate summaries, while avoiding distinctions among tail probabilities below the resolution threshold.
Across $\mathcal{E}_{\max}^{\rm loc}=5$--$8$, the annual score trajectories and the principal democracy- and inequality-stratified patterns remain qualitatively similar [Fig.~\ref{fig:dig_exogeneity_cap}].
In particular, the elevated local exogeneity among high-democracy countries in 2009 and high-inequality countries in 2020 remains statistically significant across all four caps, although the adjusted significance of the latter becomes weaker at the two larger caps.
These results indicate that the stratified patterns highlighted in the main text are not specific to the choice $\mathcal{E}_{\max}^{\rm loc}=6$.

\begin{figure}[!ht]
  \centering
  \includegraphics[width=\linewidth]{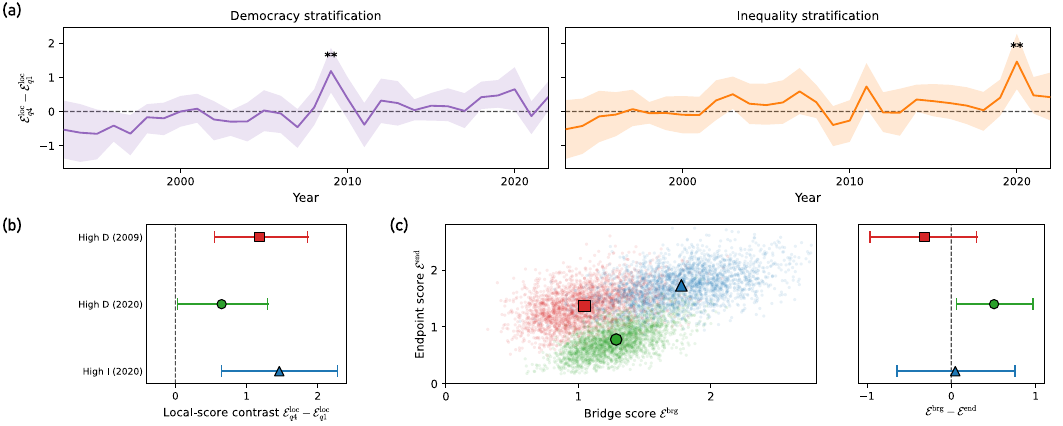}
  \caption{Stratified local exogeneity and aggregate bridge–endpoint profiles in the DIG dataset.
  (a) Annual differences in mean local exogeneity between the highest and lowest democracy and inequality quartiles, ${\mathcal{E}}^{\rm loc}_{q4}-{\mathcal{E}}^{\rm loc}_{q1}$. Lines show observed differences and shaded bands denote pointwise 95\% country-cluster bootstrap confidence intervals. Asterisks indicate Bonferroni-adjusted significance ($^{**}p_{\rm Bonf}<0.01$).
  (b) Local-score contrasts for high democracy in 2009 and 2020 and high inequality in 2020. Points show observed differences and error bars denote 95\% country-cluster bootstrap confidence intervals. The labels identify the high-stratum cases examined in (c).
  (c) Aggregate bridge–endpoint profiles of the selected high-stratum cases. Translucent points show bootstrap replicates of the joint group mean $({\mathcal{E}}^{\rm brg},{\mathcal{E}}^{\rm end})$, with larger markers indicating observed means. The right panel shows ${\mathcal{E}}^{\rm brg}-{\mathcal{E}}^{\rm end}$ with 95\% country-cluster bootstrap confidence intervals. Colors and markers identify the same cases across (b,c), and dashed lines mark zero difference. 
}
\label{fig:dig_exogeneity_group}
\end{figure}

Figure~\ref{fig:dig_exogeneity_group} provides additional detail for the stratified exogeneity patterns shown in Fig.~\ref{fig:dig_exogeneity_cap} and Fig.~3(b) of the main text and the accompanying main-text discussion.
We evaluated the annual high-minus-low contrast in $\mathcal{E}^{\rm loc}$ while resampling countries as clusters to preserve the repeated observations within each country [Fig.~\ref{fig:dig_exogeneity_group}(a)].
The resulting contrasts show more directly that the largest positive democracy-stratified contrast occurs during the Great Recession, whereas the largest positive inequality-stratified contrast occurs at the onset of the COVID-19 pandemic [Fig.~\ref{fig:dig_exogeneity_group}(a,b)].
We then examined the bootstrap distributions of the local, bridge, and endpoint exogeneity scores for three group-year cases highlighted in the main text [Fig.~\ref{fig:dig_exogeneity_group}(b,c)].
These include the 2009 high-democracy and 2020 high-inequality cases associated with pronounced between-stratum contrasts, together with the 2020 high-democracy case, for which local exogeneity is also elevated.
The 2020 high-democracy contrast excludes zero at the pointwise bootstrap level but does not remain significant after Bonferroni correction across years.
The joint distributions of $\mathcal{E}^{\rm brg}$ and $\mathcal{E}^{\rm end}$ reveal different profiles across the two 2020 groups.
The high-democracy group shows a more recovery-like balance, with bridge exogeneity tending to exceed endpoint exogeneity, whereas the high-inequality group shows elevated bridge and endpoint scores of comparable magnitude, corresponding to a more off-path-like footprint.
For the 2009 high-democracy group, the relative bridge--endpoint balance is less clearly resolved under country-level resampling, with the bootstrap interval of their difference spanning zero.
Thus, although local exogeneity is elevated in both high-democracy and high-inequality strata in 2020, their aggregate bridge--endpoint profiles differ, with a more recovery-like profile for high-democracy countries and a more off-path-like profile for high-inequality countries.

\begin{figure}[!ht]
  \centering
  \includegraphics[width=\linewidth]{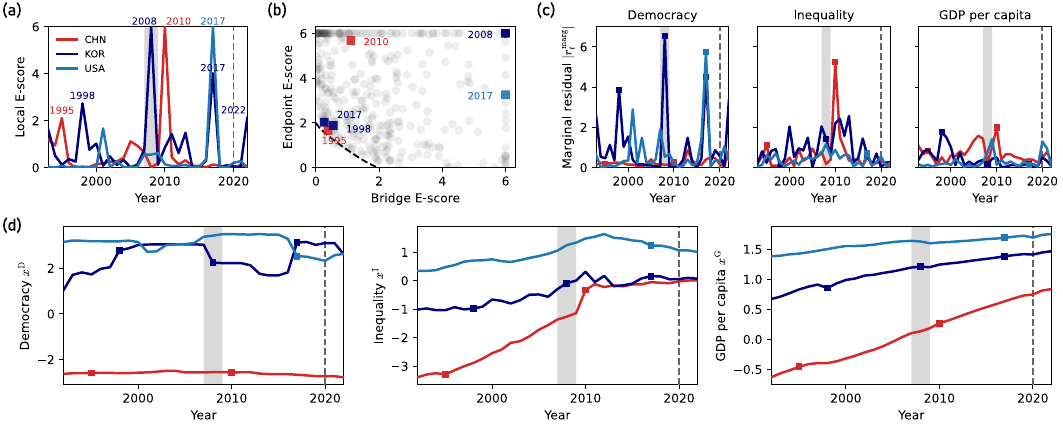}
  \caption{Country-level exogeneity diagnostics and political-economic trajectories in the DIG dataset. 
  (a) Local exogeneity scores $\mathcal{E}^{\rm loc}_t$ for CHN, KOR, and USA.
  (b) Bridge versus endpoint exogeneity scores for DIG peak transitions. Gray points denote all country-year peaks, and the dashed curve shows the lower bound implied by the local-score threshold.
  (c) Absolute marginal standardized residuals $|r^{\rm marg}_t|$ for Democracy, Inequality, and GDP per capita.
  (d) Empirical trajectories in the three state variables. 
  Colored squares in (b--d) mark the labeled peak transitions in (a).
  Peaks are defined by $\mathcal{E}^{\rm loc}_t\geq2$ for visualization; the 2022 KOR peak is omitted from (b) because no subsequent observation is available.
  Gray shading marks the Great Recession (2007--2009) and the dashed vertical line marks the onset of the COVID-19 pandemic (2020).
}
\label{fig:dig_exogeneity_country}
\end{figure}

Figure~\ref{fig:dig_exogeneity_country} further resolves the country-level exogeneity peaks shown in Fig.~3(c,d) of the main text, reproduced here in panels (a,b), by examining deviations across the three political-economic coordinates.
For each coordinate $i$, we compute the absolute marginal standardized residual
\begin{equation}
|r^{\rm marg}_{t,i}|
=
\frac{
\left|\Delta x^{\rm in}_{t,i}-F_{-,i}\Delta t_-\right|
}{
\sqrt{2D_{-,ii}\Delta t_-}
},
\label{eq:abs_marg_residual}
\end{equation}
which measures the deviation of the observed displacement from the drift-predicted displacement in units of the marginal stochastic scale [Fig.~\ref{fig:dig_exogeneity_country}(c)].
A large $|r^{\rm marg}_{t,i}|$ indicates an unusually large deviation in coordinate $i$ when considered marginally, but should not be interpreted as the contribution of that coordinate to $\mathcal{E}^{\rm loc}$.
The full local exogeneity score is obtained from the multivariate Mahalanobis discrepancy $Q^{\rm loc}$, which involves $\mathbf{D}^{-1}$ and therefore depends on the covariance structure across coordinates as well as their marginal variances.

The pattern of marginal residuals varies substantially across countries and even across peaks within the same country.
For China, the 1995 and 2010 peaks are accompanied by elevated residuals in both inequality and GDP per capita.
For Korea, the 1998 peak is prominent in democracy and GDP per capita, the 2008 peak in democracy and inequality, and the 2017 peak predominantly in democracy.
These peaks coincide with major political-economic episodes in the same periods, including the Asian financial crisis and first transfer of power to the opposition (1998), the Great Recession and return of a conservative government (2008), and the presidential impeachment followed by an early election (2017).
For the USA, the democracy residual is most prominent at the 2017 peak, while smaller democracy-related deviations are visible around 2001 and modest increases across all three components occur around the Great Recession [Fig.~\ref{fig:dig_exogeneity_country}(c,d)].
The bridge--endpoint diagnostics additionally show diverse temporal footprints across these transitions [Fig.~\ref{fig:dig_exogeneity_country}(b)].
Together, these examples show that country-level exogeneity peaks can differ substantially in both their coordinate-wise deviation patterns and their subsequent dynamical footprints, even within the same country.
Finally, we emphasize that these temporal correspondences are descriptive and do not by themselves establish the contemporaneous historical events as causal sources of the detected departures.

\subsubsection{Dynamical Irreversibility}

As reported in the main text, country-level mean irreversibility rates vary significantly across world regions.
Here, we further assessed the robustness of this regional heterogeneity using complementary statistical tests.
Welch's one-way ANOVA confirmed significant differences among regions, $F(7,30.04)=5.99$, $p=2.02\times10^{-4}$, with a descriptive effect size of $\eta^2=0.102$.
A rank-based Kruskal--Wallis test provided consistent evidence of regional heterogeneity, $H(7)=27.79$, $p=2.40\times10^{-4}$, and the Welch result remained significant after excluding each country in turn ($p_{\max}=5.59\times10^{-4}$).
These results indicate that the regional heterogeneity is robust to both the choice of omnibus test and the exclusion of any single country.

\begin{figure}[!t]
  \centering
  \includegraphics[width=\linewidth]{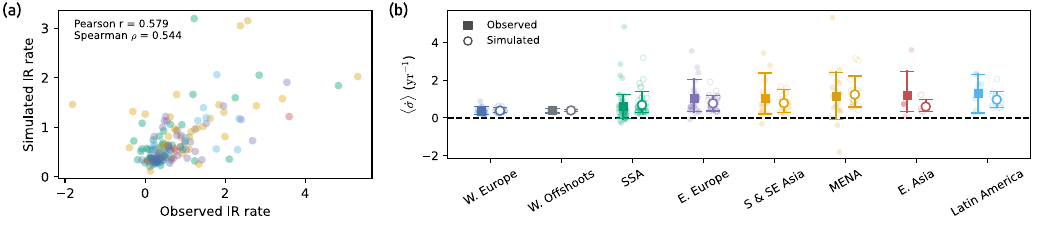}
  \caption{
  Observed and simulated irreversibility rates in modern political-economic trajectories.
  (a) Country-level mean irreversibility rates, $\langle\dot{\sigma}\rangle$, measured from the observed trajectories and from trajectories simulated under the common inferred SDE from each country's observed initial state.
  Each point represents a country, with colors denoting world regions.
  Pearson's $r$ and Spearman's $\rho$ quantify the cross-country correspondence between observed and simulated rates.
  (b) Observed and simulated irreversibility rates grouped by world region.
  Pale filled points denote observed country-level rates and pale open points denote simulated country-level rates.
  Filled squares and open circles indicate regional means, respectively, and vertical bars span the 10th--90th percentiles across countries within each region.
}
\label{fig:dig_ir_comparison}
\end{figure}

We next examined whether this regional ordering is compatible with countries evolving under the same inferred stochastic dynamics.
For each country, we generated 50 trajectories from the common inferred SDE, initialized at the country's observed starting state and propagated over the same observation period. 
Simulated trajectories were evaluated at the observed years of each country, and the country-level simulated irreversibility rate was obtained by averaging across the 50 realizations.
The resulting country-level mean irreversibility rates were positively associated with those measured from the observed trajectories (Pearson's $r=0.579$; Spearman's $\rho=0.544$) [Fig.~\ref{fig:dig_ir_comparison}(a)].
After aggregation by region, the simulated rates also retained a similar broad regional ordering, with a Spearman correlation of $\rho=0.714$ between observed and simulated regional means [Fig.~\ref{fig:dig_ir_comparison}(b)].
Because neither country nor region is included as a conditioning variable in the fitted dynamics, this correspondence indicates that part of the observed regional ordering is compatible with countries evolving from different initial positions within a common state-dependent political-economic landscape.
This does not imply that region-specific historical or institutional influences are fully represented by the shared SDE, but it shows that distinct regional patterns can arise, at least in part, without introducing separate region-specific dynamics.

\subsection{Quantitative History}

\subsubsection{Inferred Stochastic Dynamics and Uncertainty}

\begin{figure}[!b]
  \centering
  \includegraphics[width=\linewidth]{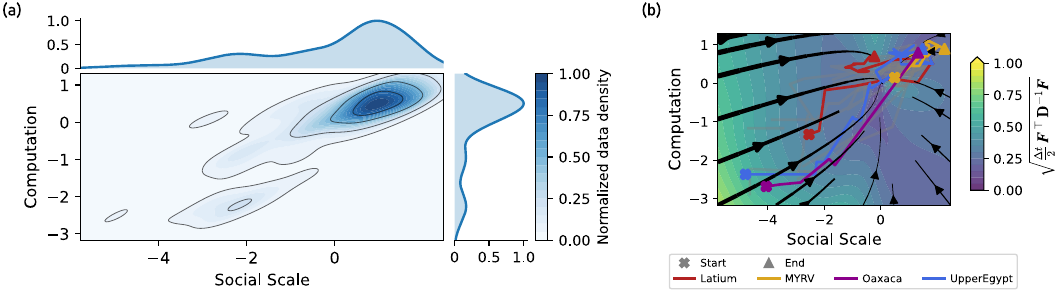}
  \vskip -0.1in
  \caption{
  (a) Empirical distribution of observed states in the Social Scale--Computation plane. 
  The central panel shows the normalized joint data density; the top and right panels show marginal densities of Social Scale and Computation, respectively.
  (b) Inferred drift-to-diffusion ratio in the historical dynamics.
  Black streamlines denote the drift field $\bm{F}(\bm{x})$. Background color shows the local drift-to-diffusion ratio, evaluated with $\Delta t$ set to one century. Smaller values indicate weaker drift relative to diffusion, whereas larger values indicate stronger drift relative to diffusion over the interval $\Delta t$.
  Colored trajectories show selected NGAs and gray trajectories show the remaining NGAs in the dataset; markers indicate the initial and final observed states.
}
\label{fig:Seshat_density_DDratio}
\end{figure}

The inferred Seshat dynamics exhibit a broad organization across the Social Scale–Computation state space, as summarized in the main text [Fig.~2(b)].
Here, we examine these dynamics in more detail through the empirical state density, the local balance between drift and diffusion, and the model-estimation uncertainty of the inferred fields.
To provide context for interpreting the inferred fields, we visualize the empirical distribution of observed NGA states in this plane [Fig.~\ref{fig:Seshat_density_DDratio}(a)], which is most densely occupied toward high social scale and high computation.

\begin{figure}[!t]
  \centering
  \includegraphics[width=\linewidth]{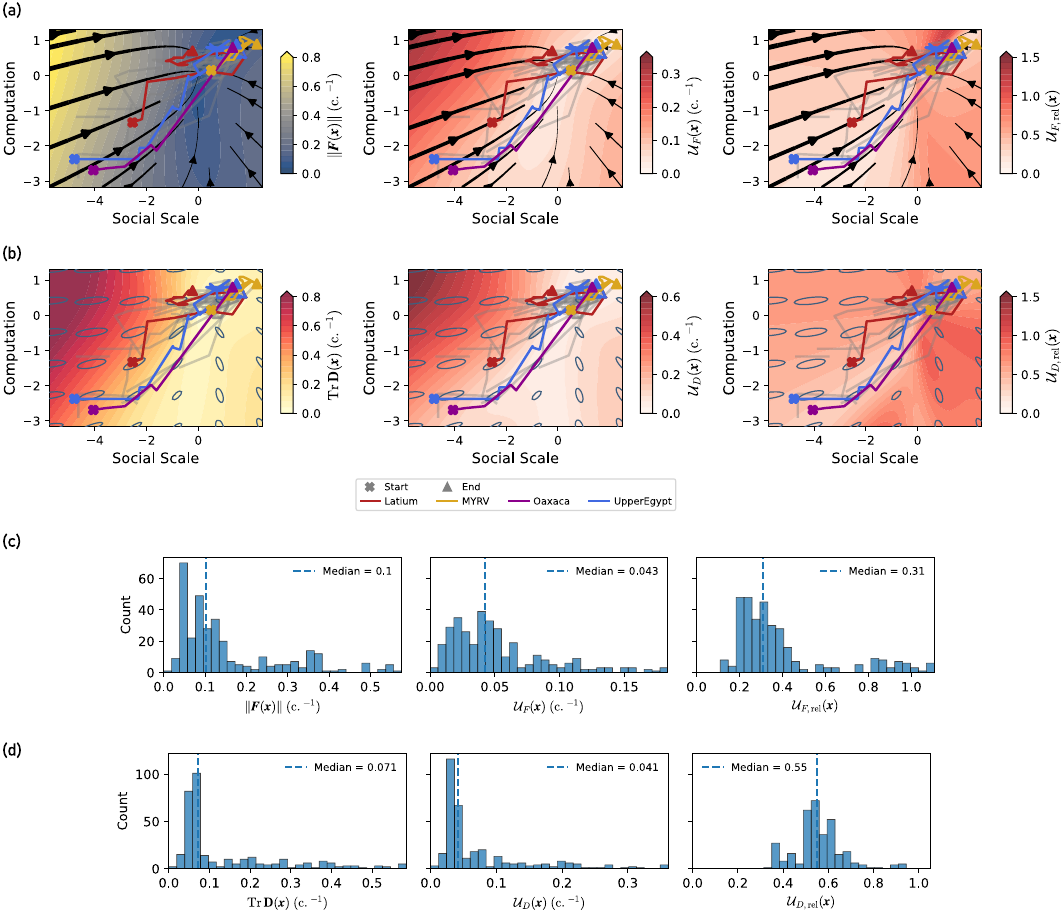}
  \caption{
  Inferred drift and diffusion fields and model-estimation uncertainty in the historical social-complexity dynamics.
  (a) Drift magnitude, $\|\bm{F}(\bm{x})\|$, absolute model-estimation uncertainty, $\mathcal{U}_{F}(\bm{x})$, and relative uncertainty, \(\mathcal{U}_{F,\mathrm{rel}}(\bm{x})=\mathcal{U}_{F}(\bm{x})/\|\bm{F}(\bm{x})\|\), on the Scale--Computation plane, from left to right. 
  Black streamlines show the ensemble-mean drift field.
  (b) Diffusion magnitude, $\mathrm{Tr}\,\mathbf{D}(\bm{x})$, absolute model-estimation uncertainty, $\mathcal{U}_{D}(\bm{x})$, and relative uncertainty, $\mathcal{U}_{D,\mathrm{rel}}(\bm{x})=\mathcal{U}_{D}(\bm{x})/\mathrm{Tr}\,\mathbf{D}(\bm{x})$, from left to right.
  Blue ellipses depict the local anisotropy and principal directions of the ensemble-mean diffusion tensor, with their axis lengths and orientations determined by its eigenvalues and eigenvectors, respectively.
  (c,d) Distributions across observed NGA-year states of the magnitude, absolute uncertainty, and relative uncertainty for (c) drift and (d) diffusion. 
  Dashed vertical lines indicate the median of each distribution. 
  Large relative uncertainty in (a,b) indicates model-estimation uncertainty that is large compared with the local field magnitude and can therefore arise from either large absolute uncertainty or a weak local drift or diffusion magnitude. 
  In (a,b), colored trajectories show selected NGAs and gray trajectories show the remaining NGAs in the dataset; markers indicate the initial and final observed states. Here, $\mathrm{c.}$ denotes one century.
}
\label{fig:Seshat_SDE_w_uct}
\end{figure}

In addition to the broad tendency toward greater social scale and computation described in the main text, both the drift and diffusion magnitudes generally decrease with increasing social scale [left panels in Fig.~\ref{fig:Seshat_SDE_w_uct}(a,b)]. 
The inferred diffusion is also anisotropic, with the major axes of the diffusion ellipses predominantly aligned with the social-scale dimension at lower social scale and rotating toward the computation dimension at higher social scale. Thus, both the magnitude and the principal direction of stochastic variability change across the state space.
To compare the drift and stochastic contributions directly over the observational timescale, we evaluate the local drift-to-diffusion ratio [Fig.~\ref{fig:Seshat_density_DDratio}(b)]. The ratio remains below unity and varies substantially across the state space, with larger values across much of the lower-social-scale region and generally smaller values at higher social scale. This indicates that the stochastic displacement scale generally exceeds the drift displacement scale over a one-century interval, while the relative contribution of drift is stronger in the lower-social-scale region.

We next examine the model-estimation uncertainty of the inferred fields, which varies substantially across the Social Scale–Computation state space [Fig.~\ref{fig:Seshat_SDE_w_uct}(a,b)].
Notably, relative uncertainty is elevated in the high-social-scale, low-computation region, which is sparsely represented in the empirical state distribution [Fig.~\ref{fig:Seshat_density_DDratio}(a); right panels in Fig.~\ref{fig:Seshat_SDE_w_uct}(a,b)].
Across the observed NGA-year states, the median relative uncertainty is $0.31$ for the drift and $0.55$ for the diffusion, with relative uncertainty remaining below unity at $96.1\%$ of the observed states for the drift and at all observed states for the diffusion [Fig.~\ref{fig:Seshat_SDE_w_uct}(c,d)].
Thus, ensemble variation is generally smaller than the local field magnitude over the observed states.

\subsubsection{Exogeneity Diagnostics}

\begin{figure}[!t]
  \centering
  \includegraphics[width=\linewidth]{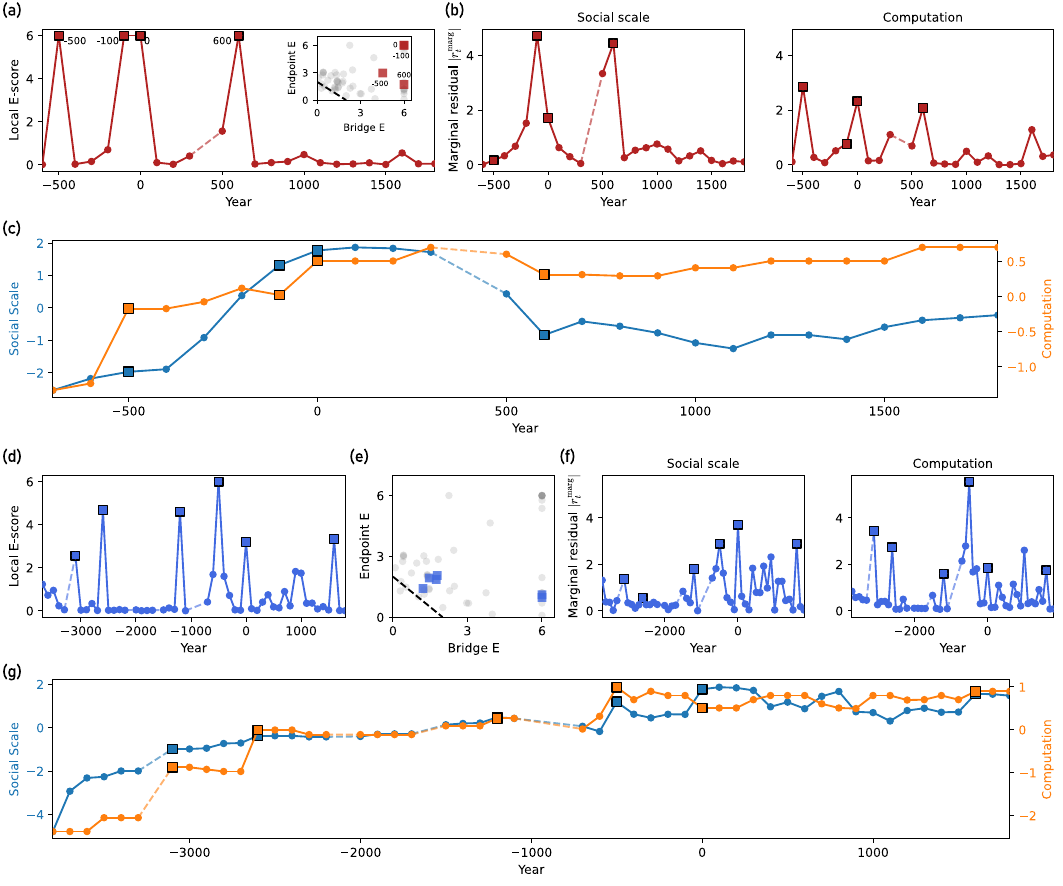}
  \caption{
  Exogeneity diagnostics for representative Seshat trajectories.
  (a–c) Latium and (d–g) Upper Egypt. 
  (a,d) Local exogeneity scores $E_t^{\mathrm{loc}}$ along each NGA trajectory. Squares mark peak observations with $\mathcal{E}_t^{\mathrm{loc}} \geq 2$; scores are capped at 6 for visualization. 
  The inset in (a) and panel (e) show bridge and endpoint exogeneity scores for peak observations. Gray points denote peak observations across the Seshat dataset, colored squares highlight peaks from the respective NGA, and the black dashed line marks the lower bound in the bridge–endpoint plane. 
  (b,f) Absolute marginal standardized residuals for the Social Scale and Computation coordinates, showing which state variable exhibits the larger coordinate-wise departure at each transition. Squares identify the same peak observations highlighted in (a,d). (c,g) Observed trajectories of Social Scale (blue, left axis) and Computation (orange, right axis), with peak observations marked by squares. 
  Dashed segments in all panels denote transitions spanning more than one century.
}
\label{fig:Seshat_exogeneity}
\end{figure}

The main text used Latium to illustrate how locally atypical transitions can exhibit distinct bridge--endpoint footprints along a long historical trajectory [Fig.~3(e)].
Here, we examine these episodes in more detail by resolving the local deviations into the Social Scale and Computation coordinates and comparing them with observed state trajectories [Fig.~\ref{fig:Seshat_exogeneity}].
We use the absolute marginal standardized residuals $| r_t^{\rm marg} |$ defined in Eq.~\eqref{eq:abs_marg_residual} to characterize the coordinate-wise magnitude of each local departure, while the bridge and endpoint scores characterize its two-step temporal footprint.
We further apply the same diagnostics to Upper Egypt, providing an additional long-run NGA trajectory with a distinct sequence of exogeneity peaks and score geometries.

In Latium, the four pronounced local exogeneity peaks around $500$ BCE, $100$ BCE, the turn of the Common Era, and $600$ CE exhibit distinct coordinate-wise deviations and temporal footprints [Fig.~\ref{fig:Seshat_exogeneity}(a--c)]. 
The $500$ BCE peak is dominated by the Computation residual, with comparatively little deviation along Social Scale, and falls near the transition from the Roman monarchy to the Republic. 
The $100$ BCE peak is instead dominated by the Social Scale residual, corresponding to a rapid increase in social scale. 
By contrast, the peak around $0$ CE shows a larger deviation in the Computation coordinate, coinciding with a sharp increase in computation and a more modest increase in social scale. 
Despite these different displacement directions, the latter two transitions lie in the high-bridge, high-endpoint region and therefore exhibit off-path-like profiles. 
Together, they form the two-stage sequence discussed in the main text, coinciding with the transition from the late Republican to early imperial period~\cite{bowman1996provincial}. 
The $600$ CE peak provides a contrasting pattern: it is accompanied by contractions in both coordinates, most prominently in social scale, while the endpoint score is substantially lower than the bridge score, yielding a more recovery-like profile. 
This episode overlaps with the disruptions associated with Lombard expansion and the accompanying political reconfiguration of Italy~\cite{humphries2001italy}. 
Together, these episodes show that pronounced local exogeneity can accompany different forms of historical change, differing both in their coordinate-wise structure and in their subsequent temporal footprints.

Upper Egypt provides a complementary example over a substantially longer historical trajectory [Fig.~\ref{fig:Seshat_exogeneity}(d--g)]. Local exogeneity is again episodic, with pronounced peaks distributed across different periods. 
The associated marginal residuals show that these departures are heterogeneous in their coordinate-wise structure: several episodes are expressed more strongly along Computation, whereas others involve larger Social Scale deviations or pronounced departures in both coordinates. 
Their bridge--endpoint scores likewise span distinct geometries, including both localized-like profiles and high-bridge, low-endpoint profiles corresponding to more recovery-like profiles [Fig.~\ref{fig:Seshat_exogeneity}(e)].
Thus, the Upper Egypt trajectory further shows that large local departures need not share the same coordinate-wise structure or two-step patterns.

\subsubsection{Dynamical Irreversibility}

\begin{figure}[!t]
  \centering
  \includegraphics[width=\linewidth]{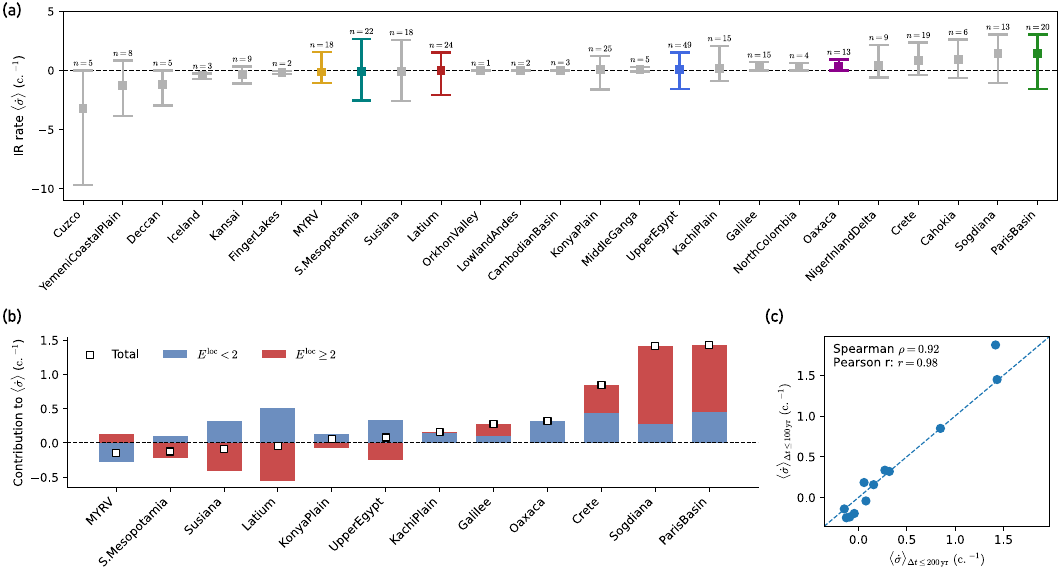}
  \caption{
  Dynamical irreversibility (IR) across Seshat trajectories.
  (a) NGA-level mean IR rates $\langle\dot{\sigma}\rangle$. 
  Squares denote the mean IR rate for each NGA, and vertical bars span the 10th–90th percentiles of transition-level irreversibility rates. The number $n$ above each NGA denotes the number of retained transitions with observation gaps $\Delta t\leq200 \,\mathrm{yr}$. Colored symbols highlight selected NGAs, and gray symbols denote the remaining NGAs.
  (b) Decomposition of $\langle\dot{\sigma}\rangle$ into contributions from transitions with $E^{\mathrm{loc}}<2$ (blue) and $E^{\mathrm{loc}}\geq2$ (red), shown for NGAs with at least ten retained transitions. 
  Each contribution is computed by summing the local IR values and normalizing by the total retained elapsed time of the NGA, so that the two contributions sum to the total mean rate, indicated by open squares.
  (c) Robustness of NGA-level mean IR rates to the observation-gap criterion. Rates obtained using the main criterion $\Delta t\leq200 \, \mathrm{yr}$ are compared with those obtained using the stricter criterion $\Delta t\leq100 \, \mathrm{yr}$ for all NGAs with at least ten retained transitions under the $\Delta t\leq200 \, \mathrm{yr}$ criterion. The dashed line denotes equality; Spearman and Pearson correlation coefficients are indicated in the panel. In all panels, IR rates are reported per century ($\mathrm{c.}^{-1}$).
}
\label{fig:Seshat_irreversibility}
\end{figure}

We extend the NGA-level comparison in the main text by showing the mean irreversibility rate $\langle\dot{\sigma}\rangle$ for all 25 Seshat trajectories, using transitions with observation gaps of at most two centuries [Fig.~\ref{fig:Seshat_irreversibility}(a)].
The rates vary substantially across NGAs, with pronounced transition-level heterogeneity within individual trajectories, as reflected by the 10th--90th percentile ranges. For many NGAs, these ranges span both positive and negative values, indicating that the long-run irreversibility rate arises from the accumulation of local transitions with opposing directional contributions rather than from uniformly signed irreversibility along the trajectory.
To avoid basing cross-NGA comparisons on only a small number of transitions, we retain the main-text criterion of at least ten retained transitions for the comparative analyses below, while showing the full set here for completeness.

To quantify the relationship between transition atypicality and directional asymmetry across NGAs, we decompose each NGA-level mean irreversibility rate $\langle \dot{\sigma} \rangle$, defined in Eq.~\eqref{eq:IR_rate}, into additive contributions from lower-exogeneity transitions with $\mathcal{E}^{\mathrm{loc}}<2$ and elevated-exogeneity transitions with $\mathcal{E}^{\mathrm{loc}}\geq2$ [Fig.~\ref{fig:Seshat_irreversibility}(b)]. 
Each contribution is obtained by summing the local irreversibility values within each subset and normalizing by the total retained elapsed time of the NGA, so that the two contributions sum to the total $\langle \dot{\sigma} \rangle$. 
Among the 12 NGAs satisfying the minimum-transition criterion, the $\mathcal{E}^{\mathrm{loc}}<2$ contribution is positive in 11, with the Middle Yellow River Valley (MYRV) as the sole exception. 
In contrast, the contribution from $\mathcal{E}^{\mathrm{loc}}\geq2$ transitions varies substantially in both sign and magnitude: it offsets the positive lower-exogeneity contribution in trajectories such as Latium, Upper Egypt, Susiana, and Southern Mesopotamia (S. Mesopotamia), while reinforcing positive irreversibility in Crete, Sogdiana, Paris Basin, and Galilee. 
Thus, elevated local exogeneity does not imply a particular sign of temporal asymmetry, as atypical transitions can contribute either positively or negatively to the NGA-level irreversibility rate.

Finally, given the sparse and irregular observation intervals of the Seshat trajectories, we assess the sensitivity of the NGA-level irreversibility rates to the maximum observation-gap criterion [Fig.~\ref{fig:Seshat_irreversibility}(c)].
The main analysis retains transitions with observation gaps $\Delta t\leq200 \, \mathrm{yr}$; here, we recompute \(\langle\dot{\sigma}\rangle\) using the stricter criterion $\Delta t\leq100 \, \mathrm{yr}$ for the same NGAs satisfying the minimum-transition criterion under the main analysis.
The resulting rates remain highly consistent with those obtained using the 200-year cutoff, with Spearman correlation $\rho=0.92$ and Pearson correlation $r=0.98$.
Thus, both the relative ordering and overall magnitudes of the NGA-level irreversibility rates are largely preserved under the stricter observation-gap criterion, indicating that the cross-NGA patterns are not primarily driven by the inclusion of two-century transitions.

\clearpage

\subsection{Cultural Analytics}

\subsubsection{Inferred Stochastic Dynamics and Uncertainty}

\begin{figure}[!t]
  \centering
  \includegraphics[width=\linewidth]{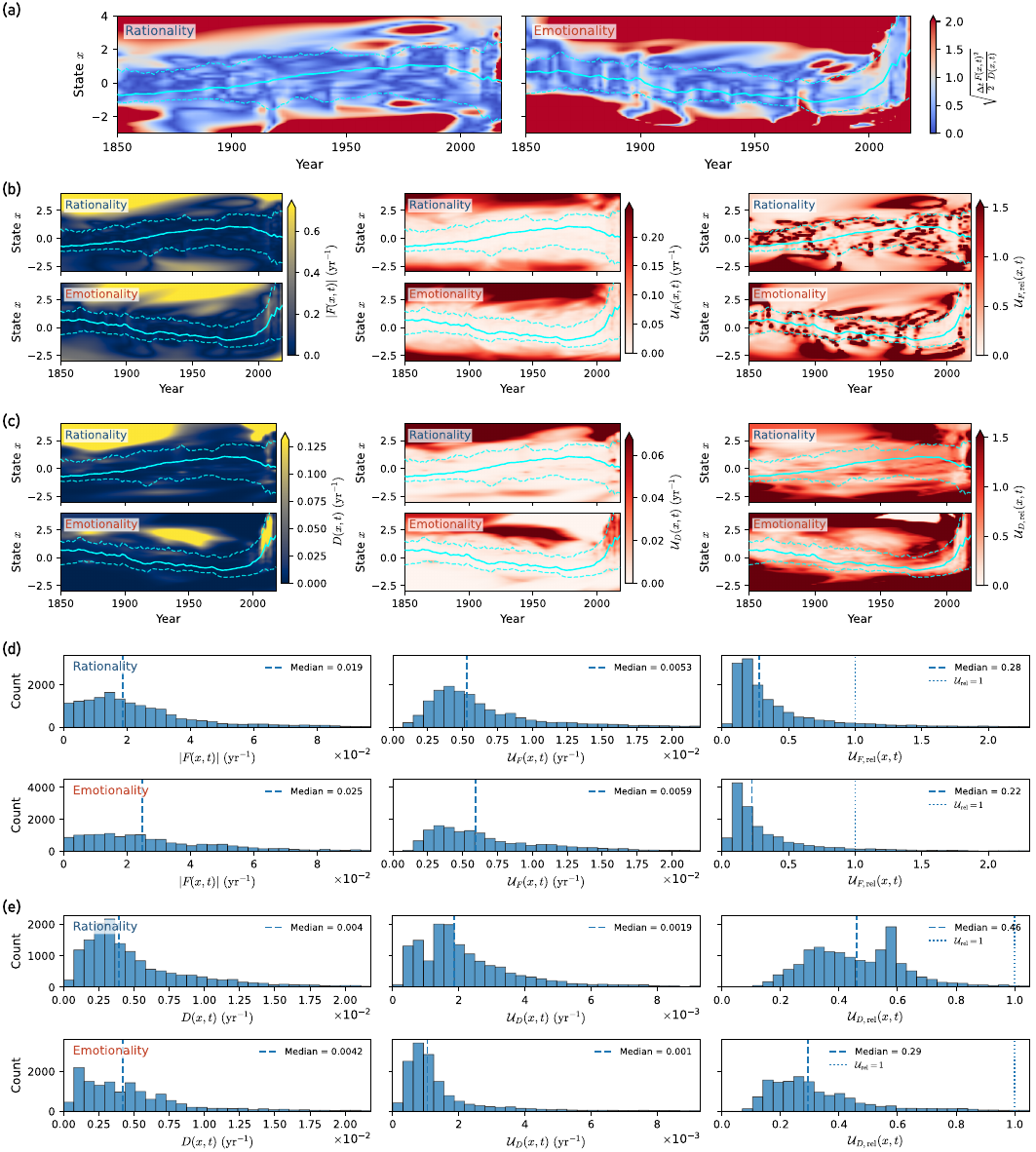}
  \caption{
  Inferred stochastic dynamics and model-estimation uncertainty in the Google Books Ngram analysis.
  (a) Drift-to-diffusion ratio for Rationality and Emotionality, evaluated over a two-year interval; a value of unity marks equal drift and stochastic displacement scales.
  (b) Drift magnitude $|F(x,t)|$, absolute model-estimation uncertainty $\mathcal{U}_F(x,t)$, and relative uncertainty $\mathcal{U}_{F,\mathrm{rel}}(x,t)=\mathcal{U}_F(x,t)/|F(x,t)|$, from left to right.
  (c) Diffusion quantities $D(x,t)$, $\mathcal{U}_D(x,t)$, and $\mathcal{U}_{D,\mathrm{rel}}(x,t)=\mathcal{U}_D(x,t)/D(x,t)$.
  (d,e) Distributions across observed word-year states of the corresponding magnitude, absolute uncertainty, and relative uncertainty for (d) drift and (e) diffusion; dashed vertical lines indicate medians and dotted lines mark $\mathcal{U}_{A,\mathrm{rel}}=1$.
  In (a--c), solid cyan lines show the mean observed state and dashed cyan lines show the 2.5th--97.5th percentile range at each observation time.
}
\label{fig:Ngram_SDE}
\end{figure}

The main text showed that the inferred time-dependent drift landscape reorganizes over the observation period, broadly tracking the reversal of the Rationality and Emotionality lexicon means around 1980.
Here, we further examine the relative contributions of systematic drift and stochastic variability across the state--time domain by evaluating the local drift-to-diffusion ratio over the two-year observation interval [Fig.~\ref{fig:Ngram_SDE}(a)].
Across much of the central 95\% range of observed states, the ratio remains below unity for both lexicons, indicating that the stochastic displacement scale generally exceeds the drift displacement scale over an observation interval.
Thus, the lexical reversal is associated with a reorganization of the time-dependent drift field while substantial stochastic variability remains present.

We next examine the inferred drift and diffusion fields together with their model-estimation uncertainty across the state--time domain [Fig.~\ref{fig:Ngram_SDE}(b,c)].
Both the drift magnitude $|F(x,t)|$ and diffusion $D(x,t)$ vary substantially with state and calendar time, while remaining relatively small across much of the central 95\% range of observed states; larger values occur mainly in more peripheral state-space regions and shift in location over time.
The absolute uncertainties are likewise heterogeneous, tending to remain smaller across much of the empirically occupied region and to increase toward more weakly sampled state--time regions.
Relative drift uncertainty exhibits localized increases even within the empirical range, reflecting in part the normalization by $|F(x,t)|$ where the drift is locally weak.
Relative diffusion uncertainty varies more smoothly across the state--time domain, with larger values generally occurring toward the more weakly sampled boundaries.

To quantify model-estimation uncertainty along the observed lexical trajectories, we examine the distributions of field magnitudes and uncertainties across all observed states [Fig.~\ref{fig:Ngram_SDE}(d,e)].
The median relative drift uncertainty is $0.28$ for Rationality and $0.22$ for Emotionality, with $\mathcal{U}_{F,\mathrm{rel}}<1$ for $86.1\%$ and $88.7\%$ of observed states, respectively.
For the diffusion, the median relative uncertainty is $0.46$ for Rationality and $0.29$ for Emotionality, with $\mathcal{U}_{D,\mathrm{rel}}<1$ for $96.2\%$ and $93.5\%$ of observed states, respectively.
Thus, the inferred local dynamics are comparatively well constrained over most empirically observed lexical states, supporting the interpretation of the time-dependent reorganization identified in the main text as a feature of the data-supported dynamics.

\subsubsection{Exogeneity Diagnostics and Dynamical Irreversibility}

The main text showed that the lexical reversal is accompanied by a pronounced increase in dynamical irreversibility [Fig.~5(c)], while the lexicon-averaged local exogeneity exhibits no comparably persistent change [Fig.~5(d)].
Here, we further examine these contrasting patterns at the word level.
We test whether the absence of a clear aggregate exogeneity pattern could conceal a broad increase in the incidence of individually atypical transitions.
We then examine whether the increase in irreversibility is broadly distributed across individual words or is instead concentrated in a smaller subset.

\begin{figure}[!hb]
  \centering
  \includegraphics[width=\linewidth]{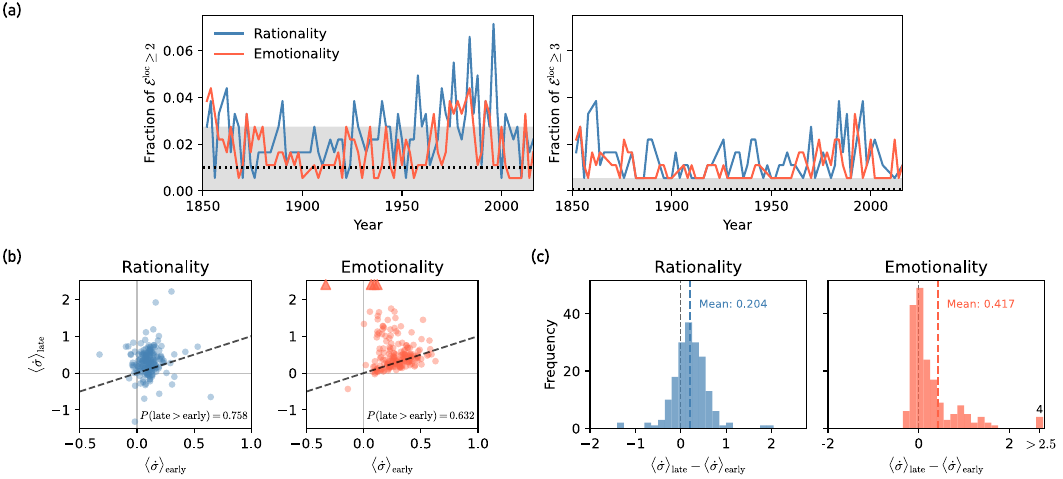}
  \caption{
  Additional exogeneity and irreversibility analyses for the lexical dynamics.
  (a) Fraction of words with elevated local exogeneity, $\mathcal{E}^{\mathrm{loc}}\geq2$ (left) and $\mathcal{E}^{\mathrm{loc}}\geq3$ (right), for the Rationality and Emotionality lexicons.
  Black dotted lines mark the nominal exceedance fractions $10^{-2}$ and $10^{-3}$, and gray bands show the 2.5th--97.5th percentile ranges of the binomial references for 182 words.
  (b) Word-level mean irreversibility rates $\langle\dot{\sigma}\rangle$ in the early (1900--1960) and late (1990--2016) periods.
  Dashed diagonals indicate equal early- and late-period rates, and annotations give the fractions of words with $\langle\dot{\sigma}\rangle_{\mathrm{late}}>\langle\dot{\sigma}\rangle_{\mathrm{early}}$. 
  Four Emotionality words above the displayed late-period range are shown as upward triangles.
  (c) Distributions of the word-level changes $\langle\dot{\sigma}\rangle_{\mathrm{late}}-\langle\dot{\sigma}\rangle_{\mathrm{early}}$.
  Black dashed lines mark zero change and colored dashed lines mark the mean change; the rightmost Emotionality bin contains the four words with changes greater than $2.5$.
}
\label{fig:Ngram_Exo_IR}
\end{figure}

To examine the word-level incidence of atypical transitions, we calculate at each observation time the fraction of words with elevated local exogeneity, using thresholds of $\mathcal{E}^{\mathrm{loc}}\geq2$ and $\mathcal{E}^{\mathrm{loc}}\geq3$ [Fig.~\ref{fig:Ngram_Exo_IR}(a)].
For both lexicons, the exceedance fractions fluctuate over time and occasionally rise above the nominal binomial reference ranges, but neither threshold reveals a sharp or sustained increase centered on the lexical reversal.
This supports the main-text conclusion that the lexical reversal is not accompanied by a distinct lexicon-wide signature in local exogeneity.

We next examine how the increase in dynamical irreversibility is distributed across individual words by comparing word-level mean irreversibility rates $\langle \dot{\sigma} \rangle$ between the early (1900--1960) and late (1990--2016) periods [Fig.~\ref{fig:Ngram_Exo_IR}(b,c)].
The late-period rate exceeds the early-period rate for $75.8\%$ of Rationality words and $63.2\%$ of Emotionality words, indicating that the aggregate increase is not confined to only a small subset of lexical trajectories.
For Rationality, the distribution of word-level changes is broadly shifted toward positive values, with similar mean and median changes ($0.204\,\mathrm{yr}^{-1}$ and $0.194\,\mathrm{yr}^{-1}$, respectively), consistent with a broadly distributed rise in irreversibility.
In contrast, Emotionality exhibits a more heterogeneous pattern: although a majority of words also increase, the mean change ($0.417\,\mathrm{yr}^{-1}$) substantially exceeds the median ($0.071\,\mathrm{yr}^{-1}$), reflecting a pronounced positive tail in which a small number of words undergo particularly large increases.
Taken together, these results show that the lexical reversal is accompanied by increased temporal asymmetry without a corresponding lexicon-wide increase in local transition atypicality, while the word-level composition of the irreversibility increase differs markedly between the two lexicons.

\section{Dataset Summary}
\label{sec:data_summary}

This section provides supplementary descriptive summaries of the datasets used in the three empirical applications.
The tables below complement the dataset construction and preprocessing procedures described in Sec.~\ref{sec:data} by reporting the observational units, temporal coverage, constituent variables, and lexical sets used in the analyses.


\begingroup
\fontsize{6}{6.8}\selectfont
\setlength{\tabcolsep}{5pt}
\renewcommand{\arraystretch}{1.05}
\setlength{\LTleft}{0pt plus 1fill}
\setlength{\LTright}{0pt plus 1fill}
\setlength{\LTcapwidth}{\textwidth}
\begin{longtable}{@{}cc*{8}{c}@{}}
\caption{Country- or territory-level summary statistics for the DIG analysis sample. 
Rows correspond to harmonized geopolitical units identified by ISO 3166-1 alpha-3 codes.
Entries report mean $\pm$ standard deviation over the available annual observations within the 1992–2022 analysis period. The listed variables are the source-level indicators used to construct the Democracy, Inequality, and economic-development coordinates; $\log_{10}\mathrm{gdppc}$ denotes the log-transformed real GDP per capita prior to standardization.}
\label{tab:country-summary-mean-sd}\\
\toprule
ISO3 & Years & \makecell{\texttt{v2x\_}\\\texttt{polyarchy}} & \makecell{\texttt{v2x\_}\\\texttt{libdem}} & \makecell{\texttt{v2x\_}\\\texttt{partipdem}} & \makecell{\texttt{v2x\_}\\\texttt{delibdem}} & \makecell{\texttt{v2x\_}\\\texttt{egaldem}} & \texttt{ghweal992j} & \texttt{gptinc992j} & \makecell{\texttt{log10\_}\\\texttt{gdppc}} \\
\midrule
\endfirsthead

\multicolumn{10}{c}{\tablename\ \thetable\ -- continued from previous page}\\
\toprule
ISO3 & Years & \makecell{\texttt{v2x\_}\\\texttt{polyarchy}} & \makecell{\texttt{v2x\_}\\\texttt{libdem}} & \makecell{\texttt{v2x\_}\\\texttt{partipdem}} & \makecell{\texttt{v2x\_}\\\texttt{delibdem}} & \makecell{\texttt{v2x\_}\\\texttt{egaldem}} & \texttt{ghweal992j} & \texttt{gptinc992j} & \makecell{\texttt{log10\_}\\\texttt{gdppc}} \\
\midrule
\endhead

\midrule
\multicolumn{10}{r}{Continued on next page}\\
\endfoot

\bottomrule
\endlastfoot
AFG & 1992--2022 & $0.24\!\pm\!0.13$ & $0.11\!\pm\!0.08$ & $0.09\!\pm\!0.06$ & $0.17\!\pm\!0.13$ & $0.13\!\pm\!0.08$ & $0.74\!\pm\!0.00$ & $0.52\!\pm\!0.01$ & $3.03\!\pm\!0.23$ \\
AGO & 1992--2022 & $0.22\!\pm\!0.08$ & $0.10\!\pm\!0.03$ & $0.07\!\pm\!0.01$ & $0.09\!\pm\!0.04$ & $0.09\!\pm\!0.03$ & $0.80\!\pm\!0.02$ & $0.66\!\pm\!0.03$ & $3.59\!\pm\!0.31$ \\
ALB & 1992--2022 & $0.49\!\pm\!0.06$ & $0.40\!\pm\!0.05$ & $0.30\!\pm\!0.04$ & $0.34\!\pm\!0.05$ & $0.36\!\pm\!0.04$ & $0.73\!\pm\!0.00$ & $0.48\!\pm\!0.01$ & $3.84\!\pm\!0.20$ \\
ARE & 1993--2022 & $0.06\!\pm\!0.03$ & $0.08\!\pm\!0.01$ & $0.02\!\pm\!0.01$ & $0.10\!\pm\!0.02$ & $0.12\!\pm\!0.02$ & $0.79\!\pm\!0.02$ & $0.65\!\pm\!0.03$ & $4.75\!\pm\!0.14$ \\
ARG & 1992--2022 & $0.80\!\pm\!0.03$ & $0.63\!\pm\!0.03$ & $0.55\!\pm\!0.03$ & $0.65\!\pm\!0.09$ & $0.63\!\pm\!0.02$ & $0.75\!\pm\!0.01$ & $0.61\!\pm\!0.05$ & $4.21\!\pm\!0.07$ \\
ARM & 1992--2022 & $0.44\!\pm\!0.14$ & $0.27\!\pm\!0.12$ & $0.25\!\pm\!0.10$ & $0.29\!\pm\!0.14$ & $0.37\!\pm\!0.12$ & $0.74\!\pm\!0.00$ & $0.51\!\pm\!0.04$ & $3.88\!\pm\!0.16$ \\
AUS & 1992--2022 & $0.88\!\pm\!0.02$ & $0.83\!\pm\!0.02$ & $0.63\!\pm\!0.02$ & $0.80\!\pm\!0.02$ & $0.78\!\pm\!0.03$ & $0.73\!\pm\!0.00$ & $0.48\!\pm\!0.01$ & $4.61\!\pm\!0.08$ \\
AUT & 1992--2022 & $0.86\!\pm\!0.01$ & $0.79\!\pm\!0.01$ & $0.63\!\pm\!0.01$ & $0.77\!\pm\!0.02$ & $0.76\!\pm\!0.01$ & $0.79\!\pm\!0.01$ & $0.42\!\pm\!0.01$ & $4.57\!\pm\!0.06$ \\
AZE & 1992--2022 & $0.23\!\pm\!0.04$ & $0.08\!\pm\!0.03$ & $0.07\!\pm\!0.01$ & $0.07\!\pm\!0.03$ & $0.15\!\pm\!0.03$ & $0.74\!\pm\!0.00$ & $0.50\!\pm\!0.04$ & $3.95\!\pm\!0.30$ \\
BDI & 1992--2022 & $0.24\!\pm\!0.08$ & $0.12\!\pm\!0.05$ & $0.17\!\pm\!0.06$ & $0.19\!\pm\!0.08$ & $0.17\!\pm\!0.04$ & $0.75\!\pm\!0.01$ & $0.58\!\pm\!0.02$ & $2.88\!\pm\!0.06$ \\
BEL & 1992--2022 & $0.89\!\pm\!0.01$ & $0.83\!\pm\!0.01$ & $0.64\!\pm\!0.01$ & $0.80\!\pm\!0.01$ & $0.83\!\pm\!0.01$ & $0.67\!\pm\!0.00$ & $0.44\!\pm\!0.01$ & $4.55\!\pm\!0.05$ \\
BEN & 1992--2022 & $0.60\!\pm\!0.08$ & $0.48\!\pm\!0.08$ & $0.36\!\pm\!0.07$ & $0.52\!\pm\!0.10$ & $0.46\!\pm\!0.05$ & $0.76\!\pm\!0.01$ & $0.60\!\pm\!0.02$ & $3.29\!\pm\!0.04$ \\
BFA & 1992--2022 & $0.53\!\pm\!0.09$ & $0.35\!\pm\!0.09$ & $0.29\!\pm\!0.07$ & $0.42\!\pm\!0.08$ & $0.36\!\pm\!0.05$ & $0.78\!\pm\!0.03$ & $0.63\!\pm\!0.04$ & $3.14\!\pm\!0.05$ \\
BGD & 1992--2022 & $0.39\!\pm\!0.11$ & $0.19\!\pm\!0.07$ & $0.17\!\pm\!0.05$ & $0.17\!\pm\!0.04$ & $0.17\!\pm\!0.05$ & $0.74\!\pm\!0.00$ & $0.54\!\pm\!0.01$ & $3.34\!\pm\!0.21$ \\
BGR & 1992--2022 & $0.68\!\pm\!0.03$ & $0.58\!\pm\!0.04$ & $0.45\!\pm\!0.02$ & $0.58\!\pm\!0.05$ & $0.54\!\pm\!0.04$ & $0.73\!\pm\!0.01$ & $0.49\!\pm\!0.03$ & $4.09\!\pm\!0.15$ \\
BHR & 2002--2022 & $0.17\!\pm\!0.04$ & $0.06\!\pm\!0.02$ & $0.06\!\pm\!0.03$ & $0.09\!\pm\!0.05$ & $0.13\!\pm\!0.02$ & $0.80\!\pm\!0.02$ & $0.66\!\pm\!0.03$ & $4.51\!\pm\!0.10$ \\
BIH & 1992--2022 & $0.52\!\pm\!0.14$ & $0.33\!\pm\!0.11$ & $0.30\!\pm\!0.08$ & $0.35\!\pm\!0.11$ & $0.36\!\pm\!0.09$ & $0.73\!\pm\!0.00$ & $0.47\!\pm\!0.02$ & $3.88\!\pm\!0.17$ \\
BLR & 1992--2022 & $0.29\!\pm\!0.11$ & $0.14\!\pm\!0.11$ & $0.12\!\pm\!0.08$ & $0.11\!\pm\!0.08$ & $0.30\!\pm\!0.07$ & $0.73\!\pm\!0.00$ & $0.45\!\pm\!0.03$ & $4.12\!\pm\!0.16$ \\
BRA & 1992--2022 & $0.83\!\pm\!0.06$ & $0.71\!\pm\!0.08$ & $0.58\!\pm\!0.07$ & $0.71\!\pm\!0.11$ & $0.56\!\pm\!0.08$ & $0.82\!\pm\!0.00$ & $0.69\!\pm\!0.01$ & $4.07\!\pm\!0.09$ \\
BWA & 1992--2022 & $0.69\!\pm\!0.04$ & $0.59\!\pm\!0.04$ & $0.45\!\pm\!0.03$ & $0.56\!\pm\!0.05$ & $0.48\!\pm\!0.04$ & $0.83\!\pm\!0.01$ & $0.74\!\pm\!0.02$ & $4.02\!\pm\!0.15$ \\
CAF & 1992--2022 & $0.33\!\pm\!0.04$ & $0.19\!\pm\!0.03$ & $0.15\!\pm\!0.02$ & $0.20\!\pm\!0.05$ & $0.19\!\pm\!0.02$ & $0.82\!\pm\!0.03$ & $0.71\!\pm\!0.04$ & $2.90\!\pm\!0.09$ \\
CAN & 1992--2022 & $0.83\!\pm\!0.01$ & $0.76\!\pm\!0.01$ & $0.58\!\pm\!0.02$ & $0.73\!\pm\!0.03$ & $0.70\!\pm\!0.02$ & $0.62\!\pm\!0.01$ & $0.49\!\pm\!0.01$ & $4.59\!\pm\!0.06$ \\
CHE & 1992--2022 & $0.89\!\pm\!0.01$ & $0.85\!\pm\!0.01$ & $0.79\!\pm\!0.01$ & $0.86\!\pm\!0.01$ & $0.82\!\pm\!0.01$ & $0.74\!\pm\!0.01$ & $0.42\!\pm\!0.01$ & $4.70\!\pm\!0.09$ \\
CHL & 1992--2022 & $0.87\!\pm\!0.02$ & $0.81\!\pm\!0.03$ & $0.57\!\pm\!0.02$ & $0.79\!\pm\!0.03$ & $0.62\!\pm\!0.05$ & $0.83\!\pm\!0.01$ & $0.73\!\pm\!0.01$ & $4.25\!\pm\!0.08$ \\
CHN & 1992--2022 & $0.09\!\pm\!0.01$ & $0.06\!\pm\!0.01$ & $0.04\!\pm\!0.01$ & $0.14\!\pm\!0.02$ & $0.10\!\pm\!0.01$ & $0.66\!\pm\!0.09$ & $0.53\!\pm\!0.04$ & $3.89\!\pm\!0.25$ \\
CIV & 1992--2022 & $0.43\!\pm\!0.07$ & $0.28\!\pm\!0.05$ & $0.29\!\pm\!0.04$ & $0.33\!\pm\!0.10$ & $0.27\!\pm\!0.04$ & $0.76\!\pm\!0.01$ & $0.59\!\pm\!0.02$ & $3.42\!\pm\!0.09$ \\
CMR & 1992--2022 & $0.30\!\pm\!0.01$ & $0.15\!\pm\!0.01$ & $0.15\!\pm\!0.02$ & $0.14\!\pm\!0.02$ & $0.19\!\pm\!0.01$ & $0.76\!\pm\!0.01$ & $0.62\!\pm\!0.01$ & $3.35\!\pm\!0.08$ \\
COD & 1992--2022 & $0.26\!\pm\!0.09$ & $0.10\!\pm\!0.05$ & $0.17\!\pm\!0.05$ & $0.15\!\pm\!0.07$ & $0.15\!\pm\!0.04$ & $0.76\!\pm\!0.01$ & $0.61\!\pm\!0.01$ & $2.78\!\pm\!0.11$ \\
COG & 1992--2022 & $0.27\!\pm\!0.08$ & $0.13\!\pm\!0.07$ & $0.17\!\pm\!0.06$ & $0.18\!\pm\!0.07$ & $0.12\!\pm\!0.04$ & $0.79\!\pm\!0.01$ & $0.66\!\pm\!0.01$ & $3.67\!\pm\!0.07$ \\
COL & 1992--2022 & $0.60\!\pm\!0.06$ & $0.48\!\pm\!0.05$ & $0.42\!\pm\!0.04$ & $0.41\!\pm\!0.09$ & $0.36\!\pm\!0.04$ & $0.83\!\pm\!0.01$ & $0.72\!\pm\!0.02$ & $4.02\!\pm\!0.09$ \\
COM & 1992--2022 & $0.39\!\pm\!0.08$ & $0.20\!\pm\!0.06$ & $0.25\!\pm\!0.04$ & $0.29\!\pm\!0.05$ & $0.28\!\pm\!0.04$ & $0.81\!\pm\!0.04$ & $0.69\!\pm\!0.05$ & $3.15\!\pm\!0.08$ \\
CPV & 1992--2022 & $0.77\!\pm\!0.02$ & $0.69\!\pm\!0.02$ & $0.46\!\pm\!0.02$ & $0.63\!\pm\!0.03$ & $0.61\!\pm\!0.03$ & $0.79\!\pm\!0.03$ & $0.66\!\pm\!0.04$ & $3.66\!\pm\!0.17$ \\
CRI & 1992--2022 & $0.90\!\pm\!0.01$ & $0.85\!\pm\!0.01$ & $0.66\!\pm\!0.03$ & $0.84\!\pm\!0.01$ & $0.80\!\pm\!0.01$ & $0.78\!\pm\!0.01$ & $0.66\!\pm\!0.02$ & $4.05\!\pm\!0.09$ \\
CYP & 1992--2022 & $0.81\!\pm\!0.03$ & $0.69\!\pm\!0.04$ & $0.54\!\pm\!0.03$ & $0.70\!\pm\!0.04$ & $0.69\!\pm\!0.05$ & $0.69\!\pm\!0.05$ & $0.47\!\pm\!0.02$ & $4.38\!\pm\!0.08$ \\
CZE & 1992--2022 & $0.87\!\pm\!0.03$ & $0.80\!\pm\!0.04$ & $0.59\!\pm\!0.03$ & $0.76\!\pm\!0.05$ & $0.79\!\pm\!0.03$ & $0.73\!\pm\!0.00$ & $0.38\!\pm\!0.01$ & $4.34\!\pm\!0.13$ \\
DEU & 1992--2022 & $0.89\!\pm\!0.01$ & $0.85\!\pm\!0.02$ & $0.65\!\pm\!0.01$ & $0.86\!\pm\!0.01$ & $0.83\!\pm\!0.01$ & $0.74\!\pm\!0.02$ & $0.46\!\pm\!0.03$ & $4.57\!\pm\!0.08$ \\
DJI & 1992--2022 & $0.25\!\pm\!0.01$ & $0.12\!\pm\!0.00$ & $0.13\!\pm\!0.01$ & $0.13\!\pm\!0.02$ & $0.16\!\pm\!0.01$ & $0.75\!\pm\!0.01$ & $0.60\!\pm\!0.01$ & $3.39\!\pm\!0.09$ \\
DNK & 1992--2022 & $0.92\!\pm\!0.01$ & $0.89\!\pm\!0.01$ & $0.72\!\pm\!0.01$ & $0.87\!\pm\!0.01$ & $0.88\!\pm\!0.01$ & $0.72\!\pm\!0.02$ & $0.41\!\pm\!0.02$ & $4.61\!\pm\!0.06$ \\
DOM & 1992--2022 & $0.59\!\pm\!0.08$ & $0.32\!\pm\!0.06$ & $0.38\!\pm\!0.05$ & $0.43\!\pm\!0.10$ & $0.31\!\pm\!0.04$ & $0.78\!\pm\!0.01$ & $0.67\!\pm\!0.02$ & $3.96\!\pm\!0.19$ \\
DZA & 1992--2022 & $0.30\!\pm\!0.05$ & $0.15\!\pm\!0.02$ & $0.12\!\pm\!0.02$ & $0.24\!\pm\!0.03$ & $0.25\!\pm\!0.02$ & $0.75\!\pm\!0.01$ & $0.57\!\pm\!0.02$ & $3.98\!\pm\!0.17$ \\
ECU & 1992--2022 & $0.65\!\pm\!0.06$ & $0.41\!\pm\!0.08$ & $0.45\!\pm\!0.04$ & $0.43\!\pm\!0.04$ & $0.39\!\pm\!0.02$ & $0.76\!\pm\!0.02$ & $0.64\!\pm\!0.04$ & $3.93\!\pm\!0.08$ \\
EGY & 1992--2022 & $0.20\!\pm\!0.03$ & $0.14\!\pm\!0.02$ & $0.08\!\pm\!0.01$ & $0.15\!\pm\!0.04$ & $0.10\!\pm\!0.01$ & $0.76\!\pm\!0.01$ & $0.58\!\pm\!0.02$ & $3.92\!\pm\!0.15$ \\
ESP & 1992--2022 & $0.88\!\pm\!0.02$ & $0.80\!\pm\!0.02$ & $0.63\!\pm\!0.02$ & $0.78\!\pm\!0.02$ & $0.79\!\pm\!0.03$ & $0.70\!\pm\!0.01$ & $0.46\!\pm\!0.01$ & $4.46\!\pm\!0.07$ \\
EST & 1992--2022 & $0.87\!\pm\!0.08$ & $0.81\!\pm\!0.08$ & $0.62\!\pm\!0.05$ & $0.76\!\pm\!0.07$ & $0.77\!\pm\!0.07$ & $0.83\!\pm\!0.01$ & $0.52\!\pm\!0.03$ & $4.31\!\pm\!0.12$ \\
ETH & 1992--2022 & $0.25\!\pm\!0.03$ & $0.11\!\pm\!0.02$ & $0.10\!\pm\!0.02$ & $0.17\!\pm\!0.04$ & $0.17\!\pm\!0.02$ & $0.75\!\pm\!0.03$ & $0.56\!\pm\!0.05$ & $3.04\!\pm\!0.17$ \\
FIN & 1992--2022 & $0.88\!\pm\!0.01$ & $0.84\!\pm\!0.01$ & $0.63\!\pm\!0.01$ & $0.82\!\pm\!0.01$ & $0.81\!\pm\!0.01$ & $0.74\!\pm\!0.01$ & $0.43\!\pm\!0.02$ & $4.54\!\pm\!0.07$ \\
FRA & 1992--2022 & $0.88\!\pm\!0.01$ & $0.80\!\pm\!0.01$ & $0.64\!\pm\!0.02$ & $0.81\!\pm\!0.01$ & $0.78\!\pm\!0.01$ & $0.71\!\pm\!0.02$ & $0.45\!\pm\!0.01$ & $4.54\!\pm\!0.04$ \\
GAB & 1992--2022 & $0.36\!\pm\!0.02$ & $0.21\!\pm\!0.01$ & $0.24\!\pm\!0.02$ & $0.30\!\pm\!0.03$ & $0.26\!\pm\!0.03$ & $0.75\!\pm\!0.01$ & $0.59\!\pm\!0.02$ & $4.12\!\pm\!0.10$ \\
GBR & 1992--2022 & $0.85\!\pm\!0.02$ & $0.79\!\pm\!0.02$ & $0.61\!\pm\!0.03$ & $0.74\!\pm\!0.02$ & $0.72\!\pm\!0.04$ & $0.71\!\pm\!0.03$ & $0.48\!\pm\!0.01$ & $4.53\!\pm\!0.05$ \\
GEO & 1992--2022 & $0.50\!\pm\!0.12$ & $0.33\!\pm\!0.13$ & $0.27\!\pm\!0.13$ & $0.35\!\pm\!0.13$ & $0.40\!\pm\!0.10$ & $0.75\!\pm\!0.01$ & $0.57\!\pm\!0.05$ & $3.86\!\pm\!0.19$ \\
GHA & 1992--2022 & $0.67\!\pm\!0.13$ & $0.57\!\pm\!0.12$ & $0.35\!\pm\!0.06$ & $0.58\!\pm\!0.13$ & $0.48\!\pm\!0.09$ & $0.75\!\pm\!0.00$ & $0.60\!\pm\!0.01$ & $3.43\!\pm\!0.13$ \\
GIN & 1992--2022 & $0.27\!\pm\!0.06$ & $0.11\!\pm\!0.03$ & $0.18\!\pm\!0.04$ & $0.12\!\pm\!0.03$ & $0.16\!\pm\!0.03$ & $0.75\!\pm\!0.02$ & $0.57\!\pm\!0.06$ & $3.09\!\pm\!0.10$ \\
GMB & 1992--2022 & $0.31\!\pm\!0.13$ & $0.19\!\pm\!0.13$ & $0.18\!\pm\!0.08$ & $0.15\!\pm\!0.13$ & $0.22\!\pm\!0.08$ & $0.77\!\pm\!0.02$ & $0.63\!\pm\!0.04$ & $3.20\!\pm\!0.06$ \\
GNB & 1992--2022 & $0.40\!\pm\!0.10$ & $0.24\!\pm\!0.07$ & $0.18\!\pm\!0.04$ & $0.21\!\pm\!0.06$ & $0.23\!\pm\!0.05$ & $0.75\!\pm\!0.01$ & $0.60\!\pm\!0.04$ & $3.13\!\pm\!0.04$ \\
GRC & 1992--2022 & $0.88\!\pm\!0.04$ & $0.78\!\pm\!0.05$ & $0.60\!\pm\!0.03$ & $0.76\!\pm\!0.04$ & $0.78\!\pm\!0.05$ & $0.75\!\pm\!0.07$ & $0.48\!\pm\!0.03$ & $4.35\!\pm\!0.07$ \\
HKG & 1992--2022 & $0.35\!\pm\!0.05$ & $0.30\!\pm\!0.05$ & $0.13\!\pm\!0.02$ & $0.30\!\pm\!0.09$ & $0.31\!\pm\!0.03$ & $0.77\!\pm\!0.01$ & $0.58\!\pm\!0.03$ & $4.60\!\pm\!0.08$ \\
HRV & 1992--2022 & $0.68\!\pm\!0.18$ & $0.55\!\pm\!0.19$ & $0.45\!\pm\!0.14$ & $0.48\!\pm\!0.20$ & $0.56\!\pm\!0.15$ & $0.73\!\pm\!0.01$ & $0.46\!\pm\!0.01$ & $4.22\!\pm\!0.14$ \\
HUN & 1992--2022 & $0.75\!\pm\!0.15$ & $0.64\!\pm\!0.16$ & $0.54\!\pm\!0.12$ & $0.51\!\pm\!0.16$ & $0.60\!\pm\!0.15$ & $0.76\!\pm\!0.02$ & $0.40\!\pm\!0.03$ & $4.24\!\pm\!0.16$ \\
IDN & 1992--2022 & $0.55\!\pm\!0.20$ & $0.40\!\pm\!0.19$ & $0.34\!\pm\!0.16$ & $0.48\!\pm\!0.20$ & $0.35\!\pm\!0.13$ & $0.74\!\pm\!0.01$ & $0.56\!\pm\!0.03$ & $3.88\!\pm\!0.14$ \\
IND & 1992--2022 & $0.65\!\pm\!0.12$ & $0.51\!\pm\!0.10$ & $0.42\!\pm\!0.09$ & $0.56\!\pm\!0.15$ & $0.41\!\pm\!0.10$ & $0.71\!\pm\!0.03$ & $0.57\!\pm\!0.06$ & $3.60\!\pm\!0.18$ \\
IRL & 1992--2022 & $0.89\!\pm\!0.01$ & $0.81\!\pm\!0.01$ & $0.64\!\pm\!0.01$ & $0.74\!\pm\!0.04$ & $0.78\!\pm\!0.01$ & $0.88\!\pm\!0.01$ & $0.46\!\pm\!0.02$ & $4.63\!\pm\!0.14$ \\
IRN & 1992--2022 & $0.18\!\pm\!0.03$ & $0.12\!\pm\!0.02$ & $0.05\!\pm\!0.01$ & $0.19\!\pm\!0.05$ & $0.18\!\pm\!0.02$ & $0.77\!\pm\!0.01$ & $0.60\!\pm\!0.03$ & $4.10\!\pm\!0.16$ \\
IRQ & 1992--2022 & $0.28\!\pm\!0.14$ & $0.18\!\pm\!0.09$ & $0.17\!\pm\!0.10$ & $0.18\!\pm\!0.14$ & $0.17\!\pm\!0.06$ & $0.78\!\pm\!0.02$ & $0.64\!\pm\!0.04$ & $3.80\!\pm\!0.31$ \\
ISL & 1992--2022 & $0.87\!\pm\!0.01$ & $0.78\!\pm\!0.01$ & $0.63\!\pm\!0.01$ & $0.74\!\pm\!0.02$ & $0.79\!\pm\!0.02$ & $0.80\!\pm\!0.06$ & $0.41\!\pm\!0.02$ & $4.55\!\pm\!0.07$ \\
ISR & 1992--2022 & $0.75\!\pm\!0.02$ & $0.68\!\pm\!0.02$ & $0.49\!\pm\!0.02$ & $0.61\!\pm\!0.02$ & $0.62\!\pm\!0.02$ & $0.77\!\pm\!0.01$ & $0.61\!\pm\!0.01$ & $4.44\!\pm\!0.06$ \\
ITA & 1992--2022 & $0.85\!\pm\!0.01$ & $0.77\!\pm\!0.02$ & $0.67\!\pm\!0.02$ & $0.79\!\pm\!0.02$ & $0.78\!\pm\!0.01$ & $0.71\!\pm\!0.06$ & $0.43\!\pm\!0.01$ & $4.52\!\pm\!0.04$ \\
JOR & 1992--2022 & $0.24\!\pm\!0.02$ & $0.20\!\pm\!0.03$ & $0.09\!\pm\!0.02$ & $0.20\!\pm\!0.01$ & $0.20\!\pm\!0.02$ & $0.76\!\pm\!0.01$ & $0.60\!\pm\!0.02$ & $3.95\!\pm\!0.08$ \\
JPN & 1992--2022 & $0.84\!\pm\!0.01$ & $0.76\!\pm\!0.01$ & $0.55\!\pm\!0.01$ & $0.75\!\pm\!0.01$ & $0.76\!\pm\!0.01$ & $0.74\!\pm\!0.00$ & $0.51\!\pm\!0.02$ & $4.54\!\pm\!0.03$ \\
KAZ & 1992--2022 & $0.26\!\pm\!0.02$ & $0.13\!\pm\!0.02$ & $0.09\!\pm\!0.01$ & $0.11\!\pm\!0.03$ & $0.20\!\pm\!0.01$ & $0.74\!\pm\!0.00$ & $0.53\!\pm\!0.03$ & $4.19\!\pm\!0.20$ \\
KEN & 1992--2022 & $0.44\!\pm\!0.08$ & $0.32\!\pm\!0.08$ & $0.28\!\pm\!0.07$ & $0.39\!\pm\!0.05$ & $0.28\!\pm\!0.05$ & $0.77\!\pm\!0.01$ & $0.63\!\pm\!0.04$ & $3.37\!\pm\!0.10$ \\
KGZ & 1992--2022 & $0.36\!\pm\!0.07$ & $0.22\!\pm\!0.06$ & $0.19\!\pm\!0.07$ & $0.25\!\pm\!0.08$ & $0.27\!\pm\!0.04$ & $0.74\!\pm\!0.00$ & $0.50\!\pm\!0.04$ & $3.59\!\pm\!0.09$ \\
KHM & 1992--2022 & $0.29\!\pm\!0.06$ & $0.11\!\pm\!0.03$ & $0.12\!\pm\!0.02$ & $0.17\!\pm\!0.06$ & $0.11\!\pm\!0.02$ & $0.78\!\pm\!0.03$ & $0.64\!\pm\!0.05$ & $3.36\!\pm\!0.16$ \\
KOR & 1992--2022 & $0.79\!\pm\!0.06$ & $0.70\!\pm\!0.08$ & $0.53\!\pm\!0.07$ & $0.70\!\pm\!0.09$ & $0.68\!\pm\!0.06$ & $0.80\!\pm\!0.01$ & $0.44\!\pm\!0.06$ & $4.44\!\pm\!0.12$ \\
KWT & 1992--2022 & $0.30\!\pm\!0.03$ & $0.28\!\pm\!0.01$ & $0.10\!\pm\!0.01$ & $0.29\!\pm\!0.02$ & $0.24\!\pm\!0.01$ & $0.77\!\pm\!0.02$ & $0.62\!\pm\!0.05$ & $4.71\!\pm\!0.21$ \\
LAO & 1992--2022 & $0.10\!\pm\!0.02$ & $0.08\!\pm\!0.01$ & $0.10\!\pm\!0.01$ & $0.04\!\pm\!0.01$ & $0.11\!\pm\!0.01$ & $0.75\!\pm\!0.00$ & $0.59\!\pm\!0.01$ & $3.53\!\pm\!0.22$ \\
LBN & 1992--2022 & $0.40\!\pm\!0.07$ & $0.25\!\pm\!0.04$ & $0.19\!\pm\!0.06$ & $0.30\!\pm\!0.06$ & $0.26\!\pm\!0.03$ & $0.78\!\pm\!0.02$ & $0.64\!\pm\!0.02$ & $3.99\!\pm\!0.16$ \\
LBR & 1992--2022 & $0.46\!\pm\!0.19$ & $0.33\!\pm\!0.15$ & $0.26\!\pm\!0.10$ & $0.40\!\pm\!0.17$ & $0.29\!\pm\!0.11$ & $0.74\!\pm\!0.00$ & $0.55\!\pm\!0.01$ & $2.89\!\pm\!0.12$ \\
LKA & 1992--2022 & $0.49\!\pm\!0.08$ & $0.33\!\pm\!0.07$ & $0.29\!\pm\!0.06$ & $0.29\!\pm\!0.08$ & $0.33\!\pm\!0.04$ & $0.75\!\pm\!0.01$ & $0.59\!\pm\!0.02$ & $3.87\!\pm\!0.15$ \\
LSO & 1992--2022 & $0.51\!\pm\!0.14$ & $0.39\!\pm\!0.10$ & $0.29\!\pm\!0.10$ & $0.36\!\pm\!0.11$ & $0.43\!\pm\!0.09$ & $0.79\!\pm\!0.03$ & $0.67\!\pm\!0.05$ & $3.34\!\pm\!0.05$ \\
LTU & 1992--2022 & $0.82\!\pm\!0.01$ & $0.76\!\pm\!0.01$ & $0.61\!\pm\!0.01$ & $0.70\!\pm\!0.03$ & $0.73\!\pm\!0.02$ & $0.73\!\pm\!0.01$ & $0.48\!\pm\!0.02$ & $4.22\!\pm\!0.19$ \\
LUX & 1992--2022 & $0.87\!\pm\!0.00$ & $0.79\!\pm\!0.00$ & $0.62\!\pm\!0.02$ & $0.82\!\pm\!0.01$ & $0.82\!\pm\!0.00$ & $0.76\!\pm\!0.01$ & $0.48\!\pm\!0.02$ & $4.70\!\pm\!0.05$ \\
LVA & 1992--2022 & $0.81\!\pm\!0.03$ & $0.71\!\pm\!0.04$ & $0.58\!\pm\!0.06$ & $0.65\!\pm\!0.05$ & $0.68\!\pm\!0.04$ & $0.78\!\pm\!0.02$ & $0.49\!\pm\!0.03$ & $4.20\!\pm\!0.17$ \\
MAR & 1992--2022 & $0.24\!\pm\!0.03$ & $0.22\!\pm\!0.02$ & $0.14\!\pm\!0.03$ & $0.21\!\pm\!0.05$ & $0.17\!\pm\!0.03$ & $0.75\!\pm\!0.00$ & $0.58\!\pm\!0.00$ & $3.79\!\pm\!0.11$ \\
MDA & 1992--2022 & $0.56\!\pm\!0.09$ & $0.44\!\pm\!0.09$ & $0.34\!\pm\!0.08$ & $0.37\!\pm\!0.12$ & $0.41\!\pm\!0.07$ & $0.73\!\pm\!0.00$ & $0.50\!\pm\!0.01$ & $3.66\!\pm\!0.12$ \\
MDG & 1992--2022 & $0.45\!\pm\!0.10$ & $0.24\!\pm\!0.07$ & $0.28\!\pm\!0.06$ & $0.24\!\pm\!0.06$ & $0.21\!\pm\!0.05$ & $0.76\!\pm\!0.01$ & $0.61\!\pm\!0.02$ & $3.11\!\pm\!0.04$ \\
MEX & 1992--2022 & $0.62\!\pm\!0.08$ & $0.41\!\pm\!0.09$ & $0.38\!\pm\!0.06$ & $0.46\!\pm\!0.09$ & $0.32\!\pm\!0.05$ & $0.82\!\pm\!0.02$ & $0.72\!\pm\!0.02$ & $4.14\!\pm\!0.07$ \\
MKD & 1992--2022 & $0.50\!\pm\!0.09$ & $0.37\!\pm\!0.06$ & $0.33\!\pm\!0.07$ & $0.31\!\pm\!0.09$ & $0.37\!\pm\!0.06$ & $0.73\!\pm\!0.00$ & $0.46\!\pm\!0.02$ & $3.99\!\pm\!0.10$ \\
MLI & 1992--2022 & $0.53\!\pm\!0.11$ & $0.36\!\pm\!0.08$ & $0.35\!\pm\!0.07$ & $0.44\!\pm\!0.07$ & $0.36\!\pm\!0.06$ & $0.75\!\pm\!0.03$ & $0.59\!\pm\!0.05$ & $3.14\!\pm\!0.06$ \\
MLT & 1992--2022 & $0.79\!\pm\!0.02$ & $0.61\!\pm\!0.02$ & $0.54\!\pm\!0.02$ & $0.65\!\pm\!0.02$ & $0.68\!\pm\!0.02$ & $0.60\!\pm\!0.03$ & $0.44\!\pm\!0.02$ & $4.36\!\pm\!0.11$ \\
MMR & 1992--2022 & $0.18\!\pm\!0.13$ & $0.08\!\pm\!0.09$ & $0.11\!\pm\!0.08$ & $0.12\!\pm\!0.13$ & $0.11\!\pm\!0.06$ & $0.76\!\pm\!0.01$ & $0.60\!\pm\!0.03$ & $3.47\!\pm\!0.23$ \\
MNE & 1998--2022 & $0.47\!\pm\!0.04$ & $0.37\!\pm\!0.04$ & $0.30\!\pm\!0.03$ & $0.38\!\pm\!0.03$ & $0.37\!\pm\!0.03$ & $0.73\!\pm\!0.00$ & $0.49\!\pm\!0.03$ & $4.09\!\pm\!0.20$ \\
MNG & 1992--2022 & $0.67\!\pm\!0.03$ & $0.55\!\pm\!0.03$ & $0.38\!\pm\!0.03$ & $0.56\!\pm\!0.04$ & $0.54\!\pm\!0.05$ & $0.74\!\pm\!0.00$ & $0.56\!\pm\!0.01$ & $3.75\!\pm\!0.32$ \\
MOZ & 1992--2022 & $0.41\!\pm\!0.07$ & $0.27\!\pm\!0.04$ & $0.24\!\pm\!0.06$ & $0.29\!\pm\!0.05$ & $0.27\!\pm\!0.04$ & $0.82\!\pm\!0.01$ & $0.69\!\pm\!0.02$ & $3.06\!\pm\!0.05$ \\
MRT & 1992--2022 & $0.36\!\pm\!0.07$ & $0.16\!\pm\!0.04$ & $0.16\!\pm\!0.05$ & $0.22\!\pm\!0.06$ & $0.17\!\pm\!0.03$ & $0.74\!\pm\!0.01$ & $0.56\!\pm\!0.05$ & $3.41\!\pm\!0.13$ \\
MUS & 1992--2022 & $0.71\!\pm\!0.06$ & $0.61\!\pm\!0.07$ & $0.45\!\pm\!0.04$ & $0.64\!\pm\!0.07$ & $0.57\!\pm\!0.05$ & $0.75\!\pm\!0.00$ & $0.57\!\pm\!0.01$ & $4.20\!\pm\!0.08$ \\
MWI & 1992--2022 & $0.49\!\pm\!0.09$ & $0.39\!\pm\!0.09$ & $0.27\!\pm\!0.08$ & $0.38\!\pm\!0.09$ & $0.30\!\pm\!0.05$ & $0.80\!\pm\!0.03$ & $0.69\!\pm\!0.09$ & $3.01\!\pm\!0.05$ \\
MYS & 1992--2022 & $0.32\!\pm\!0.06$ & $0.21\!\pm\!0.05$ & $0.22\!\pm\!0.04$ & $0.23\!\pm\!0.05$ & $0.24\!\pm\!0.04$ & $0.76\!\pm\!0.01$ & $0.55\!\pm\!0.04$ & $4.23\!\pm\!0.13$ \\
NAM & 1992--2022 & $0.65\!\pm\!0.03$ & $0.54\!\pm\!0.02$ & $0.41\!\pm\!0.02$ & $0.52\!\pm\!0.03$ & $0.42\!\pm\!0.04$ & $0.84\!\pm\!0.00$ & $0.77\!\pm\!0.03$ & $3.85\!\pm\!0.08$ \\
NER & 1992--2022 & $0.51\!\pm\!0.10$ & $0.38\!\pm\!0.09$ & $0.31\!\pm\!0.09$ & $0.43\!\pm\!0.10$ & $0.36\!\pm\!0.06$ & $0.76\!\pm\!0.01$ & $0.59\!\pm\!0.04$ & $2.91\!\pm\!0.05$ \\
NGA & 1992--2022 & $0.42\!\pm\!0.13$ & $0.28\!\pm\!0.10$ & $0.29\!\pm\!0.07$ & $0.31\!\pm\!0.12$ & $0.24\!\pm\!0.06$ & $0.74\!\pm\!0.00$ & $0.57\!\pm\!0.02$ & $3.53\!\pm\!0.19$ \\
NLD & 1992--2022 & $0.87\!\pm\!0.01$ & $0.82\!\pm\!0.01$ & $0.60\!\pm\!0.02$ & $0.80\!\pm\!0.01$ & $0.79\!\pm\!0.01$ & $0.68\!\pm\!0.04$ & $0.40\!\pm\!0.01$ & $4.60\!\pm\!0.07$ \\
NOR & 1992--2022 & $0.89\!\pm\!0.01$ & $0.85\!\pm\!0.01$ & $0.65\!\pm\!0.01$ & $0.86\!\pm\!0.01$ & $0.85\!\pm\!0.01$ & $0.73\!\pm\!0.01$ & $0.40\!\pm\!0.03$ & $4.81\!\pm\!0.13$ \\
NPL & 1992--2022 & $0.45\!\pm\!0.15$ & $0.36\!\pm\!0.13$ & $0.26\!\pm\!0.10$ & $0.34\!\pm\!0.09$ & $0.32\!\pm\!0.08$ & $0.74\!\pm\!0.00$ & $0.54\!\pm\!0.01$ & $3.30\!\pm\!0.11$ \\
NZL & 1992--2022 & $0.88\!\pm\!0.01$ & $0.83\!\pm\!0.01$ & $0.69\!\pm\!0.02$ & $0.75\!\pm\!0.03$ & $0.77\!\pm\!0.02$ & $0.73\!\pm\!0.00$ & $0.48\!\pm\!0.05$ & $4.48\!\pm\!0.07$ \\
OMN & 1992--2022 & $0.12\!\pm\!0.07$ & $0.12\!\pm\!0.02$ & $0.09\!\pm\!0.03$ & $0.06\!\pm\!0.01$ & $0.16\!\pm\!0.02$ & $0.81\!\pm\!0.00$ & $0.68\!\pm\!0.00$ & $4.50\!\pm\!0.20$ \\
PAK & 1992--2022 & $0.38\!\pm\!0.07$ & $0.22\!\pm\!0.05$ & $0.24\!\pm\!0.05$ & $0.26\!\pm\!0.05$ & $0.14\!\pm\!0.02$ & $0.74\!\pm\!0.00$ & $0.54\!\pm\!0.01$ & $3.60\!\pm\!0.10$ \\
PER & 1992--2022 & $0.66\!\pm\!0.22$ & $0.51\!\pm\!0.25$ & $0.45\!\pm\!0.17$ & $0.51\!\pm\!0.22$ & $0.40\!\pm\!0.15$ & $0.81\!\pm\!0.02$ & $0.69\!\pm\!0.03$ & $3.91\!\pm\!0.15$ \\
PHL & 1992--2022 & $0.52\!\pm\!0.05$ & $0.38\!\pm\!0.05$ & $0.36\!\pm\!0.04$ & $0.45\!\pm\!0.05$ & $0.27\!\pm\!0.04$ & $0.76\!\pm\!0.01$ & $0.61\!\pm\!0.02$ & $3.72\!\pm\!0.13$ \\
POL & 1992--2022 & $0.83\!\pm\!0.10$ & $0.74\!\pm\!0.14$ & $0.59\!\pm\!0.09$ & $0.67\!\pm\!0.14$ & $0.73\!\pm\!0.09$ & $0.85\!\pm\!0.00$ & $0.46\!\pm\!0.03$ & $4.23\!\pm\!0.18$ \\
PRT & 1992--2022 & $0.88\!\pm\!0.01$ & $0.82\!\pm\!0.02$ & $0.62\!\pm\!0.03$ & $0.80\!\pm\!0.03$ & $0.79\!\pm\!0.02$ & $0.75\!\pm\!0.01$ & $0.49\!\pm\!0.01$ & $4.38\!\pm\!0.05$ \\
PSE & 1992--2022 & $0.32\!\pm\!0.14$ & $0.20\!\pm\!0.07$ & $0.19\!\pm\!0.06$ & $0.23\!\pm\!0.09$ & $0.26\!\pm\!0.08$ & $0.78\!\pm\!0.01$ & $0.64\!\pm\!0.01$ & $3.69\!\pm\!0.05$ \\
QAT & 1992--2022 & $0.04\!\pm\!0.02$ & $0.09\!\pm\!0.01$ & $0.02\!\pm\!0.01$ & $0.07\!\pm\!0.02$ & $0.11\!\pm\!0.00$ & $0.80\!\pm\!0.01$ & $0.67\!\pm\!0.00$ & $4.87\!\pm\!0.35$ \\
ROU & 1992--2022 & $0.64\!\pm\!0.06$ & $0.49\!\pm\!0.08$ & $0.43\!\pm\!0.07$ & $0.41\!\pm\!0.07$ & $0.50\!\pm\!0.04$ & $0.73\!\pm\!0.00$ & $0.51\!\pm\!0.05$ & $4.09\!\pm\!0.24$ \\
RUS & 1992--2022 & $0.34\!\pm\!0.09$ & $0.17\!\pm\!0.08$ & $0.21\!\pm\!0.06$ & $0.23\!\pm\!0.10$ & $0.25\!\pm\!0.05$ & $0.76\!\pm\!0.05$ & $0.57\!\pm\!0.04$ & $4.21\!\pm\!0.19$ \\
RWA & 1992--2022 & $0.18\!\pm\!0.05$ & $0.12\!\pm\!0.02$ & $0.16\!\pm\!0.04$ & $0.18\!\pm\!0.05$ & $0.19\!\pm\!0.03$ & $0.81\!\pm\!0.02$ & $0.67\!\pm\!0.03$ & $3.13\!\pm\!0.12$ \\
SAU & 1992--2022 & $0.02\!\pm\!0.00$ & $0.04\!\pm\!0.00$ & $0.02\!\pm\!0.00$ & $0.05\!\pm\!0.01$ & $0.11\!\pm\!0.00$ & $0.80\!\pm\!0.00$ & $0.64\!\pm\!0.00$ & $4.53\!\pm\!0.17$ \\
SDN & 1992--2022 & $0.18\!\pm\!0.05$ & $0.05\!\pm\!0.02$ & $0.10\!\pm\!0.03$ & $0.06\!\pm\!0.03$ & $0.06\!\pm\!0.01$ & $0.74\!\pm\!0.00$ & $0.55\!\pm\!0.00$ & $3.38\!\pm\!0.14$ \\
SEN & 1992--2022 & $0.68\!\pm\!0.04$ & $0.52\!\pm\!0.05$ & $0.41\!\pm\!0.04$ & $0.58\!\pm\!0.04$ & $0.49\!\pm\!0.04$ & $0.75\!\pm\!0.01$ & $0.60\!\pm\!0.02$ & $3.33\!\pm\!0.05$ \\
SGP & 1992--2022 & $0.39\!\pm\!0.02$ & $0.31\!\pm\!0.01$ & $0.10\!\pm\!0.01$ & $0.33\!\pm\!0.02$ & $0.35\!\pm\!0.02$ & $0.75\!\pm\!0.01$ & $0.52\!\pm\!0.04$ & $4.69\!\pm\!0.15$ \\
SLE & 1992--2022 & $0.42\!\pm\!0.15$ & $0.26\!\pm\!0.14$ & $0.28\!\pm\!0.10$ & $0.35\!\pm\!0.11$ & $0.25\!\pm\!0.10$ & $0.75\!\pm\!0.01$ & $0.58\!\pm\!0.02$ & $3.13\!\pm\!0.11$ \\
SLV & 1992--2022 & $0.54\!\pm\!0.09$ & $0.34\!\pm\!0.10$ & $0.33\!\pm\!0.06$ & $0.35\!\pm\!0.09$ & $0.25\!\pm\!0.05$ & $0.76\!\pm\!0.01$ & $0.65\!\pm\!0.02$ & $3.81\!\pm\!0.11$ \\
SRB & 1992--2022 & $0.46\!\pm\!0.15$ & $0.35\!\pm\!0.14$ & $0.30\!\pm\!0.11$ & $0.34\!\pm\!0.17$ & $0.36\!\pm\!0.10$ & $0.73\!\pm\!0.00$ & $0.51\!\pm\!0.04$ & $3.98\!\pm\!0.18$ \\
STP & 1992--2022 & $0.66\!\pm\!0.02$ & $0.54\!\pm\!0.03$ & $0.42\!\pm\!0.01$ & $0.48\!\pm\!0.02$ & $0.47\!\pm\!0.02$ & $0.74\!\pm\!0.01$ & $0.54\!\pm\!0.04$ & $3.45\!\pm\!0.10$ \\
SVK & 1993--2022 & $0.82\!\pm\!0.05$ & $0.71\!\pm\!0.07$ & $0.61\!\pm\!0.06$ & $0.62\!\pm\!0.11$ & $0.65\!\pm\!0.06$ & $0.61\!\pm\!0.04$ & $0.39\!\pm\!0.02$ & $4.27\!\pm\!0.13$ \\
SVN & 1992--2022 & $0.84\!\pm\!0.03$ & $0.76\!\pm\!0.04$ & $0.67\!\pm\!0.06$ & $0.74\!\pm\!0.06$ & $0.75\!\pm\!0.04$ & $0.68\!\pm\!0.03$ & $0.41\!\pm\!0.02$ & $4.38\!\pm\!0.09$ \\
SWE & 1992--2022 & $0.91\!\pm\!0.01$ & $0.88\!\pm\!0.01$ & $0.67\!\pm\!0.01$ & $0.87\!\pm\!0.02$ & $0.86\!\pm\!0.02$ & $0.88\!\pm\!0.10$ & $0.40\!\pm\!0.01$ & $4.58\!\pm\!0.08$ \\
SWZ & 1992--2022 & $0.13\!\pm\!0.01$ & $0.10\!\pm\!0.01$ & $0.13\!\pm\!0.01$ & $0.08\!\pm\!0.01$ & $0.08\!\pm\!0.01$ & $0.83\!\pm\!0.01$ & $0.71\!\pm\!0.02$ & $3.79\!\pm\!0.11$ \\
SYC & 1992--2022 & $0.47\!\pm\!0.12$ & $0.37\!\pm\!0.12$ & $0.19\!\pm\!0.06$ & $0.36\!\pm\!0.12$ & $0.39\!\pm\!0.08$ & $0.77\!\pm\!0.00$ & $0.61\!\pm\!0.05$ & $4.25\!\pm\!0.17$ \\
SYR & 1992--2022 & $0.14\!\pm\!0.01$ & $0.04\!\pm\!0.01$ & $0.05\!\pm\!0.00$ & $0.02\!\pm\!0.00$ & $0.10\!\pm\!0.02$ & $0.80\!\pm\!0.01$ & $0.65\!\pm\!0.01$ & $3.76\!\pm\!0.20$ \\
TCD & 1992--2022 & $0.26\!\pm\!0.04$ & $0.09\!\pm\!0.01$ & $0.13\!\pm\!0.01$ & $0.11\!\pm\!0.01$ & $0.10\!\pm\!0.01$ & $0.75\!\pm\!0.00$ & $0.59\!\pm\!0.01$ & $3.06\!\pm\!0.16$ \\
TGO & 1992--2022 & $0.37\!\pm\!0.07$ & $0.20\!\pm\!0.04$ & $0.17\!\pm\!0.04$ & $0.29\!\pm\!0.07$ & $0.29\!\pm\!0.04$ & $0.76\!\pm\!0.00$ & $0.61\!\pm\!0.01$ & $3.09\!\pm\!0.05$ \\
THA & 1992--2022 & $0.36\!\pm\!0.15$ & $0.28\!\pm\!0.11$ & $0.17\!\pm\!0.06$ & $0.30\!\pm\!0.18$ & $0.24\!\pm\!0.07$ & $0.79\!\pm\!0.01$ & $0.65\!\pm\!0.02$ & $4.09\!\pm\!0.09$ \\
TJK & 1992--2022 & $0.21\!\pm\!0.03$ & $0.06\!\pm\!0.02$ & $0.08\!\pm\!0.02$ & $0.08\!\pm\!0.02$ & $0.12\!\pm\!0.03$ & $0.74\!\pm\!0.00$ & $0.51\!\pm\!0.05$ & $3.44\!\pm\!0.17$ \\
TKM & 1992--2022 & $0.13\!\pm\!0.03$ & $0.02\!\pm\!0.01$ & $0.05\!\pm\!0.02$ & $0.02\!\pm\!0.01$ & $0.10\!\pm\!0.02$ & $0.75\!\pm\!0.00$ & $0.58\!\pm\!0.06$ & $3.95\!\pm\!0.30$ \\
TUN & 1992--2022 & $0.37\!\pm\!0.25$ & $0.27\!\pm\!0.24$ & $0.23\!\pm\!0.14$ & $0.31\!\pm\!0.24$ & $0.33\!\pm\!0.18$ & $0.79\!\pm\!0.02$ & $0.64\!\pm\!0.02$ & $3.95\!\pm\!0.10$ \\
TUR & 1992--2022 & $0.50\!\pm\!0.13$ & $0.34\!\pm\!0.14$ & $0.32\!\pm\!0.08$ & $0.33\!\pm\!0.15$ & $0.35\!\pm\!0.08$ & $0.78\!\pm\!0.02$ & $0.62\!\pm\!0.02$ & $4.19\!\pm\!0.15$ \\
TWN & 1992--2022 & $0.72\!\pm\!0.17$ & $0.61\!\pm\!0.15$ & $0.53\!\pm\!0.15$ & $0.63\!\pm\!0.14$ & $0.63\!\pm\!0.14$ & $0.75\!\pm\!0.01$ & $0.59\!\pm\!0.04$ & $4.51\!\pm\!0.14$ \\
TZA & 1992--2022 & $0.45\!\pm\!0.04$ & $0.38\!\pm\!0.03$ & $0.28\!\pm\!0.04$ & $0.35\!\pm\!0.03$ & $0.34\!\pm\!0.01$ & $0.75\!\pm\!0.01$ & $0.59\!\pm\!0.02$ & $3.23\!\pm\!0.19$ \\
UGA & 1992--2022 & $0.30\!\pm\!0.02$ & $0.24\!\pm\!0.02$ & $0.21\!\pm\!0.03$ & $0.27\!\pm\!0.01$ & $0.20\!\pm\!0.02$ & $0.77\!\pm\!0.01$ & $0.62\!\pm\!0.01$ & $3.18\!\pm\!0.12$ \\
UKR & 1992--2022 & $0.48\!\pm\!0.07$ & $0.29\!\pm\!0.06$ & $0.30\!\pm\!0.04$ & $0.35\!\pm\!0.07$ & $0.35\!\pm\!0.06$ & $0.73\!\pm\!0.00$ & $0.44\!\pm\!0.04$ & $3.89\!\pm\!0.13$ \\
URY & 1992--2022 & $0.88\!\pm\!0.02$ & $0.81\!\pm\!0.03$ & $0.73\!\pm\!0.03$ & $0.82\!\pm\!0.04$ & $0.77\!\pm\!0.04$ & $0.84\!\pm\!0.01$ & $0.58\!\pm\!0.02$ & $4.18\!\pm\!0.09$ \\
USA & 1992--2022 & $0.86\!\pm\!0.03$ & $0.80\!\pm\!0.04$ & $0.62\!\pm\!0.03$ & $0.79\!\pm\!0.08$ & $0.69\!\pm\!0.04$ & $0.84\!\pm\!0.02$ & $0.56\!\pm\!0.02$ & $4.68\!\pm\!0.05$ \\
UZB & 1992--2022 & $0.17\!\pm\!0.02$ & $0.04\!\pm\!0.02$ & $0.05\!\pm\!0.01$ & $0.07\!\pm\!0.05$ & $0.13\!\pm\!0.01$ & $0.74\!\pm\!0.00$ & $0.55\!\pm\!0.04$ & $3.82\!\pm\!0.15$ \\
VNM & 1992--2022 & $0.15\!\pm\!0.02$ & $0.10\!\pm\!0.01$ & $0.17\!\pm\!0.00$ & $0.18\!\pm\!0.01$ & $0.20\!\pm\!0.01$ & $0.74\!\pm\!0.00$ & $0.58\!\pm\!0.01$ & $3.59\!\pm\!0.19$ \\
YEM & 1992--2022 & $0.25\!\pm\!0.08$ & $0.12\!\pm\!0.05$ & $0.11\!\pm\!0.04$ & $0.15\!\pm\!0.08$ & $0.08\!\pm\!0.04$ & $0.82\!\pm\!0.01$ & $0.69\!\pm\!0.01$ & $3.55\!\pm\!0.12$ \\
ZAF & 1992--2022 & $0.69\!\pm\!0.14$ & $0.58\!\pm\!0.13$ & $0.47\!\pm\!0.09$ & $0.61\!\pm\!0.11$ & $0.46\!\pm\!0.12$ & $0.97\!\pm\!0.04$ & $0.69\!\pm\!0.06$ & $3.97\!\pm\!0.11$ \\
ZMB & 1992--2022 & $0.47\!\pm\!0.05$ & $0.38\!\pm\!0.04$ & $0.31\!\pm\!0.03$ & $0.39\!\pm\!0.08$ & $0.31\!\pm\!0.03$ & $0.81\!\pm\!0.02$ & $0.68\!\pm\!0.02$ & $3.34\!\pm\!0.18$ \\
ZWE & 1992--2022 & $0.28\!\pm\!0.02$ & $0.19\!\pm\!0.03$ & $0.22\!\pm\!0.01$ & $0.23\!\pm\!0.02$ & $0.18\!\pm\!0.01$ & $0.77\!\pm\!0.00$ & $0.63\!\pm\!0.02$ & $3.25\!\pm\!0.08$ \\
\end{longtable}
\endgroup

\clearpage

\clearpage
\begin{table*}[!t]
\centering
\caption{Seshat variables used to construct the social-scale and computation coordinates. Definitions of the individual variables are provided in Ref.~\cite{turchin2015seshat}.}
\label{tab:seshat_variables}
\small
\setlength{\tabcolsep}{4pt}

\begin{tabular}{cccc}
\hline
\textbf{State component} &
\textbf{Domain} &
\textbf{No.} &
\parbox[c]{0.60\textwidth}{\centering\textbf{Constituent variables}} \\
\hline
Social scale
& ---
& 3
& \parbox[t]{0.60\textwidth}{
polity population; polity territory; population of the largest settlement.
} \\[3ex]

\hline

Computation
& Government
& 11
& \parbox[t]{0.60\textwidth}{
professional military officers; professional soldiers; professional priesthoods;
full-time bureaucrats; professional lawyers; examination systems; courts; judges;
formal legal codes; merit promotions; specialized government buildings.
}  \\[10ex]

Computation
& Infrastructure
& 12
& \parbox[t]{0.60\textwidth}{
irrigation systems; drinking-water supplies; markets; food-storage sites; roads;
bridges; canals; ports; mines or quarries; couriers; postal stations;
general postal services.
}  \\[10ex]

Computation
& Information
& 15
& \parbox[t]{0.60\textwidth}{
mnemonic devices; scripts; lists, tables, and classifications;
phonetic/alphabetic writings; non-phonetic writings; written records;
nonwritten records; calendars; scientific literatures; sacred texts; histories;
religious literatures; fictions; practical literatures; philosophies.
} \\[10ex]

Computation
& Monetary systems
& 6
& \parbox[t]{0.60\textwidth}{
articles; tokens; precious metals; foreign coins; indigenous coins;
\\ paper currencies.
} \\[3ex]
\hline
\end{tabular}
\end{table*}

\begin{table}[htbp]
    \centering
    \caption{NGA-level summary of the retained Seshat dataset. Rows correspond to Natural Geographic Areas (NGAs). The reported years denote the first and last retained NGA-year observations and do not imply continuous coverage between them. 
    Social Scale and Computation entries give the state-coordinate values at the first and last retained observations, respectively. $n_{\mathrm{points}}$ denotes the total number of retained NGA-year states in each trajectory before any transition-specific filtering used in downstream diagnostics.
    }
    \label{tab:aggregated_seshat}
    \small
    \setlength{\tabcolsep}{14pt}
    \begin{tabular}{lcccr}
        \toprule
        NGA name & Years & Social scale & Computation & $n$ points \\
        \midrule
        Upper Egypt & 3800 BCE--1800 CE & -4.786 $\rightarrow$ 1.477 & -2.372 $\rightarrow$ 0.895 & 51 \\
        Galilee & 2000 BCE--100 BCE & -2.214 $\rightarrow$ 0.720 & -0.879 $\rightarrow$ 0.367 & 17 \\
        Konya Plain & 1600 BCE--1800 CE & -1.002 $\rightarrow$ 1.461 & 0.035 $\rightarrow$ 0.992 & 27 \\
        Crete & 1400 BCE--1800 CE & -2.113 $\rightarrow$ 1.477 & -1.242 $\rightarrow$ 0.895 & 22 \\
        Middle Yellow River Valley & 1000 BCE--1800 CE & 0.502 $\rightarrow$ 2.264 & 0.148 $\rightarrow$ 0.895 & 20 \\
        Latium & 700 BCE--1800 CE & -2.536 $\rightarrow$ -0.221 & -1.333 $\rightarrow$ 0.701 & 25 \\
        Valley of Oaxaca & 700 BCE--1700 CE & -4.052 $\rightarrow$ 1.326 & -2.689 $\rightarrow$ 0.798 & 15 \\
        Susiana & 500 BCE--1700 CE & 1.189 $\rightarrow$ 0.882 & 0.992 $\rightarrow$ 0.507 & 20 \\
        Kachi Plain & 500 BCE--1700 CE & 1.189 $\rightarrow$ 1.674 & 0.992 $\rightarrow$ 0.595 & 17 \\
        Sogdiana & 500 BCE--1500 CE & 1.189 $\rightarrow$ 1.270 & 0.992 $\rightarrow$ 0.561 & 16 \\
        Southern Mesopotamia & 500 BCE--1800 CE & 1.189 $\rightarrow$ 1.477 & 0.992 $\rightarrow$ 0.895 & 23 \\
        Deccan & 300 BCE--1800 CE & 1.515 $\rightarrow$ 1.470 & 0.410 $\rightarrow$ 0.943 & 7 \\
        Paris Basin & 300 BCE--1800 CE & -3.841 $\rightarrow$ 1.366 & -0.950 $\rightarrow$ 0.798 & 21 \\
        Middle Ganga & 300 BCE--1800 CE & 1.515 $\rightarrow$ 1.674 & 0.410 $\rightarrow$ 0.701 & 8 \\
        Yemeni Coastal Plain & 100 CE--1600 CE & -0.242 $\rightarrow$ 1.560 & -0.058 $\rightarrow$ 0.895 & 11 \\
        Niger Inland Delta & 400 CE--1400 CE & -2.731 $\rightarrow$ -0.073 & -2.157 $\rightarrow$ 0.275 & 10 \\
        Cambodian Basin & 600 CE--1800 CE & -1.646 $\rightarrow$ -0.002 & 0.001 $\rightarrow$ 0.281 & 6 \\
        Cuzco & 700 CE--1700 CE & -0.699 $\rightarrow$ 1.326 & -1.479 $\rightarrow$ 0.798 & 7 \\
        Cahokia & 800 CE--1400 CE & -4.630 $\rightarrow$ -2.923 & -2.883 $\rightarrow$ -2.485 & 7 \\
        Kansai & 800 CE--1800 CE & 0.297 $\rightarrow$ 1.246 & 0.577 $\rightarrow$ 0.992 & 10 \\
        Iceland & 1000 CE--1300 CE & -2.922 $\rightarrow$ -2.580 & -2.302 $\rightarrow$ -0.849 & 4 \\
        Orkhon Valley & 1100 CE--1800 CE & -2.360 $\rightarrow$ 2.264 & -2.399 $\rightarrow$ 0.895 & 3 \\
        North Colombia & 1100 CE--1500 CE & -4.685 $\rightarrow$ -3.451 & -1.172 $\rightarrow$ -1.172 & 5 \\
        Finger Lakes & 1600 CE--1800 CE & -2.387 $\rightarrow$ -2.688 & -2.496 $\rightarrow$ -2.302 & 3 \\
        Lowland Andes & 1600 CE--1800 CE & -5.241 $\rightarrow$ -5.241 & -2.689 $\rightarrow$ -2.689 & 3 \\
        \bottomrule
    \end{tabular}
\end{table}

\clearpage

\begin{table}[!ht]
\centering
\caption{
Seed words defining the Emotionality pole of the semantic axis and the final Emotionality lexicon used in the Google Ngram analysis.
The seed set contains 26 words and is used only to construct the semantic axis. The final lexicon contains 182 lowercase English 1-grams selected using the persistence, lexical-filtering, frequency-coverage, and balancing procedures described in Sec.~\ref{sec:data}.
}
\label{tab:emotionality_lexicon}

\begingroup
\scriptsize
\setlength{\tabcolsep}{2.5pt}
\renewcommand{\arraystretch}{1.05}

\begin{tabular*}{\textwidth}
{@{\extracolsep{\fill}}llllllll@{}}
\toprule
\multicolumn{8}{l}{\textit{Seed words defining the Emotionality pole} ($n=26$)} \\
\addlinespace[2pt]
anger & angry & anxious & ashamed & delightful & depressed & disgusting & embarrassed \\
emotion & excited & fear & feeling & friendly & grateful & happy & hate \\
jealous & joy & lonely & love & pleased & proud & sad & shame \\
surprise & thankful & & & & & & \\
\addlinespace[2pt]
\midrule
\addlinespace[2pt]
\multicolumn{8}{l}{\textit{Final Emotionality lexicon} ($n=182$)} \\
\addlinespace[2pt]
afraid & ah & alarm & alive & always & anxiety & anxious & ask \\
assured & away & badly & beat & beside & bid & bitter & blame \\
boy & breath & broke & brother & burst & busy & catch & caught \\
comfort & confident & crowd & cry & dark & darkness & daughter & dear \\
depressed & desire & distress & disturbed & drink & drove & dying & eat \\
enjoy & ever & everybody & excited & excitement & eyes & face & father \\
fear & feel & feeling & fell & felt & filled & forever & forget \\
forgotten & fortune & friend & friendly & generous & gentle & girl & glad \\
glory & grateful & grew & guest & happy & hear & heart & hearts \\
honest & hope & hurt & husband & ill & innocent & joy & knew \\
ladies & lady & listen & look & looking & looks & lose & love \\
loving & mad & marry & miss & moment & morning & mother & never \\
night & obliged & oh & pain & painful & pale & passion & people \\
personally & pleasant & please & pleased & pleasure & poor & praise & pride \\
proud & quiet & ran & remember & sad & save & saw & say \\
saying & seeing & shut & sick & sight & silence & silent & sing \\
sister & sleep & smile & soon & sorry & spirits & spoke & stay \\
stayed & strange & struck & sudden & suddenly & suffer & suffering & sure \\
surprise & surprised & suspicion & sweet & sympathy & talk & talking & tell \\
telling & tender & terrible & thank & thanks & think & till & tired \\
told & touched & uncle & unto & visit & voice & waiting & watched \\
watching & welcome & widow & wife & wild & wish & wished & woman \\
wonder & worse & wounded & yes & yesterday & youth & & \\
\bottomrule
\end{tabular*}

\endgroup
\end{table}

\begin{table}[!ht]
\centering
\caption{
Seed words defining the Rationality pole of the semantic axis and the final Rationality lexicon used in the Google Ngram analysis. The seed set contains 29 words and is used only to construct the semantic axis. The final lexicon contains 182 lowercase English 1-grams selected using the persistence, lexical-filtering, and frequency-coverage procedures described in Sec.~\ref{sec:data}.
}
\label{tab:rationality_lexicon}

\begingroup
\scriptsize
\setlength{\tabcolsep}{2.5pt}
\renewcommand{\arraystretch}{1.05}

\begin{tabular*}{\textwidth}
{@{\extracolsep{\fill}}llllllll@{}}
\toprule
\multicolumn{8}{l}{\textit{Seed words defining the Rationality pole} ($n=29$)} \\
\addlinespace[2pt]
analysis & biology & chemical & chemistry & conclusion & data & determine & evidence \\
experiment & intellect & limit & logic & mathematical & mathematics & medicine & method \\
objective & philosophy & physics & principle & quantitative & research & science & scientific \\
statics & study & system & theoretical & theory & & & \\
\addlinespace[2pt]
\midrule
\addlinespace[2pt]
\multicolumn{8}{l}{\textit{Final Rationality lexicon} ($n=182$)} \\
\addlinespace[2pt]
academy & according & actual & addition & advanced & agency & aim & analysis \\
annual & application & applied & area & article & arts & association & available \\
based & basis & branch & bureau & central & century & chapter & chart \\
chemical & chemistry & collection & combination & commercial & complex & connection & consistent \\
construction & corporation & critical & data & definition & dental & department & description \\
designed & determine & development & direct & discussion & division & documentary & economy \\
education & electrical & electricity & enforcement & engineering & equipment & essay & established \\
ethics & evidence & evolution & examination & examine & experiment & explanation & extension \\
federal & field & finance & financial & fluid & formula & foundation & function \\
general & graduate & historical & history & included & industry & internal & international \\
investigate & investigation & journal & knowledge & laboratory & law & lecture & legal \\
library & limit & limited & literature & logic & logical & main & management \\
manual & marine & material & maximum & medical & medicine & method & minimum \\
model & modern & natural & nature & necessary & new & objective & observation \\
organization & paper & particular & perspective & phase & philosophy & physical & physics \\
plant & political & practical & practice & primary & principle & principles & problem \\
procedure & process & production & psychological & purpose & range & reference & regarding \\
relative & research & review & sample & scale & scheme & scholarship & science \\
scientific & scope & section & sequence & series & shown & similar & social \\
society & solid & solution & specific & standard & structure & studied & study \\
surgical & system & teaching & technical & test & tested & testing & theory \\
total & training & transmission & type & unit & university & use & used \\
useful & value & variety & various & view & volume & & \\
\bottomrule
\end{tabular*}

\endgroup
\end{table}

\clearpage

\bibliography{main}